\documentclass[runningheads,envcountsame]{llncs}

\makeatletter
\RequirePackage[bookmarks,unicode,colorlinks=true]{hyperref}%
   \def\@citecolor{blue}%
   \def\@urlcolor{blue}%
   \def\@linkcolor{blue}%

\def\orcidID#1{\smash{\href{http://orcid.org/#1}{\protect\raisebox{-1.25pt}{\protect\includegraphics{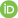}}}}}
\makeatother

\usepackage{xcolor,colortbl}
\usepackage{mathtools}
\usepackage{multirow}
\usepackage{stmaryrd}
\usepackage{cancel}
\usepackage{amsmath,amssymb}
\usepackage{thmtools, thm-restate}
\usepackage[shortlabels]{enumitem}
\usepackage{microtype}
\usepackage{wrapfig}
\usepackage{pbox}
\usepackage{marvosym}
\usepackage{listings}
\usepackage[ruled, vlined]{algorithm2e}
\hypersetup{
  pdftitle={Annotated DPs for Full $\mathtt{AST}$
    of Probabilistic Term
          Rewriting}, colorlinks=true, linkcolor=blue, citecolor=olive, filecolor=magenta,
  urlcolor=cyan
}
\usepackage{todonotes}
\usepackage{placeins}
\usepackage{float}
\usepackage{tikz}
\usepackage{caption}
\usepackage{subcaption}
\usepackage{mymatrix}
\usepackage{array, longtable}
\usepackage{nicefrac,xfrac}
\usetikzlibrary{shapes,calc,arrows,automata,decorations.pathmorphing,backgrounds,arrows.meta,shapes.geometric}

\RequirePackage{makecell}

\usepackage[capitalize,nameinlink]{cleveref}
\usepackage{IEEEtrantools}

\newenvironment{myproof}{
	\noindent{\it Proof.}
}{\qed
	\medskip
}
\def\namedlabel#1#2{\begingroup
    #2%
    \def\@currentlabel{#2}%
    \phantomsection\label{#1}\endgroup
}

\newcommand{\disabledcomment}[1]{}
\newcommand{\oldcomment}[1]{}

\newcommand{\aprove}{\textsf{AProVE}\xspace}

\renewcommand{\emph}[1]{\index{#1}\textit{#1}}

\renewcommand{\emptyset}{\varnothing}

\newcommand{\IN}{\mathbb{N}}
\newcommand{\IR}{\mathbb{R}}

\newcommand{\F}[1]{\mathfrak{#1}}

\makeatletter
\def\moverlay{\mathpalette\mov@rlay}
\def\mov@rlay#1#2{\leavevmode\vtop{%
   \baselineskip\z@skip \lineskiplimit-\maxdimen
   \ialign{\hfil$\m@th#1##$\hfil\cr#2\crcr}}}
\newcommand{\charfusion}[3][\mathord]{
    #1{\ifx#1\mathop\vphantom{#2}\fi
        \mathpalette\mov@rlay{#2\cr#3}
      }
    \ifx#1\mathop\expandafter\displaylimits\fi}
\makeatother

\newcommand{\Var}{\mathcal{V}}

\renewcommand{\P}{\mathcal{P}}

\newcommand{\Proc}{\operatorname{Proc}}

\newcommand{\PosDPoss}{\pos_{\text{Poss}}}

\newcommand{\Junk}{\mathrm{Junk}}

\newcommand{\TSet}[2]{\mathcal{T}\left(#1,#2\right)}

\newcommand{\VSet}{\mathcal{V}}

\newcommand{\R}{\mathcal{R}}

\newcommand{\DPair}[1]{\mathcal{DP}(#1)}

\newcommand{\FDist}{\operatorname{FDist}}
\newcommand{\Supp}{\operatorname{Supp}}

\newcommand{\rootsym}{\operatorname{root}}

\newcommand{\rules}{\operatorname{Rules}}
\newcommand{\urules}{\mathcal{U}}

\newcommand{\Pol}{\operatorname{Pol}}

\newcommand{\nonprob}{\normalfont{\text{np}}}
\newcommand{\nonprobDP}{\normalfont{\text{dp}}}
\newcommand{\PP}{\mathcal{P}}
\newcommand{\InI}{\mathcal{I}}
\newcommand{\JJ}{\mathcal{J}}

\newcommand{\tplus}{\mathsf{plus}}
\renewcommand{\ts}{\mathsf{s}}
\newcommand{\tz}{\mathsf{0}}

\newcommand{\tf}{\mathsf{f}}
\newcommand{\tg}{\mathsf{g}}
\renewcommand{\th}{\mathsf{h}}
\newcommand{\ta}{\mathsf{a}}
\newcommand{\tb}{\mathsf{b}}
\newcommand{\tc}{\mathsf{c}}
\newcommand{\td}{\mathsf{d}}

\newcommand{\tcons}{\mathsf{cons}}
\newcommand{\tnil}{\mathsf{nil}}

\newcommand{\ttrue}{\mathsf{true}}
\newcommand{\tfalse}{\mathsf{false}}

\newcommand{\tD}{\mathsf{D}}
\newcommand{\tF}{\mathsf{F}}
\newcommand{\tG}{\mathsf{G}}

\newcommand{\tA}{\mathsf{A}}

\newcommand{\tapp}{\mathsf{app}}

\newcommand{\tinit}{\mathsf{init}}
\newcommand{\tInit}{\mathsf{Init}}
\newcommand{\tsum}{\mathsf{sum}}
\newcommand{\taddNum}{\mathsf{addNum}}
\newcommand{\tcreateL}{\mathsf{createL}}
\newcommand{\ttree}{\mathsf{tree}}
\newcommand{\tleaf}{\mathsf{leaf}}
\newcommand{\tconcat}{\mathsf{concat}}
\newcommand{\tcreateT}{\mathsf{createT}}
\newcommand{\tlessleaves}{\mathsf{lessleaves}}

\newcommand{\tloopOne}{\mathsf{loop1}}
\newcommand{\tloopTwo}{\mathsf{loop2}}
\newcommand{\tdouble}{\mathsf{double}}
\newcommand{\ttriple}{\mathsf{triple}}

\newcommand{\tLoopOne}{\mathsf{L1}}
\newcommand{\tLoopTwo}{\mathsf{L2}}
\newcommand{\tDouble}{\mathsf{D}}
\newcommand{\tTriple}{\mathsf{T}}

\newcommand{\xs}{\mathit{xs}}
\newcommand{\ys}{\mathit{ys}}

\newcommand{\ctleaf}{\mathtt{Leaf}}

\newcommand{\val}{\mathit{\mathcal{S}\!um}}

\crefname{definition}{Def.}{Def.}
\crefname{example}{Ex.}{Ex.}
\crefname{counterexample}{Counterex.}{Counterex.}
\crefname{appendix}{App.}{App.}
\crefname{ex}{Ex.}{Ex.}
\crefname{theorem}{Thm.}{Thm.}
\crefname{lemma}{Lemma}{Lemmas}
\crefname{remark}{Remark}{Remarks}
\crefname{section}{Sect.}{Sect.}
\crefname{subsection}{Sect.}{Sect.}
\crefname{subsubsection}{Sect.}{Sect.}
\crefname{line}{Line}{Lines}
\crefname{corollary}{Cor.}{Cor.}
\crefname{figure}{Fig.}{Fig.}
\crefname{enumi}{}{}
\crefname{algorithm}{Alg.}{Alg.}

\makeatletter
\NewDocumentCommand{\dparrow}{+O{} +O{0.5cm}}{%
\begin{tikzpicture}[baseline=-0.5ex] {
\node[inner sep=0](@1) at (0,0) {};
\node[inner sep=0](@2) at (#2,0) {};
\draw [arrows={-Triangle[open]},shorten >= 1pt,shorten <= 1pt](@1)--(@2) node[pos=.5,above,inner sep=1pt] {\ensuremath{#1}};}
\end{tikzpicture}\xspace
}

\NewDocumentCommand{\myto}{+O{} +O{0.5cm}}{%
\begin{tikzpicture}[baseline=-0.5ex] {
\node[inner sep=0](@1) at (0,0) {};
\node[inner sep=0](@2) at (#2,0) {};
\draw [arrows={-to}](@1)--(@2) node[pos=.5,above,inner sep=1pt] {\ensuremath{#1}};}
\end{tikzpicture}\xspace
}

\NewDocumentCommand{\paraarrow}{+O{} +O{0.4cm}}{%
\begin{tikzpicture}[baseline=-0.5ex] {
\node[inner sep=0](@1) at (0,0) {};
\node[inner sep=0](@2) at (#2,0) {};
\node[inner sep=0](@3) at (0.07,0) {};
\draw [arrows={-to}](@1)--(@2) node[pos=.5,above,inner sep=1pt] {\ensuremath{#1}};
\draw [arrows={-to}](@1)--(@3);}
\end{tikzpicture}\xspace
}

\NewDocumentCommand{\paradparrow}{+O{} +O{0.4cm}}{%
\begin{tikzpicture}[baseline=-0.5ex] {
\node[inner sep=0](@1) at (0,0) {};
\node[inner sep=0](@2) at (#2,0) {};
\node[inner sep=0](@3) at (0.07,0) {};
\draw [arrows={-Triangle[open]}](@1)--(@2) node[pos=.5,above,inner sep=1pt] {\ensuremath{#1}};
\draw [arrows={-to}](@1)--(@3);}
\end{tikzpicture}\xspace
}

\newcommand{\oset}[2]{%
  {\mathop{#2}\limits^{\vbox to 1\ex@{\kern-\tw@\ex@
   \hbox{\scriptsize #1}\vss}}}}

\newcommand{\osetthree}[2]{%
  {\mathop{#2}\limits^{\vbox to 3\ex@{\kern-\tw@\ex@
   \hbox{\scriptsize #1}\vss}}}}

\newcommand{\osetfive}[2]{%
  {\mathop{#2}\limits^{\vbox to 5\ex@{\kern-\tw@\ex@
   \hbox{\scriptsize #1}\vss}}}}

\newcommand{\osetminus}[2]{%
  {\mathop{#2}\limits^{\vbox to -2\ex@{\kern-\tw@\ex@
   \hbox{\scriptsize #1}\vss}}}}
\makeatother

\newcommand{\ito}{\mathrel{\smash{\stackrel{\raisebox{3.4pt}{\tiny $\mathsf{i}\:$}}{\smash{\rightarrow}}}}}

\newcommand{\itorstar}{\mathrel{\ito_{\R}^{*}}}

\newcommand{\itor}{\mathrel{\ito_{\R}}}

\newcommand{\itodr}{\mathrel{\ito_{\mathcal{D},\R}}}

\newcommand{\fs}[1]{\mathsf{#1}}
\newcommand{\fun}[1]{\mathrm{#1}}
\renewcommand{\phi}{\varphi}
\renewcommand{\emptyset}{\varnothing}
\newcommand{\true}{\fs{true}}
\newcommand{\false}{\fs{false}}
\newcommand{\pos}{\fun{Pos}}
\newcommand{\posT}{\fun{Pos}_{\SignatureA}}
\newcommand{\posD}{\fun{Pos}_{\SignatureD}}
\newcommand{\posDT}{\fun{Pos}_{\SignatureD \cup \SignatureA}}
\newcommand{\anno}{\#} %{fun{\mathcal{A}}}
\newcommand{\annoEps}{\anno_{\varepsilon}}
\newcommand{\annoD}{\anno_{\SignatureD}}
\newcommand{\disannoPos}[1]{\flat_{#1}^{\uparrow}}
\newcommand{\TT}{\mathcal{T}}
\newcommand{\SignatureADC}{\Sigma^\#}
\newcommand{\SignatureC}{\mathcal{C}}
\newcommand{\SignatureD}{\mathcal{D}}
\newcommand{\SignatureA}{\mathcal{D}^\#}
\newcommand{\NF}{\mathtt{NF}}
\newcommand{\ANF}{\mathtt{ANF}}
\newcommand{\NN}{\mathbb{N}}
\newcommand{\ruleArr}[3]{
  \mathrel{
    \xrightarrow{{}_{\scriptstyle #1}}
    \!\!{}^{#2}_{#3}
  }
}
\newcommand{\tored}[3]{
  \mathrel{
    \xhookrightarrow{{}_{\scriptstyle #1}}
    \!\!{}^{#2}_{#3}
  }
}
\newcommand{\itored}[3]{
  \mathrel{
    \smash{\stackrel{\raisebox{3.4pt}{\tiny $\mathsf{i}\:$}}{\smash{\hookrightarrow}}}^{#2}_{#3}
  }
}

\newcommand{\defemph}[1]{{\rm #1}}

\SetKwInOut{Input}{input}
\SetKwInOut{Output}{output}

\definecolor{Gray}{gray}{0.85}
\definecolor{LightCyan}{rgb}{0.88,1,1}

\newcolumntype{a}{>{\columncolor{Gray}}c}
\newcolumntype{b}{>{\columncolor{white}}c}

 \newcommand{\makeproof}[2]{}
 \newcommand{\paper}[1]{}
 \newcommand{\report}[1]{#1}
  \newcommand{\final}[1]{\paper{#1}}

 \report{
  \setlength{\textwidth}{125mm}
  \setlength{\textheight}{198mm}
}

 \newcommand{\notnew}[1]{}
 
\paper{\usepackage[ hyperref=true, backend=bibtex, firstinits=true, maxbibnames=99, sortcites, style=numeric-comp ]{biblatex}
\addbibresource{biblio.bib}}

\title{Annotated Dependency Pairs for\\Full
Almost-Sure Termination of\\Probabilistic Term Rewriting\thanks{funded by the
  DFG Research Training Group 2236 UnRAVeL}}
\titlerunning{Annotated DPs for Full $\mathtt{AST}$
    of Probabilistic Term
          Rewriting}
\author{Jan-Christoph Kassing\final{$^{(\href{mailto:kassing@cs.rwth-aachen.de}{\mbox{\Letter}})}$}\orcidID{0009-0001-9972-2470} \and Jürgen Giesl\final{$^{(\href{mailto:giesl@cs.rwth-aachen.de}{\mbox{\Letter}})}$}\orcidID{0000-0003-0283-8520}}
\institute{RWTH Aachen University, Aachen, Germany\\
 \email{\{kassing,giesl\}@cs.rwth-aachen.de}}
\authorrunning{J.-C.\ Kassing, J.\ Giesl}

\begin{document}
\allowdisplaybreaks

\maketitle \begin{abstract}
	Dependency pairs (DPs) are one of the most powerful techniques for automated termination
    analysis of term rewrite systems.
    Recently, we adapted the DP framework to the probabilistic setting 
    to prove almost-sure termination ($\mathtt{AST}$) via annotated DPs (ADPs).
    However, this adaption only handled $\mathtt{AST}$ w.r.t.\
    the \emph{innermost} eval\-uation strategy. 
    In this paper, we improve the ADP framework to prove $\mathtt{AST}$ for \emph{full}
    rewriting. Moreover, we refine the framework for
    rewrite sequences that start with \emph{basic} terms containing
    a single defined function symbol.
    We implemented and evaluated the new framework in our tool \aprove.
\end{abstract}

\section{Introduction}\label{sec-introduction}

Term rewrite systems (TRSs)
are used for automated termination analysis of many programming languages. 
There  exist numerous powerful tools to prove termination of TRSs, e.g.,
\cite{JAR-AProVE2017,ttt2_sys,natt_sys_2014,gutierrez_mu-term_2020}.
Dependency pairs (DPs, see e.g.,
\cite{arts2000termination,gieslLPAR04dpframework,giesl2006mechanizing,hirokawa2005automating,DBLP:journals/iandc/HirokawaM07})
are one of the main concepts used in all these tools.

In \cite{BournezRTA02,bournez2005proving,avanzini2020probabilistic,Faggian2019ProbabilisticRN},
TRSs were extended to the probabilistic setting.
Probabilistic programs describe randomized algorithms and probability
distributions, with applications in many areas, see, e.g., \cite{Gordon14}.
Instead of only considering ordinary termination (i.e., absence of 
infinite evaluation sequences),
in the probabilistic setting one is interested in \emph{almost-sure termination} ($\mathtt{AST}$), 
where infinite evaluation sequences are allowed, but their probability is $0$.
A strictly stronger notion is \emph{positive} $\mathtt{AST}$ ($\mathtt{PAST}$), which requires that
the expected runtime is finite \cite{bournez2005proving,DBLP:conf/mfcs/Saheb-Djahromi78}.

There exist numerous techniques to prove $\mathtt{(P)AST}$ of
imperative programs on numbers (like the probabilistic guarded command language
\textsf{pGCL} \cite{Kozen85,McIverMorgan05}), e.g.,
\cite{kaminski2018weakest,mciver2017new,TACAS21,lexrsm,FoundationsTerminationMartingale2020,FoundationsExpectedRuntime2020,rsm,cade19,dblp:journals/pacmpl/huang0cg19,amber,ecoimp,absynth}. In contrast,
\emph{probabilistic TRSs} (PTRSs) are especially suitable for modeling and analyzing functional programs and
algorithms operating on (user-defined) 
data structures like lists, trees, etc. Up to now,
there exist only few automatic approaches to analyze $\mathtt{(P)AST}$ of probabilistic programs with complex non-tail recursive
structure \cite{beutner2021probabilistic,Dallago2017ProbSizedTyping,lago_intersection_2021}. The
approaches that are suitable for algorithms on recursive data structures
\cite{wang2020autoexpcost,LeutgebCAV2022amor,KatoenPOPL23}
are mostly specialized for specific data structures and cannot easily be adjusted to
other (possibly user-defined) ones, or are not yet fully automated.
In contrast, \pagebreak[3] our goal is
a fully automatic termination analysis for arbitrary PTRSs.

For PTRSs,  orderings based on interpretations were adapted to prove
$\mathtt{PAST}$ of\linebreak \emph{full} rewriting (w.r.t.\ any evaluation strategy)
in \cite{avanzini2020probabilistic},
and we presented a related\linebreak technique to prove $\mathtt{AST}$ in 
\cite{kassinggiesl2023iAST}.
However, already for non-probabilistic TRSs, such
a direct application of orderings is limited in power. 
To obtain a powerful approach, one should
combine orderings in a modular way, as in the DP framework.

Indeed, based on initial work in \cite{kassinggiesl2023iAST},
in \cite{FLOPS2024} we adapted the DP framework to the probabilistic setting to prove
innermost AST ($\mathtt{iAST}$) of PTRSs
via so-called \emph{annotated dependency pairs} (ADPs).
However, this adaption is restricted to \emph{innermost} rewriting, 
i.e., one only considers sequences that rewrite at innermost positions of terms.
Already for non-probabilistic TRSs, innermost termination is easier\linebreak 
to prove than \emph{full} termination,
and this remains true in the probabilistic setting.

\begin{wrapfigure}[14]{r}{0.29\textwidth}
    \begin{minipage}{0.29\textwidth}
        \vspace*{-0.8cm}
        \hspace*{-.1cm}%
        \begin{algorithm}[H]
            \DontPrintSemicolon
            \SetInd{-.3cm}{.3cm}
            \DecMargin{1cm}
            \caption{}
            \label{alg1}
            \hspace*{-.5cm}   $x \gets 0$\;
            \hspace*{-.5cm}   \While{$x = 0$}
            {
                $\{$\;
                $\phantom{\{} x \gets 0 \oplus_{\nicefrac{1}{2}} x \gets 1;\hspace*{-3.8cm}$ \hspace*{-3.8cm}\; 
                $\phantom{\{} y \gets 2 \cdot y;$\;
                $\} \square \{$\;
                $\phantom{\{} x \gets 0 \oplus_{\nicefrac{1}{3}} x \gets 1;$\hspace*{-3.8cm}\; 
                $\phantom{\{} y \gets 3 \cdot y;$\;
        
                $\}$
            }
            \hspace*{-.5cm} \While{$y > 0$}
            {
                $y \gets y - 1;$\;
            }
        \end{algorithm}
    \end{minipage}
\end{wrapfigure}
In the current paper, we adapt the definition of 
ADPs to use them for any evaluation strategy.
As our running example, we transform \Cref{alg1}  on the right (written in \textsf{pGCL}) into
an equivalent PTRS and show how\linebreak our new ADP framework proves $\mathtt{AST}$.
Here, $\oplus_{\nicefrac{1}{2}}$ denotes probabilistic choice, and $\square$ denotes demonic non-determinism.
Note that there are proof rules (e.g., \cite{mciver2017new}) and tools (e.g., \cite{amber}) that can
prove $\mathtt{AST}$ for both loops of \cref{alg1} individually, and hence for the whole algorithm.
Moreover, the tool \textsf{Caesar} \cite{DBLP:journals/pacmpl/SchroerBKKM23}
can prove $\mathtt{AST}$ if one provides
super-martingales for the two loops. However, to the best of our knowledge
there exist no automatic techniques to handle similar algorithms on
arbitrary algebraic data structures, i.e., (non-deterministic) algorithms 
that first create a random data object $y$ in a first loop 
and then access or modify it in a second loop, 
whereas this is possible with our new ADP framework.\footnote{Such examples can be found in our benchmark set, 
see \Cref{Evaluation} and \Cref{Examples}.} 
Note that while \Cref{alg1} is $\mathtt{AST}$, its expected runtime is
infinite, i.e., it is not $\mathtt{PAST}$.\footnote{This already holds for the program
where only the first possibility of the first \textbf{while}-loop is considered (i.e., where $y$
is always doubled in its body). Then for the
initial value $y = 1$, the expected number of
iterations of the second \textbf{while}-loop which decrements $y$ is $\tfrac{1}{2} \cdot
2 + \tfrac{1}{4} \cdot 4 + \tfrac{1}{8} \cdot 8 + \ldots = 1 + 1 + 1 + \ldots = \infty$.} 

In \cite{FoSSaCS2024}, we developed the first criteria for classes of PTRSs where $\mathtt{iAST}$ implies
$\mathtt{AST}$. So for PTRSs from these classes, one can use our ADP framework for $\mathtt{iAST}$ in order to
conclude $\mathtt{AST}$. However, these criteria exclude non-probabilistic
non-determinism,
i.e., they require that the rules of the PTRS must be non-overlapping. In addition, they impose
linearity restrictions on both sides of the rewrite rules.
In contrast, our novel ADP framework can be applied to overlapping PTRSs and
it also weakens the linearity requirements considerably.

We start with preliminaries on (probabilistic) term rewriting in \cref{Probabilistic Term Rewriting}.
In \cref{Probabilistic Annotated Dependency Pairs} we recapitulate annotated dependency pairs 
for innermost $\mathtt{AST}$ \cite{FLOPS2024},
explain why they cannot prove $\mathtt{AST}$ for \emph{full} rewriting, and adapt them accordingly.
We present the probabilistic ADP framework in \cref{The Probabilistic ADP Framework}, 
illustrate its main processors, and show
how to adapt them from $\mathtt{iAST}$ to $\mathtt{AST}$.
Finally, in \cref{Evaluation} we evaluate the implementation 
of our approach in the tool
\aprove{} \cite{JAR-AProVE2017}. 
We refer to \Cref{Examples} to illustrate our approach on examples with non-numerical data structures
like lists or trees, and to\report{ \Cref{appendix}}\paper{ \cite{report}} for all proofs.

\section{Preliminaries}\label{Probabilistic Term Rewriting}

\Cref{sect:Term Rewriting,sect:Probabilistic Term Rewriting,sect:Existing Techniques for Proving Full AST} recapitulate classic \cite{baader_nipkow_1999} 
and probabilistic 
\cite{avanzini2020probabilistic,BournezRTA02,bournez2005proving,kassinggiesl2023iAST} term
rewrit\-ing, and
results on PTRSs where
$\mathtt{iAST}$ and $\mathtt{AST}$ are equivalent, respectively.

\subsection{Term Rewriting}\label{sect:Term Rewriting}

We regard a (finite) signature $\Sigma = \biguplus_{n \in \IN} \Sigma_n$ and a set of variables $\VSet$.
The set of \emph{terms} $\TSet{\Sigma}{ \VSet}$ (or simply $\TT$) is the smallest set with $\VSet \subseteq \TSet{\Sigma}{ \VSet}$, 
and if $f \in \Sigma_n$ and $t_1, \dots, t_n \in \TSet{\Sigma}{ \VSet}$ then $f(t_1,\dots,t_n) \in \TSet{\Sigma}{ \VSet}$.
We say that $s$ is a \emph{subterm}
of $t$ (denoted $s \trianglelefteq t$) if $s = t$, or $t = f(t_1, \ldots, t_n)$ and $s \trianglelefteq t_i$ for some $1 \leq i \leq n$.
It is a \emph{proper} subterm (denoted $s \vartriangleleft t$) if $s \trianglelefteq t$ and $s \neq t$.
A \emph{substitution} is a function $\sigma:\VSet \to \TT$ with $\sigma(x) = x$ for all but finitely many $x \in \VSet$.
We often write $x\sigma$ instead of $\sigma(x)$.
Substitutions can also be applied to terms: If $t=f(t_1,\dots,t_n)\in \TT$ then $t \sigma
= f(t_1\sigma,\dots,t_n \sigma)$.
For a term $t \in \TT$, the set of \emph{positions} $\pos(t)$ 
is the smallest subset of $\IN^*$ satisfying $\varepsilon \in \pos(t)$, 
and if $t=f(t_1,\dots,t_n)$ then for all $1 \leq i \leq n$ and all $\pi \in \pos(t_i)$ we have $i.\pi \in \pos(t)$.
If $\pi \in \pos(t)$ then $t|_{\pi}$ denotes the subterm at position $\pi$,
where we have $t|_\varepsilon = t$ for the  \emph{root position} $\varepsilon$ and
$f(t_1,\dots,t_n)|_{i.\pi} = t_i|_\pi$. The \emph{root symbol} at position $\varepsilon$
is also denoted by  $\rootsym(t)$.
If $r \in \TT$ and $\pi \in \pos(t)$ then $t[r]_{\pi}$ denotes the term that results from replacing the subterm $t|_{\pi}$ with the term $r$.

A \emph{rewrite rule} $\ell \to r \in
\TT \times \TT$ is a pair
with $\ell \not\in \VSet$ and $\VSet(r) \subseteq \VSet(\ell)$.
A \emph{term rewrite system} (TRS) is
a (finite) set of rewrite rules.
For example, $\R_{\td}$ with the only rule $\td(x) \to \tc(x,x)$ is a TRS.
A TRS $\R$ induces a \emph{rewrite rela\-tion} ${\to_{\R}} \subseteq \TT
\times \TT$ where $s \to_{\R} t$ holds if there are an $\ell \to r \in \R$, a substitution $\sigma$, and a $\pi \in
\pos(s)$
such that $s|_{\pi}=\ell\sigma$
and $t = s[r\sigma]_{\pi}$.
A term $s$ is in \emph{normal form} w.r.t.\ $\R$
(denoted $s \in \NF_{\R}$)
if there is no term $t$ with $s \to_{\R} t$, and in \emph{argument normal form} w.r.t.\ $\R$
(denoted $s \in \ANF_{\R}$)
if $s' \in \NF_{\R}$ for all proper subterms $s' \vartriangleleft s$.
A rewrite step $s \to_{\R} t$ is \emph{innermost} (denoted $s \itor t$) if the used \emph{redex} $\ell \sigma$
is in argument normal form.
For example,  $\td(\td(\tz)) \ito_{\R_{\td}} \td(\tc(\tz,\tz))$, but
$\td(\td(\tz)) \to_{\R_{\td}} \tc(\td(\tz),\td(\tz))$ is not an innermost step.
A TRS $\R$ is \emph{(innermost) terminating} if ($\itor$) $\to_{\R}$ is well founded.

Two rules $\ell_1 \to r_1, \ell_2 \to r_2 \in \R$ with renamed variables such that
$\VSet(\ell_1) \cap \VSet(\ell_2)\linebreak = \emptyset$ are \emph{overlapping}
if there exists a non-variable position $\pi$ of $\ell_1$ such that $\ell_1|_{\pi}$ and
$\ell_2$ are unifiable, i.e., there exists a substitution $\sigma$ such that $\ell_1|_{\pi} \sigma = \ell_2 \sigma$.
If $(\ell_1 \to r_1) = (\ell_2 \to r_2)$, then we require that $\pi \neq \varepsilon$.
$\R$ is \emph{non-overlapping} if it has no overlapping rules (e.g.,
$\R_{\td}$ is non-overlapping).
A TRS is \emph{left-linear} (\emph{right-linear}) if
every variable occurs at most once in the left-hand side (right-hand side) of a rule.
Finally, a TRS is \emph{non-duplicating} if for every rule,
every variable occurs at most as often in the right-hand side as in the left-hand side.
As an example, $\R_{\td}$ is left-linear, not right-linear, and hence duplicating.

\subsection{Probabilistic Term Rewriting}\label{sect:Probabilistic Term Rewriting}

In contrast to TRSs,
a \emph{probabilistic TRS} (PTRS)
\cite{avanzini2020probabilistic,BournezRTA02,bournez2005proving,kassinggiesl2023iAST}
has (finite) multi-distributions on the right-hand sides of its rewrite rules.
A finite \emph{multi-distribution} $\mu$ on a set $A \neq \emptyset$ is a finite multiset
of pairs $(p:a)$, where $0 < p \leq 1$ is a probability and $a \in A$, such that  $\sum
_{(p:a) \in \mu} \, p = 1$.
$\FDist(A)$ is the set of all finite multi-distributions on $A$.
For $\mu\in\FDist(A)$, its \emph{support} is the multiset $\Supp(\mu)\!=\!\{a \mid (p\!:\!a)\!\in\!\mu$ for some $p\}$.
A \emph{probabilistic rewrite rule} $\ell \to \mu \in
\TT \times \FDist(\TT)$ is a pair such that
$\ell \not\in \VSet$ and 
$\VSet(r) \subseteq \VSet(\ell)$ for every $r \in \Supp(\mu)$.
Examples for probabilistic rewrite rules are

\vspace*{-.6cm}

\hspace*{-.8cm}\begin{minipage}[t]{4.4cm}
    \begin{align}
        \tg &\to \{\nicefrac{3}{4}:\td(\tg), \, \nicefrac{1}{4}:\tz\} \label{rule-01}
    \end{align}
\end{minipage}
\hspace*{.1cm}
\begin{minipage}[t]{4.2cm}
    \begin{align}
      \td(x) &\to \{1:\tc(x,x)\} \label{rule-02}\\
        \td(\td(x)) &\to \{1:\tc(x,\tg)\} \label{rule-02nd}
    \end{align}
\end{minipage}
\hspace*{.1cm}
\begin{minipage}[t]{3.3cm}
  \begin{align}
 \nonumber   \\
        \td(x) &\to \{1:\tz\} \label{rule-03}
    \end{align}
\end{minipage}

\vspace*{.15cm}

\noindent
A \emph{probabilistic TRS} is a finite set $\R$ of probabilistic rewrite rules,
e.g., $\R_1 = \{\eqref{rule-01}\}$, $\R_2 = \{\eqref{rule-01}, \eqref{rule-02}\}$, or
$\R_3 = \{\eqref{rule-01}, \eqref{rule-02nd}, \eqref{rule-03}\}$. 
Similar to TRSs, a PTRS $\R$ induces a \emph{rewrite relation}
${\to_{\R}} \subseteq \TT \times \FDist(\TT)$
where 	$s \to_{\R} \{p_1:t_1, \ldots, p_k:t_k\}$  if there are
an $\ell \to \{p_1:r_1, \ldots, p_k:r_k\} \in \R$,
a substitution $\sigma$,
and a $\pi \in\linebreak \pos(s)$
such that $s|_{\pi}=\ell\sigma$ and $t_j = s[r_j\sigma]_{\pi}$ for all $1 \leq j \leq k$.
The step is \emph{inner\-most} (denoted $s \itor \{p_1:r_1, \ldots, p_k:r_k\}$) 
if $\ell \sigma \in \ANF_{\R}$.
So the PTRS $\R_1$ can be\linebreak interpreted as a biased coin flip that terminates 
in each step with a chance of $\nicefrac{1}{4}$.

To track all possible rewrite sequences (up to non-determinism) with their corresponding
probabilities,
as in \cite{FLOPS2024}
we \emph{lift} $\to_{\R}$ to \emph{rewrite sequence trees (RSTs)}.
The nodes $v$ of an $\R$-RST are labeled by pairs $(p_v:t_v)$ of a
probability $p_v$ and a term $t_v$, where
the root is always labeled with the probability $1$.
For each node $v$ with the successors $w_1, \ldots, w_k$, the edge relation represents a probabilistic rewrite step, 
i.e., $t_v \to_{\R} \{\tfrac{p_{w_1}}{p_v}:t_{w_1}, \ldots, \tfrac{p_{w_k}}{p_v}:t_{w_k}\}$. 
An  $\R$-RST is an \emph{innermost} $\R$-RST if the edge relation represents only innermost steps.
For an $\R$-RST $\F{T}$ we define $|\F{T}|_{\ctleaf} = \sum_{v \in \ctleaf} \, p_v$, 
where $\ctleaf$ is the set of all its leaves, and we say that a PTRS $\R$ is 
\emph{almost-surely terminating ($\mathtt{AST}$)} (\emph{almost-surely innermost terminating ($\mathtt{iAST}$)}) if $|\F{T}|_{\ctleaf} = 1$ holds for all
$\R$-RSTs (innermost $\R$-RSTs) $\F{T}$.
While $|\F{T}|_{\ctleaf} = 1$ for every finite RST $\F{T}$, 
for infinite RSTs $\F{T}$ we may have  $|\F{T}|_{\ctleaf} <1$ or
even $|\F{T}|_{\ctleaf} = 0$ if  $\F{T}$\paper{ has no leaf at all.}
\begin{wrapfigure}[6]{r}{0.38\textwidth}
    \scriptsize
    \vspace*{-0.75cm}
          \begin{tikzpicture}
          \tikzstyle{adam}=[thick,draw=black!100,fill=white!100,minimum size=4mm, shape=rectangle split, rectangle split parts=2,rectangle split horizontal]
          \tikzstyle{empty}=[rectangle,thick,minimum size=4mm]
          
          \node[adam] at (-4, 0)  (a) {$1$\nodepart{two}$\tg$};
          \node[adam] at (-5, -0.7)  (b) {$\nicefrac{3}{4}$\nodepart{two}$\td(\tg)$};
          \node[adam,label=below:{$\quad \mathtt{NF}_{\R_{1}}$}] at (-3, -0.7)  (c) {$\nicefrac{1}{4}$\nodepart{two}$\tz$};
          \node[adam] at (-6, -1.4)  (d) {$\nicefrac{9}{16}$\nodepart{two}$\td(\td(\tg))$};
          \node[adam,label=below:{$\mathtt{NF}_{\R_{1}}$}] at (-4, -1.4)  (e) {$\nicefrac{3}{16}$\nodepart{two}$\td(\tz)$};
          \node[empty] at (-6.5, -2)  (f) {$\ldots$};
          \node[empty] at (-5.5, -2)  (g) {$\ldots$};
          
          \draw (a) edge[->] (b);
          \draw (a) edge[->] (c);
          \draw (b) edge[->] (d);
          \draw (b) edge[->] (e);
          \draw (d) edge[->] (f);
          \draw (d) edge[->] (g);
          \end{tikzpicture}
     \end{wrapfigure}  
\report{has no leaf at all. }This notion of $\mathtt{AST}$ is equivalent to
the ones in \cite{bournez2005proving,avanzini2020probabilistic,kassinggiesl2023iAST}, 
where $\mathtt{AST}$ is defined via a lifting of $\to_{\R}$
to multisets or via stochastic processes.
The infinite $\R_1$-RST $\F{T}$ on the side
has $|\F{T}|_{\ctleaf} = 1$.
As this holds for all $\R_1$-RSTs, $\R_1$ is $\mathtt{AST}$.

\begin{example}\label{example:running-1}
    $\R_2$ is not $\mathtt{AST}$.
    If we always apply $\eqref{rule-02}$
    directly after $\eqref{rule-01}$, this corresponds to
    the rule 
    $\tg \to \{\nicefrac{3}{4}:\tc(\tg, \tg), \nicefrac{1}{4}:\tz\}$,
    which represents
    a random walk on the number of $\tg$'s in a term biased towards non-termination (as
    $\tfrac{3}{4} > \tfrac{1}{4}$).
    $\R_3$ is not $\mathtt{AST}$ either, because if we always apply  \eqref{rule-02nd}
    after two applications of \eqref{rule-01}, this corresponds to 
     $\tg \to \{\nicefrac{9}{16}:\tc(\tg,\tg), \; \nicefrac{3}{16}:\tz, \; \nicefrac{1}{4}:\tz \}$,
    which is also biased towards non-termination (as
    $\tfrac{9}{16} > \tfrac{3}{16} + \tfrac{1}{4}$).

    However, in innermost evaluations,
    the $\td$-rule \eqref{rule-02} can only duplicate normal forms, and hence $\R_2$ is
    $\mathtt{iAST}$, see \cite{FoSSaCS2024}.
    $\R_3$ is $\mathtt{iAST}$ as well, as \eqref{rule-02nd} 
    is not applicable in\linebreak  innermost evaluations.
    For both $\R_2$ and $\R_3$, $\mathtt{iAST}$
    can also be proved automatical\-ly by 
    our implementation of the ADP
    framework for $\mathtt{iAST}$ in \aprove{} \cite{kassinggiesl2023iAST,FLOPS2024}.
\end{example}

\begin{example}\label{example:running-2}
    The following PTRS $\R_{\textsf{alg}}$ corresponds to \Cref{alg1}. 
    Here, the non-deter\-minism is modeled by the non-deterministic choice 
    between the overlapping rules
    \eqref{loopOne-1} and \eqref{loopOne-2}.
    In \Cref{The Probabilistic ADP Framework}, we will prove that $\R_{\textsf{alg}}$ is $\mathtt{AST}$ via our
    new notion of ADPs.

    \vspace*{-.3cm}
    
    {\small
    \begin{align}
        \tloopOne(y) &\to \{\nicefrac{1}{2}:\tloopOne(\tdouble(y)), \;
        \nicefrac{1}{2}:\tloopTwo(\tdouble(y))\} \label{loopOne-1}\\
        \tloopOne(y) &\to \{\nicefrac{1}{3}:\tloopOne(\ttriple(y)), \;
        \nicefrac{2}{3}:\tloopTwo(\ttriple(y))\}\! \label{loopOne-2}
    \end{align}
    \begin{minipage}[t]{6cm}
        \vspace*{-0.8cm}
        \begin{align*}
            \tloopTwo(\ts(y)) &\to \{1:\tloopTwo(y)\}\\
            \tdouble(\ts(y)) &\to \{1:\ts(\ts(\tdouble(y)))\}\\
            \tdouble(\tz) &\to \{1:\tz\}\!
        \end{align*}
    \end{minipage}
    \begin{minipage}[t]{6cm}
        \vspace*{-0.6cm}
        \begin{align*}
            \ttriple(\ts(y)) &\to \{1:\ts(\ts(\ts(\ttriple(y))))\}\\
            \ttriple(\tz) &\to \{1:\tz\}\!
        \end{align*}
    \end{minipage}}
\end{example}

A PTRS $\R$ is \emph{right-linear} iff the TRS
$\{\ell \to r \mid \ell \to \mu \in \R, r \in \Supp(\mu)\}$
is right-linear.
Left-linearity and being non-overlapping can be lifted to PTRSs directly,
as their rules also have just a single term on their left-hand sides.

For a PTRS $\R$, we decompose its signature $\Sigma = \SignatureC \uplus \SignatureD$ such
that $f \in \SignatureD$ iff $f = \rootsym(\ell)$ for some $\ell \to \mu \in \R$. 
The symbols in $\SignatureC$ and $\SignatureD$ are called \emph{constructors} and
\emph{defined symbols}, respectively. 
For $\R_2$ we have $\SignatureC = \{\tc, \tz\}$ and $\SignatureD = \{\tg, \td\}$.
A term $t \in \TT$ is \emph{basic} if $t = f(t_1, \ldots, t_n)$ with
$f \in \SignatureD$ and $t_i \in \TSet{\SignatureC}{\VSet}$ for all $1 \leq i \leq n$.
So a basic term represents an algorithm $f$ applied to arguments $t_i$
which only represent data and do not contain executable functions.

Finally, we define \emph{spareness} \cite{frohn_analyzing_nodate}, 
which prevents the duplication of redexes if the evaluation starts with a basic term.
A rewrite step $\ell\sigma \to_\R \mu\sigma$ is \emph{spare}
if $\sigma(x) \in \NF_\R$ for ev\-ery $x \in \Var$ that occurs more than once in some $r \in \Supp(\mu)$.
An $\R$-RST is spare if all rewrite steps corresponding to its edges are spare.
A PTRS $\R$ is spare if each $\R$-RST that starts with $\{1 : t\}$ for a basic term $t$ is spare.
So for example, $\R_2$ is not spare, because the basic term $\tg$ starts a rewrite sequence where
the redex $\tg$ is duplicated by Rule \eqref{rule-02}.
Computable sufficient conditions for
spareness were presented in \cite{frohn_analyzing_nodate}.

\subsection{Existing Techniques for Proving Full $\mathtt{AST}$}\label{sect:Existing Techniques for Proving Full AST}

In order to prove $\mathtt{AST}$ automatically, one can either use orderings directly
on the whole PTRS \cite{avanzini2020probabilistic,kassinggiesl2023iAST}, or check
whether the PTRS $\R$ belongs to a class where it is known
that $\R$ is $\mathtt{AST}$ iff $\R$ is $\mathtt{iAST}$. 
Then, it suffices to analyze $\mathtt{iAST}$,
and to this end,
one can use the existing ADP framework \cite{FLOPS2024}.
In \cite{FoSSaCS2024}, we introduced the following first criteria for classes of PTRSs
where $\mathtt{iAST}$ is equivalent to $\mathtt{AST}$.

\begin{restatable}[From $\mathtt{iAST}$ to $\mathtt{AST}$ (1) \cite{FoSSaCS2024}]{theorem}{ASTAndIASTPropertyOne}\label{properties-eq-AST-iAST-1}
    If a PTRS $\R$ is non-overlapping, left-linear, and right-linear, 
    then $\R$ is $\mathtt{AST}$ iff $\R$ is $\mathtt{iAST}$.
\end{restatable}

Moreover, if one restricts the analysis to basic start terms, 
then we can weaken right-linearity to spareness.
In the following, ``$\mathtt{b(i)AST}$'' (basic $\mathtt{(i)AST}$) means that 
one only considers rewrite sequences that start with
$\{1:t\}$ for basic terms $t$. 

\begin{restatable}[From $\mathtt{iAST}$ to $\mathtt{AST}$ (2) \cite{FoSSaCS2024}]{theorem}{ASTAndIASTPropertyTwo}\label{properties-eq-AST-iAST-2}
    If a PTRS $\R$ is non-overlapping, left-linear, and spare, 
    then $\R$ is $\mathtt{bAST}$ iff \pagebreak[3] $\R$ is $\mathtt{biAST}$.
\end{restatable}

Since $\mathtt{iAST}$ obviously implies
$\mathtt{biAST}$, under the conditions of \Cref{properties-eq-AST-iAST-2}
it suffices to analyze $\mathtt{iAST}$ to prove
$\mathtt{bAST}$.\footnote{Instead of restricting start terms to
basic terms, one could allow start terms in argu\-ment normal form (denoted $\mathtt{ANF}$-$\mathtt{AST})$. Both
\Cref{properties-eq-AST-iAST-2} as well as our results on the ADP\linebreak framework in
\Cref{Probabilistic Annotated Dependency Pairs,The Probabilistic ADP Framework}
also hold
for $\mathtt{ANF}$-$\mathtt{AST}$ ($\mathtt{ANF}$-$\mathtt{iAST}$) instead of $\mathtt{bAST}$
($\mathtt{biAST}$). 
While $\mathtt{ANF}$-$\mathtt{iAST}$ is equivalent to $\mathtt{iAST}$, 
the requirement of
start terms in $\ANF_\R$ is a real restriction for $\mathtt{AST}$.
Already in the non-probabilistic setting
there are non-terminating TRSs $\R$ where all terms in $\ANF_\R$
are terminating
(e.g., the well-known example of \cite{DBLP:journals/ipl/Toyama87} with the rules $\tf(\ta,\tb,x) \to \tf(x,x,x)$,
$\th(x,y) \to x$, and $\th(x,y) \to y$).
}
In addition to 
\Cref{properties-eq-AST-iAST-1,properties-eq-AST-iAST-2}, \cite{FoSSaCS2024} presented
ano\-ther criterion to weaken the left-linearity
condition. 
We do not recapitulate it here,\linebreak as
our novel approach in \Cref{Probabilistic Annotated Dependency Pairs,The Probabilistic ADP Framework} will not require
left-linearity anyway.

$\R_{\textsf{alg}}$ from \Cref{example:running-2} is left- and
right-linear, but overlapping.
Hence, $\R_{\textsf{alg}}$ does not belong to any known class of PTRSs where
$\mathtt{iAST}$ is equivalent to
$\mathtt{AST}$.
Thus, to prove $\mathtt{AST}$ of such
PTRSs, one  needs a new approach, e.g., as 
in \Cref{Probabilistic Annotated Dependency Pairs,The Probabilistic ADP Framework}.

\section{Probabilistic Annotated Dependency Pairs}\label{Probabilistic Annotated Dependency Pairs}

In \Cref{ADPs and Chains for iAST}
we recapitulate annotated dependency pairs (ADPs) \cite{FLOPS2024}
which adapt DPs in order to prove 
$\mathtt{iAST}$.
Then in \Cref{ADPs and Chains for AST} we
introduce our novel adaption of 
ADPs for full probabilistic rewriting w.r.t.\ any evaluation strategy.

\subsection{ADPs and Chains - Innermost Rewriting}\label{ADPs and Chains for iAST}

Instead of comparing left- and right-hand sides of rules to prove termination,
ADPs only consider the subterms with defined root symbols
in the right-hand sides,
as only these subterms might be evaluated further.
In the probabilistic setting, we use annotations to mark which subterms in
right-hand sides could potentially lead to a non-$\mathtt{(i)AST}$ evaluation.
For every $f \in \SignatureD$, we introduce a fresh 
\emph{annotated symbol} $f^{\#}$ of the same arity. 
Let $\SignatureA$ denote the set of all annotated symbols, $\SignatureADC = \SignatureA \uplus \Sigma$, and $\TT^{\#} = \TSet{\SignatureADC}{\VSet}$.
To ease readability, we often use capital letters like $F$ instead of $f^\#$.
For any $t = f(t_1,\ldots,t_n) \in \TT$ with $f \in \SignatureD$, let
$t^{\#} = f^{\#}(t_1,\ldots,t_n)$.
For $t \in \TT^{\#}$ 
and $\mathcal{X} \subseteq \SignatureADC \cup \VSet$, let $\pos_{\mathcal{X}}(t)$ be all positions of $t$
with symbols or variables from $\mathcal{X}$.
For a set of positions $\Phi \subseteq \posDT(t)$, let
$\anno_\Phi(t)$ be the variant\linebreak of $t$ where the symbols at positions from $\Phi$ in $t$
are annotated, and all other anno\-tations are removed.
Thus, $\posT(\anno_\Phi(t)) = \Phi$, and
$\anno_\emptyset(t)$ removes all annotations\linebreak from $t$, where we often write
$\flat(t)$ instead of $\anno_\emptyset(t)$.
Moreover, let $\disannoPos{\pi}(t)$ result from
removing all annotations from $t$ that are strictly above the position $\pi$.
So for $\R_2$,
we have $\anno_{\{1\}}(\td(\tg)) =
\anno_{\{1\}}(\tD(\tG)) = \td(\tG)$, $\flat(\tD(\tG)) = \td(\tg)$,
and $\disannoPos{1}(\tD(\tG)) = \td(\tG)$.
To transform the rules of a PTRS into ADPs, initially we annotate all $f \in \SignatureD$ occurring
in right-hand sides.

Every ADP also has a flag $m \in \{\ttrue, \tfalse\}$
to indicate whether this ADP may be applied
to rewrite at a position below an annotated symbol
in non-$\mathtt{(i)AST}$ evaluations.
This flag will be modified and used
by the processors in \Cref{The Probabilistic ADP Framework}.

\begin{definition}[ADPs]\label{def:canonical-ADPs}
  An \defemph{annotated dependency pair (ADP)} has the
  form $\ell \ruleArr{}{}{} \{ p_1:r_{1}, \ldots, p_k: r_k\}^{m}$,
    where $\ell \in \TT$ with $\ell \notin \VSet$, $m \in \{\ttrue, \tfalse\}$, 
    and for all $1 \leq j \leq k$ we have $r_{j} \!\in\! \TT^{\#}$ \pagebreak[3] with $\VSet(r_j) \subseteq \VSet(\ell)$.

    For a rule $\ell \to \mu = \{ p_1 : r_1, \ldots, p_k : r_k \}$, its
    \defemph{canonical annotated dependency pair} is 
    $\DPair{\ell \to \mu}  = \ell \to \{ p_1 : \anno_{\pos_{\SignatureD}(r_1)}(r_1), \ldots, p_k : \anno_{\pos_{\SignatureD}(r_k)}(r_k)\}^{\ttrue}$.
    The canonical
    ADPs of a PTRS $\R$ are $\DPair{\R} = \{\DPair{\ell \to \mu} \mid \ell \to \mu \in \R\}$.
\end{definition}

\begin{example}\label{example:running-1-ADPs}
    We obtain 
    $\DPair{\R_2} = \{  \eqref{run1-ADP-1},
    \eqref{run1-ADP-2} \}$ and  $\DPair{\R_3} = \{  \eqref{run1-ADP-1},
    \eqref{run1-ADP-2nd},
    \eqref{run1-ADP-3}
    \}$ with

    \vspace*{-.4cm}

   \hspace*{-.7cm}{\small \begin{minipage}[t]{4.4cm}
        \begin{align}
            \tg &\to \{\nicefrac{3}{4}:\tD(\tG),  \nicefrac{1}{4}:\tz\}^{\ttrue} \label{run1-ADP-1}
        \end{align}
    \end{minipage}\hspace*{-.2cm}
       \begin{minipage}[t]{4.5cm}
        \begin{align}
          \td(x) &\to \{1:\tc(x,x)\}^{\ttrue} \label{run1-ADP-2}\\
          \td(\td(x)) &\to \{1:\tc(x,\tG)\}^{\ttrue} \label{run1-ADP-2nd}
                \end{align}
    \end{minipage}
       \begin{minipage}[t]{3.4cm}
         \begin{align}
        \nonumber   \\
           \td(x) &\to \{1:\tz\}^{\ttrue} \label{run1-ADP-3}
     \end{align}
    \end{minipage}}
\end{example}

\begin{example}\label{example:running-2-ADPs}
    For $\R_{\textsf{alg}}$, the canonical ADPs are
    {\small
    \begin{align}
        \tloopOne(y) &\to \{\nicefrac{1}{2}:\tLoopOne(\tDouble(y)), \; \nicefrac{1}{2}:\tLoopTwo(\tDouble(y))\}^{\ttrue} \label{run2-ADP-1}\\
        \tloopOne(y) &\to \{\nicefrac{1}{3}:\tLoopOne(\tTriple(y)), \; \nicefrac{2}{3}:\tLoopTwo(\tTriple(y))\}^{\ttrue} \label{run2-ADP-2}\!
    \end{align}
    \begin{minipage}[t]{6cm}
        \vspace*{-0.8cm}
        \begin{align}
            \tloopTwo(\ts(y)) &\to \{1:\tLoopTwo(y)\}^{\ttrue} \label{run2-ADP-3}\\
            \tdouble(\ts(y)) &\to \{1:\ts(\ts(\tDouble(y)))\}^{\ttrue} \label{run2-ADP-4}\\
            \tdouble(\tz) &\to \{1:\tz\}^{\ttrue} \label{run2-ADP-5}\!
        \end{align}
    \end{minipage}
    \begin{minipage}[t]{6cm}
        \vspace*{-0.6cm}
        \begin{align}
            \ttriple(\ts(y)) &\to \{1:\ts(\ts(\ts(\tTriple(y))))\}^{\ttrue} \label{run2-ADP-6}\\
            \ttriple(\tz) &\to \{1:\tz\}^{\ttrue} \label{run2-ADP-7}\!
        \end{align}
    \end{minipage}}
    \vspace*{.1cm}

    \noindent
\end{example}

\noindent
We use the following rewrite relation
in the ADP framework for $\mathtt{iAST}$.

\begin{definition}[Innermost Rewriting with ADPs, $\itored{}{}{\PP}$]\label{def:ADPs-and-Rewriting}
  Let $\PP$ be a finite set of ADPs (a so-called \defemph{ADP problem}).
  We define $t \in \ANF_\PP$ if
 there are no $t' \vartriangleleft t$,
 $\ell \to \mu^m \in \PP$, and substitution $\sigma$ with $\ell \sigma = 
    \flat(t')$ (i.e., no left-hand side $\ell$ matches a proper subterm $t'$ of $t$ when removing its annotations).

    A term $s \in \TT^{\#}$ \defemph{rewrites innermost} with $\PP$ to
    $\mu = \{p_1:t_1,\ldots,p_k:t_k\}$ (deno\-ted  $s \itored{}{}{\PP} \mu$)
    if there are $\ell \ruleArr{}{}{} \{ p_1:r_{1}, \ldots, p_k: r_k\}^{m} \in
    \PP$, a substitution $\sigma$, and a\linebreak $\pi \in \pos_{\SignatureD \cup
    \SignatureA}(s)$ such that $\flat(s|_\pi)=\ell\sigma \in \ANF_\PP$, and  
    for all $1 \leq j \leq k$ we have:
    \begin{equation*}
        \begin{array}{rll@{\;\;\;}l@{\;\;}l@{\;\;}l@{\;\;}l@{\hspace*{1cm}}l}
        t_j &=                  &s[r_j\sigma]_{\pi}         & \text{if} & \pi \in \posT(s)    & \text{and} & m = \ttrue   & (\mathbf{at})\\
        t_j &= \disannoPos{\pi}(&s[r_j\sigma]_{\pi})        & \text{if} & \pi \in \posT(s)    & \text{and} & m = \tfalse  & (\mathbf{af})\\
        t_j &=                  &s[\flat(r_j)\sigma]_{\pi}  & \text{if} & \pi \not\in\posT(s) & \text{and} & m = \ttrue   & (\mathbf{nt})\\
        t_j &= \disannoPos{\pi}(&s[\flat(r_j)\sigma]_{\pi}) & \text{if} & \pi
        \not\in\posT(s) & \text{and} & m = \tfalse  & (\mathbf{nf}) \!
        \end{array}
    \end{equation*}
  \end{definition}

Rewriting with $\PP$ is like ordinary probabilistic term rewriting while considering and modifying annotations
that indicate where a non-$\mathtt{iAST}$ evaluation may arise.
A step of the form $(\mathbf{at})$ (for \underline{\textbf{a}}nnotation
and \underline{\textbf{t}}\textsf{rue})
is performed at the position of an\linebreak annotation, i.e.,
this can potentially lead to a non-$\mathtt{iAST}$ evaluation.
Hence,
all annotations from the right-hand side $r_j$ of the used ADP are kept
during the rewrite\linebreak step.
However, 
annotations of subterms that
correspond to variables of the ADP are removed, as
these subterms are in normal form due to the innermost strategy.
An example is the rewrite step $\tD(\tG) \itored{}{}{\DPair{\R_3}}
\{\nicefrac{3}{4}:\tD(\tD(\tG)),  \nicefrac{1}{4}:\tD(\tz)\}$ using the\linebreak ADP
$\eqref{run1-ADP-1}$.
A step of the form $(\mathbf{af})$ (for \underline{\textbf{a}}nnotation
and \underline{\textbf{f}}\textsf{alse})
is similar but due to the flag $m = \tfalse$ 
this ADP cannot be used below an annotation in a non-$\mathtt{iAST}$ evaluation.
Hence,
we remove all annotations above the used redex.
So using an ADP of the form $\tg \to \{\nicefrac{3}{4}:\tD(\tG),
\nicefrac{1}{4}:\tz\}^{\tfalse}$ on the term $\tD(\tG)$ would
yield $\tD(\tG) \itored{}{}{}
\{\nicefrac{3}{4}:\td(\tD(\tG)), \nicefrac{1}{4}:\td(\tz)\}$, i.e., we remove the
annotation of
$\tD$ at the root.\linebreak
A step of the form $(\mathbf{nt})$ (for \underline{\textbf{n}}o annotation
and \underline{\textbf{t}}\textsf{rue})
is performed at the position of a subterm without annotation.
Hence, the subterm cannot lead to a non-$\mathtt{iAST}$ evaluation, but this rewrite
step may 
be needed for an annotation at a\linebreak position above.
As an example, one could
rewrite the non-annotated subterm $\tg$ in $\tD(\tg) \itored{}{}{\DPair{\R_3}}
\{\nicefrac{3}{4}:\tD(\td(\tg)),  \nicefrac{1}{4}:\tD(\tz)\}$ using the ADP
$\eqref{run1-ADP-1}$.
Finally, a step of\linebreak the form $(\mathbf{nf})$  (for \underline{\textbf{n}}o annotation
and \underline{\textbf{f}}\textsf{alse})
is irrelevant for non-$\mathtt{iAST}$ evaluations, 
because the redex is not annotated and due to $m =
\tfalse$,
afterwards\linebreak one cannot rewrite
an annotated term at a position above. 
For example, if one had the ADP
$\tg \to \{\nicefrac{3}{4}:\tD(\tG),  \nicefrac{1}{4}:\tz\}^{\false}$, then we would obtain
$\tD(\tg) \itored{}{}{} \{\nicefrac{3}{4}:\td(\td(\tg)),  \nicefrac{1}{4}:\td(\tz)\}$.
The case
$(\mathbf{nf})$
is only needed to ensure that normal forms always remain the same, even if
we remove or add annotations in rules.

Due to the annotations, we now consider specific RSTs, 
called \emph{chain trees} \cite{kassinggiesl2023iAST,FLOPS2024}.
Chain trees are defined analogously to RSTs, but the crucial requirement is that every
infinite path of the tree must contain infinitely many steps of the forms $(\mathbf{at})$ or $(\mathbf{af})$,
as we specifically want to analyze the rewrite steps at annotated positions.
We say that $\F{T}=(V,E,L,A)$ is a $\PP$\emph{-innermost chain tree} (iCT) if
\begin{enumerate}
    \item $(V, E)$ is a (possibly infinite) directed tree with nodes
    $V \neq \emptyset$ and directed edges $E \subseteq V \times V$  where $vE = \{ w \mid (v,w) \in E \}$ is  finite for every $v \in V$.
    \item $L:V\rightarrow(0,1]\times\TT^{\#}$ labels every node $v$ by a probability $p_v$ and a term $t_v$.
    For the root $v \in V$ of the tree, we have $p_v = 1$.
    \item $A \subseteq V \setminus \ctleaf$ (where $\ctleaf$ are all leaves) is a subset
    of the inner nodes to indicate that we use $(\mathbf{at})$ or $(\mathbf{af})$ for the next step.
    $N = V \setminus (\ctleaf \cup A)$ are all other inner nodes, i.e.,
    where we rewrite using $(\mathbf{nt})$ or $(\mathbf{nf})$.
    \item  If $vE = \{w_1, \ldots, w_k\}$, then $t_v  \itored{}{}{\PP}
    \{\tfrac{p_{w_1}}{p_v}:t_{w_1}, \ldots, \tfrac{p_{w_k}}{p_v}:t_{w_k}\}$, where we use 
    Case $(\mathbf{at})$ or $(\mathbf{af})$ if $v \in A$, and where we use Case
    $(\mathbf{nt})$ or $(\mathbf{nf})$ if $v \in N$.
    \item Every infinite path in $\F{T}$ contains infinitely many nodes from $A$. 
\end{enumerate}

\begin{wrapfigure}[6]{r}{0.38\textwidth}
    \scriptsize
    \vspace*{-0.8cm}
          \begin{tikzpicture}
          \tikzstyle{adam}=[thick,draw=black!100,fill=white!100,minimum size=4mm, shape=rectangle split, rectangle split parts=2,rectangle split horizontal]
          \tikzstyle{empty}=[rectangle,thick,minimum size=4mm]
          
          \node[adam] at (-4, 0)  (a) {$1$\nodepart{two}$\tG$};
          \node[adam] at (-5, -0.7)  (b) {$\nicefrac{3}{4}$\nodepart{two}$\tD(\tG)$};
          \node[adam] at (-3, -0.7)  (c) {$\nicefrac{1}{4}$\nodepart{two}$\tz$};
          \node[adam] at (-6, -1.4)  (d) {$\nicefrac{9}{16}$\nodepart{two}$\tD(\tD(\tG))$};
          \node[adam] at (-4, -1.4)  (e) {$\nicefrac{3}{16}$\nodepart{two}$\tD(\tz)$};
          \node[empty] at (-6.5, -2)  (f) {$\ldots$};
          \node[empty] at (-5.5, -2)  (g) {$\ldots$};
          
          \draw (a) edge[->] (b);
          \draw (a) edge[->] (c);
          \draw (b) edge[->] (d);
          \draw (b) edge[->] (e);
          \draw (d) edge[->] (f);
          \draw (d) edge[->] (g);
          \end{tikzpicture}
     \end{wrapfigure}  
Let $|\F{T}|_{\ctleaf} = \sum_{v \in \ctleaf} \, p_v$.
Then a PTRS $\PP$ is $\mathtt{iAST}$ if $|\F{T}|_{\ctleaf} = 1$ for all $\PP$-iCTs $\F{T}$.
The corresponding $\DPair{\R_1}$-chain tree for the $\R_1$-RST from \Cref{sect:Probabilistic Term Rewriting} 
is shown on the right.
Here, we again have $|\F{T}|_{\ctleaf} = 1$.
With these definitions, in \cite{FLOPS2024} we obtained the following result. 

\begin{restatable}[{\small Chain Criterion for $\mathtt{iAST}$}]{theorem}{ProbChainCriterionInnermost}\label{theorem:prob-chain-criterion-innermost}
 \hspace*{-.3cm}   A PTRS $\R$ is $\mathtt{iAST}$ iff $\DPair{\R}$ is $\mathtt{iAST}$.
\end{restatable}

\noindent So for $\mathtt{iAST}$, one can analyze the canonical ADPs instead of the original PTRS.

\subsection{ADPs and Chains - Full Rewriting}\label{ADPs and Chains for AST}

When adapting ADPs from innermost to full rewriting, the most crucial part is to define
how to handle annotations if we rewrite above them.
For innermost ADPs, we removed the annotations below the position of the redex,
as such terms are always in normal form.
However, this is not the case for full rewriting.

\begin{example}\label{example:ADPs-full-rewriting-fail}
    Reconsider $\R_3$ and its canonical ADPs $\DPair{\R_3} = \{
    \eqref{run1-ADP-1}, \eqref{run1-ADP-2nd}, \eqref{run1-ADP-3} \}$ from \Cref{example:running-1-ADPs}.
    As noted in \Cref{example:running-1}, $\R_3$ is $\mathtt{iAST}$, but not $\mathtt{AST}$.
    To adapt \Cref{def:ADPs-and-Rewriting} to full rewriting, clearly we have to omit the
    requirement that the redex is in $\ANF$. However, this is not sufficient for
    soundness for full $\mathtt{AST}$:
    Applying two rewrite steps with \eqref{run1-ADP-1} to $\tG$
    would result in a chain tree with the leaves
    $\nicefrac{9}{16}:\tD(\tD(\tG))$, $\nicefrac{3}{16}:\tD(\tz)$ (which can be extended by
    the child
    $\nicefrac{3}{16}:\tz$), and $\nicefrac{1}{4}:\tz$. 
    However, by
    \Cref{def:ADPs-and-Rewriting}, every application of the ADP \eqref{run1-ADP-2nd}
    removes the annotations of its 
    arguments. So when applying \eqref{run1-ADP-2nd} to $\tD(\tD(\tG))$,
    we obtain 
    $\{1:\tc(\tg,\tG)\}$.
    But this would mean that the number of $\tG$-symbols is never increased. However, for all such chain trees $\F{T}$ we have $|\F{T}|_{\ctleaf}
    = 1$, i.e.,
    we would falsely conclude that $\R_3$ is $\mathtt{AST}$.
\end{example}

\Cref{example:ADPs-full-rewriting-fail} shows that for full rewriting,
we have to keep certain annotations below the used redex. 
After rewriting above a subterm like $\tG$ (which starts a
non-$\mathtt{AST}$ evaluation), it should still be possible to continue the evaluation of
$\tG$ if this subterm was ``completely inside'' the substitution of the applied rewrite step.

We use \emph{variable reposition functions (VRFs)} to relate positions of
variables in
the left-hand side
of an ADP to those positions of the
same variables in the right-hand sides
where we want to keep the annotations of the instantiated variables.
So for an ADP $\ell \to \mu$ with $\ell|_\pi = x$, we indicate which occurrence of $x$ in $r \in \Supp(\mu)$ should
keep the annotations if one rewrites an instance of $\ell$ where the subterm at position
$\pi$ 
contains annotations.\footnote{VRFs 
were introduced in \cite{IJCAR2024} when adapting ADPs to full \emph{relative} rewriting. 
However, due to the probabilistic setting,
our definition is slightly different.}

\begin{definition}[Variable Reposition Functions]\label{def:Var-Repos-Func}
    Let $\ell \to \{p_1:r_1, \ldots, p_k:r_k\}^m$ be an ADP.
	A family of functions $\varphi_j: \pos_{\VSet}(\ell) \to \pos_{\VSet}(r_j) \uplus \{\bot\}$ 
    with $1 \leq j \leq k$ is called a family of
    \defemph{variable reposition functions  (VRF)} for the ADP
    iff for all $1 \leq j \leq k$ we have 
    $\ell|_\pi = r_j|_{\varphi_j(\pi)}$ whenever $\varphi_j(\pi) \neq \bot$.
\end{definition}

\noindent
Now we can define arbitrary (possibly non-innermost)
rewriting with ADPs.

\begin{definition}[Rewriting with ADPs, $\tored{}{}{\PP}$]\label{def:ADPs-and-Rewriting-full}
    ADPs and canonical ADPs are defined as in the innermost case.
    Let $\PP$ be an ADP problem.
    A term $s \in \TT^{\#}$ \defemph{rewrites} with $\PP$ to
    $\mu = \{p_1:t_1,\ldots,p_k:t_k\}$ (denoted  $s \tored{}{}{\PP} \mu$)
    if there are an $\ell \ruleArr{}{}{} \{ p_1:r_{1}, \ldots, p_k: r_k\}^{m} \in
    \PP$, a VRF $(\varphi_j)_{1 \leq j \leq k}$ for this ADP, a substitution $\sigma$,
    and a $\pi \in \pos_{\SignatureD \cup 
      \SignatureA}(s)$ such that $\flat(s|_\pi)=\ell\sigma$,
    and for all $1 \leq j \leq k$ we have:
    \begin{equation*}
          \begin{array}{rll@{\;\;\;}l@{\;\;}l@{\;\;}l@{\;\;}l@{\hspace*{1cm}}l}
             t_j &=                  &s[\anno_{\Phi_j}(r_j\sigma)]_{\pi}         & \text{if} & \pi \in \posT(s)    & \text{and} & m = \ttrue   & (\mathbf{at})\\
        t_j &= \disannoPos{\pi}(&s[\anno_{\Phi_j}(r_j\sigma)]_{\pi})        & \text{if} & \pi \in \posT(s)    & \text{and} & m = \tfalse  & (\mathbf{af})\\
        t_j &=                  &s[\anno_{\Psi_j}(r_j\sigma)]_{\pi}  & \text{if} & \pi \not\in\posT(s) & \text{and} & m = \ttrue   & (\mathbf{nt})\!
        \end{array}
    \end{equation*}
    {\small Here, $\Psi_j\!=\!\{\varphi_j(\rho).\tau \mid \rho \!\in\! \pos_{\VSet}(\ell), \,
    \varphi_j(\rho)\!\neq\!\bot,  \, \rho.\tau \!\in\! \posT(s|_{\pi}) \}$
    and $\Phi_j\!=\!\posT(r_j) \cup \Psi_j$.}
\end{definition}

So $\Psi_j$ considers all positions $\rho.\tau$ of annotated symbols in $s|_{\pi}$ that
are below
positions $\rho$ of variables in $\ell$. If $\varphi_j$
maps $\rho$ to a variable position
$\rho'$ in
$r_j$, then the annotations below $\pi.\rho$ in $s$ are kept in the resulting subterm at
position $\pi.\rho'$
after the rewriting.
As an example, consider $\tD(\tD(\tG)) \tored{}{}{\DPair{\R_3}} \{1:\tc(\tG,\tG)\}$.
Here, we use the ADP
$\td(\td(x)) \to \{1:\tc(x,\tG)\}^{\ttrue}\,$ \eqref{run1-ADP-2nd}, with $\pi=\varepsilon$, $\sigma(x) = \tg$, and the VRF $\varphi_1(1.1) = 1$.
We get $\flat(\tD(\tD(\tG))|_{\varepsilon}) = \td(\td(\tg)) = \ell \sigma$, $1.1 \in
\pos_{\VSet}(\ell)$, $1.1.\varepsilon \in \posT(s|_{\pi})$, and thus $\Psi_1 =
\{\varphi_1(1.1).\varepsilon\} = \{1\}$ and $\Phi_1  = \posT(r_1) \cup \Psi_1= \{1,2\}$.

The case $(\mathbf{nf})$ from \Cref{def:ADPs-and-Rewriting} is missing in \Cref{def:ADPs-and-Rewriting-full},
as we do not consider (argument) normal forms anymore. \pagebreak[3]
ADPs without annotations in the right-hand side 
and with the flag $\tfalse$ are
not needed for non-$\mathtt{AST}$ chain trees and thus, they
could simply be removed from ADP problems.

Note that our VRFs in \Cref{def:Var-Repos-Func}
map a position of the left-hand side $\ell$
to at most one position in each right-hand side $r_j$ of an ADP, 
even if the ADP is duplicating.
A probabilistic rule or ADP $\ell \to \mu$ is \emph{non-duplicating} if all rules in 
$\{\ell \to r \mid r \in \Supp(\mu)\}$ are, and a PTRS or ADP problem
is non-duplicating if all of its rules are (disregarding the flag for ADPs).
For example, for the duplicating ADP
$\td(x) \to \{1:\tc(x,x)\}^{\ttrue}\,$ \eqref{run1-ADP-2},
we have three different VRFs which
map position $1$ to either $\bot$, $1$, or $2$, but we
cannot map it to both  positions $1$ and $2$.

Therefore, our VRFs cannot handle duplicating rules and ADPs correctly.
With VRFs as in \Cref{def:Var-Repos-Func}, 
$\DPair{\R_2}$ would be considered to be $\mathtt{AST}$, as
$\tD(\tG)$ only rewrites to $\{ 1: \tc(\tG,\tg) \}$ or $\{ 1: \tc(\tg,\tG) \}$,
but the annotation cannot be duplicated. 
Hence, the chain criterion would be unsound for duplicating PTRSs like $\R_2$.

To handle duplicating rules, one can adapt the direct application of 
orderings
to prove $\mathtt{AST}$ from \cite{kassinggiesl2023iAST}
and try to remove the duplicating rules of the PTRS
before constructing the canonical ADPs.

Alternatively, one could modify the definition of 
the rewrite relation $\tored{}{}{\PP}$ and use
\emph{generalized} VRFs (GVRFs)
which can duplicate annotations instead of VRFs. This would yield
a sound and complete chain criterion for full $\mathtt{AST}$ of possibly
duplicating PTRSs,
but then one would also have to consider this modified definition of $\tored{}{}{\PP}$ for the processors of the
ADP
framework in \Cref{The Probabilistic ADP Framework}. Unfortunately,
almost all processors would become unsound when defining the rewrite relation
$\tored{}{}{\PP}$ via
GVRFs
(see Ex.\ \ref{example:dup-problem}, \ref{UsRuleUnsoundNonSpare}, and
\ref{RPPexample-iAST}).
Therefore, we use VRFs instead and
restrict ourselves to non-duplicating
PTRSs for the soundness of the chain criterion.\footnote{A related restriction is needed
in the setting of (non-probabilistic) relative termination due to the VRFs  \cite{IJCAR2024}.}

\emph{Chain trees (CTs)} are now defined like iCTs,
where instead of $\itored{}{}{\PP}$ we only require steps with
$\tored{}{}{\PP}$. Then an ADP problem $\P$ is $\mathtt{AST}$ if
$|\F{T}|_{\ctleaf} = 1$ for all $\PP$-CTs $\F{T}$.
This leads to 
 our desired chain criterion
for $\mathtt{AST}$.

\begin{restatable}[Chain Criterion for
      $\mathtt{AST}$]{theorem}{ProbChainCriterionFull}\label{theorem:prob-chain-criterion-full} 
A non-duplicating PTRS $\R$ is $\mathtt{AST}$ iff $\DPair{\R}$ is $\mathtt{AST}$.
\end{restatable}

The above chain criterion allows us to
analyze full $\mathtt{AST}$ for a significantly larger
class of PTRSs than 
\Cref{properties-eq-AST-iAST-1}: 
we do not impose
non-overlappingness and left-linearity anymore, and
only require non-duplication
instead of
right-linearity.

Similar to \Cref{properties-eq-AST-iAST-2}, 
the ADP framework becomes more powerful if
we restrict ourselves to
basic start terms. Then it suffices if
the PTRS is spare (instead of non-duplicating),
since then redexes are never duplicated.
In fact, \emph{weak} spareness is sufficient, which subsumes both spareness and
non-duplication.
A rewrite step $\ell\sigma \to_\R \mu\sigma$ is \emph{weakly spare} 
if $\sigma(x) \in \NF_\R$ for every $x \in \Var$
where $x$ occurs less often in $\ell$ than in some
$r \in \Supp(\mu)$. 
An $\R$-RST is weakly spare if all rewrite steps corresponding to its edges are weakly
spare.
A PTRS $\R$ is weakly spare if each $\R$-RST that starts with $\{1 : t\}$ for a basic term $t$ is
weakly spare. The sufficient conditions
for spareness in \cite{frohn_analyzing_nodate} can easily be adapted to 
weak spareness.

In the ADP framework for $\mathtt{bAST}$, we only have
to prove that no term starting a\linebreak  non-$\mathtt{AST}$ evaluation
can be reached from a basic start term. 
Here we use \emph{basic ADP\linebreak problems} $(\mathcal{I}, \PP)$,
where $\mathcal{I}$ and $\PP$ are finite sets of ADPs.
$\PP$
are again the ADPs which we analyze for $\mathtt{AST}$ and
the \emph{reachability component}
$\mathcal{I}$ contains so-called \emph{initial} ADPs.
A  basic ADP problem $(\mathcal{I},\PP)$ is
$\mathtt{bAST}$
 if 
$|\F{T}|_{\ctleaf} = 1$ holds for all those
$(\mathcal{I} \cup \PP)$-CTs $\F{T}$ that start with a term $t^\#$
where $t \in \TT$
is basic, and where ADPs from $\mathcal{I}\setminus \PP$ 
are only used finitely often within the tree $\F{T}$. 
Thus, every basic ADP problem $(\mathcal{I},\PP)$ can be replaced
by $(\mathcal{I}\setminus\PP,\PP)$.
For a PTRS $\R$, the \emph{canonical basic ADP problem} is $(\emptyset,\DPair{\R})$.

\begin{restatable}[Chain Criterion for $\mathtt{bAST}$]{theorem}{ProbChainCriterionFullBasic}\label{theorem:prob-chain-criterion-full-basic}
    A weakly spare PTRS $\R$ is $\mathtt{bAST}$ iff $(\emptyset,\DPair{\R})$ is $\mathtt{bAST}$.
\end{restatable}

\begin{remark} \label{remark:minimality}
    In the chain criterion for
    non-probabilistic DPs, it suffices to regard only
    instantiations where all terms below an annotated symbol are terminating. The
    reason is the \emph{minimality property}
    of non-probabilistic term rewriting, i.e.,
    whenever a term starts an infinite rewrite sequence, then
    it also starts an infinite sequence where all proper subterms of
    every used redex are terminating.
    However, 
    in the probabilistic setting 
    the minimality property does not hold \cite{FoSSaCS2024}.
    For $\R_3$,  $\tg$ starts a
    non-$\mathtt{AST}$ RST, but in this RST, one has to apply Rule \eqref{rule-02nd}  to
    the redex $\td(\td(\tg))$,
    although
    it contains the proper subterm $\tg$ that starts a non-$\mathtt{AST}$ RST.
\end{remark}

\section{The Probabilistic ADP Framework for Full Rewriting}\label{The Probabilistic ADP Framework}

The idea of the DP framework for non-probabilistic TRSs
is to apply \emph{processors}
repeatedly
which
transform a DP problem into simpler sub-problems \cite{gieslLPAR04dpframework,giesl2006mechanizing}.
Since different techniques can be applied to different sub-problems, this results in a\linebreak
\emph{modular} approach for termination analysis.
This idea is also used in the ADP framework.
An \emph{ADP processor} $\Proc$ has the form $\Proc(\PP) = \{\PP_1, \ldots,\PP_n\}$ for
ADP problems
$\PP, \PP_1, \ldots, \PP_n$.
Let $\mathcal{Z} \in \{\mathtt{AST}, \mathtt{iAST}\}$.
$\Proc$ is \emph{sound} for $\mathcal{Z}$ if $\PP$ is $\mathcal{Z}$ whenever 
$\PP_i$ is $\mathcal{Z}$ for all $1 \leq i \leq n$. 
It is \emph{complete} for $\mathcal{Z}$ if $\PP_i$ is $\mathcal{Z}$ for all 
$1 \leq i \leq n$ whenever $\PP$ is $\mathcal{Z}$.
The definitions for $\mathtt{bAST}$ are analogous, but with basic ADP problems $(\InI,\PP)$.
Thus, one starts with the canonical (basic)
ADP problem and applies sound 
(and preferably complete) ADP processors repeatedly until there are no more remaining 
ADP problems. This implies that the canonical (basic) ADP problem is
$\mathcal{Z}$ and by the chain criterion, the original PTRS is $\mathcal{Z}$ as well.

An ADP problem without annotations is always $\mathtt{AST}$, because then
no rewrite step increases the number of annotations (recall that VRFs 
cannot duplicate annotations). Hence, then any term with 
$n$ annotations only starts rewrite sequences with at most $n$ steps of the form
$(\mathbf{at})$ or $(\mathbf{af})$, i.e., all
$\PP$-CTs are finite.

In the following, we recapitulate the main processors for
$\mathtt{iAST}$
from \cite{FLOPS2024} and adapt them to our new framework for $\mathtt{AST}$ and
$\mathtt{bAST}$.

\subsection{Dependency Graph Processor}

The innermost $\PP$-\emph{dependency graph} is a control flow graph whose nodes are the
ADPs from $\PP$. It
indicates whether an ADP $\alpha$ may lead to an application of another 
ADP $\alpha'$ on an annotated subterm introduced by $\alpha$. 
This possibility is not related to the probabilities. 
Hence, here we use the \emph{non-probabilistic variant}
$\nonprob(\PP) = \{\ell \to \flat(r_j) \mid \ell \to \{p_1:r_1, \ldots, p_k:r_k\}^{\ttrue} \in \PP, 1
\leq j \leq k\}$, which is an ordinary TRS over the original signature $\Sigma$.
For $\nonprob(\PP)$ we only consider rules
with the flag $\ttrue$, since only they are needed for rewriting
below annotations.
We define $t \trianglelefteq_{\#} s$ if
there is a $\pi \in \posT(s)$ and $t = \flat(s|_\pi)$, i.e., $t$ results from a subterm of
$s$ with annotated root symbol by removing its annotations.

\begin{definition}[Innermost Dependency Graph]\label{def:iDG}
   $\!\!$The \defemph{innermost $\PP$-dependen\-cy graph} has the set of nodes $\PP$, and 
    there is an edge from $\ell_1 \ruleArr{}{}{} \{ p_1:r_{1}, \ldots, p_k: r_k\}^{m}$ to $\ell_2 \to \ldots$ 
    if there are substitutions
    $\sigma_1, \sigma_2$ and a $t \trianglelefteq_{\#} r_{j}$ for some $1 \leq j \leq k$ 
    such that $t^\# \sigma_1 \ito_{\nonprob(\PP)}^*
    \ell_2^\# \sigma_2$ and both $\ell_1 \sigma_1$ and $\ell_2 \sigma_2$ are in $\ANF_{\PP}$.
\end{definition}

So there is an edge from an ADP $\alpha$ to an ADP $\alpha'$ if after a
$\itored{}{}{\PP}$-step of the\linebreak form 
$(\mathbf{at})$ or $(\mathbf{af})$ with  $\alpha$ at position $\pi$
there may eventually come another
$\itored{}{}{\PP}$-step\linebreak of the form $(\mathbf{at})$ or $(\mathbf{af})$ with $\alpha'$ on or below $\pi$.
Since every infinite path in an iCT\linebreak contains infinitely
many nodes from $A$, every such path traverses a
cycle of the innermost dependency graph infinitely often.
Thus, it suffices to consider its strongly\linebreak connected components (SCCs)\footnote{A
set $\mathcal{P}'$ of ADPs is  an \emph{SCC} if it is a maximal cycle,
i.e., a maximal set where for any $\alpha, \alpha'$
in $\mathcal{P}'$ there is
a non-empty path from $\alpha$ to $\alpha'$ only traversing nodes from
$\mathcal{P}'$.}
separately.
In our framework, this means that we remove the annotations from all ADPs except those
in the SCC that we want to analyze.
Since checking whether there exist $\sigma_1, \sigma_2$
as in \Cref{def:iDG}
is undecidable, 
to automate the following processor, the same over-approximation
techniques as for the non-probabilistic dependency graph can be used, see, e.g., \cite{arts2000termination,giesl2006mechanizing,hirokawa2005automating}.
In the following, $\flat(\PP)$ denotes the ADP problem $\PP$ where all annotations are removed.

\begin{restatable}[Dependency Graph Processor for $\mathtt{iAST}$]{theorem}{ProbIDepGraphProc}\label{theorem:prob-iDGP}
    For the SCCs $\PP_1,\report{\linebreak} \ldots, \PP_n$ of the
    innermost $\PP$-dependency graph, the processor
    $\Proc_{\mathtt{DG}}(\PP)=\{\PP_1 \cup \flat(\PP \setminus \PP_1), \ldots, \PP_n \cup \flat(\PP \setminus \PP_n)\}$ is sound and complete for $\mathtt{iAST}$.
\end{restatable}

\begin{example}
    Consider the PTRS $\R_{2}$ and its canonical ADPs from
    \Cref{example:running-1-ADPs}. The in-
\end{example}
    
    \vspace*{-0.16cm}
    \begin{wrapfigure}[3]{r}{0.08\textwidth}
        \scriptsize
        \vspace*{-0.8cm}
        \hspace*{-.10cm}\begin{tikzpicture}
                \node[shape=rectangle,draw=black!100, minimum size=3mm] (A) at (0,0) {\eqref{run1-ADP-1}};
                \node[shape=rectangle,draw=black!100, minimum size=3mm] (B) at (0,-0.7) {\eqref{run1-ADP-2}};
            
                \path [->] (A) edge (B);
                \path [->, loop right] (A) edge (A);            
            \end{tikzpicture}      
    \end{wrapfigure}
    \noindent nermost $\DPair{\R_2}$-dependency graph is on the right.
    As the only SCC $\{ \eqref{run1-ADP-1} \}$ does not contain
    \eqref{run1-ADP-2}, we can remove all annotations from
    \eqref{run1-ADP-2}.
    How\-ever, \eqref{run1-ADP-2} has no annotations. Thus,  $\Proc_{\mathtt{DG}}$
 does not change $\DPair{\R_2}$.

\vspace*{-.1cm}
 
\subsubsection{Adaption for $\mathtt{AST}$:}

To handle full rewriting, we have to change the definition of the
dependency graph as we can now also perform non-innermost steps.

\begin{definition}[Dependency Graph]
    The \defemph{$\PP$-dependency graph} has the nodes $\PP$ and 
    there is an edge from $\ell_1 \ruleArr{}{}{} \{ p_1:r_{1}, \ldots, p_k: r_k\}^{m}$ to $\ell_2 \to \ldots$ 
    if there are substitutions
    $\sigma_1, \sigma_2$ and a $t \trianglelefteq_{\#} r_{j}$ for some $1 \leq j \leq k$ 
    with $t^\# \sigma_1 \to_{\nonprob(\PP)}^* \ell_2^\# \sigma_2$.
\end{definition}

\vspace*{-.15cm}

\begin{restatable}[Dependency Graph Processor for $\mathtt{AST}$]{theorem}{ProbDepGraphProc}\label{theorem:prob-DGP}
    For the SCCs $\PP_1, \ldots,\report{\linebreak} \PP_n$ of the
    $\PP$-dependency graph,
    $\Proc_{\mathtt{DG}}(\PP)=\{\PP_1 \cup \flat(\PP \setminus \PP_1), \ldots,\PP_n \cup \flat(\PP \setminus \PP_n)\}$ is sound and complete for $\mathtt{AST}$.
\end{restatable}

\begin{example}\label{Ex:DepGraphAlg}
    Consider $\R_{\mathsf{alg}}$ and its canonical ADPs from \Cref{example:running-2-ADPs}. 
    The $\DPair{\R_{\mathsf{alg}}}$-de-\linebreak pen\-dency graph is given below.
  Its SCCs are $\{\eqref{run2-ADP-1}, \eqref{run2-ADP-2}\}$,  $\{\eqref{run2-ADP-3}\}$, $\{\eqref{run2-ADP-4}\}$, 
    $\{\eqref{run2-ADP-6}\}$.
  \end{example}
    
    %\vspace*{-0.17cm}
    \begin{wrapfigure}[8]{r}{0.18\textwidth}
        \scriptsize
        \vspace*{-0.1cm}
        \hspace*{-.25cm}\begin{tikzpicture}
                \node[shape=rectangle,draw=black!100, minimum size=3mm] (A) at (0,0.5) {\eqref{run2-ADP-1}};
                \node[shape=rectangle,draw=black!100, minimum size=3mm] (B) at (0,-0.5) {\eqref{run2-ADP-2}};
                \node[shape=rectangle,draw=black!100, minimum size=3mm] (C) at (0,1.5) {\eqref{run2-ADP-4}};
                \node[shape=rectangle,draw=black!100, minimum size=3mm] (D) at (0,-1.5) {\eqref{run2-ADP-6}};
                \node[shape=rectangle,draw=black!100, minimum size=3mm] (E) at (0.7,1) {\eqref{run2-ADP-5}};
                \node[shape=rectangle,draw=black!100, minimum size=3mm] (F) at (0.7,0) {\eqref{run2-ADP-3}};
                \node[shape=rectangle,draw=black!100, minimum size=3mm] (G) at (0.7,-1) {\eqref{run2-ADP-7}};
            
                \path [->] (A) edge (B);
                \path [->] (A) edge (C); 
                \path [->, out=350, in=90] (A) edge (F); 
                \path [->, out=10, in=270] (A) edge (E); 
                \path [->, loop left] (A) edge (A); 
                \path [->] (B) edge (A);     
                \path [->] (B) edge (D); 
                \path [->, out=10, in=270] (B) edge (F);  
                \path [->, out=350, in=90] (B) edge (G);  
                \path [->, loop left] (B) edge (B);    
                \path [->, loop left] (C) edge (C);  
                \path [->, out=0, in=90] (C) edge (E);    
                \path [->, loop left] (D) edge (D); 
                \path [->, out=0, in=270] (D) edge (G); 
                \path [->, loop right] (F) edge (F);        
            \end{tikzpicture}      
    \end{wrapfigure}
    \noindent 
         For each SCC we create a separate ADP problem, where all annota\-tions  outside 
    the SCC are removed.
    This leads to the ADP problems
    $\{\eqref{run2-ADP-1}, \eqref{run2-ADP-2}, \flat(\ref{run2-ADP-3})$ - $\flat(\ref{run2-ADP-7})\}$,
    $\{\eqref{run2-ADP-3}, \flat(\ref{run2-ADP-1}), \flat(\ref{run2-ADP-2}),
    \flat(\ref{run2-ADP-4})$ - $\flat(\ref{run2-ADP-7})\}$,
    $\{\eqref{run2-ADP-4}, \flat(\ref{run2-ADP-1})$ - $\flat(\ref{run2-ADP-3}),
    \flat(\ref{run2-ADP-5})$ - $\flat(\ref{run2-ADP-7})\}$, and
    $\{\eqref{run2-ADP-6}, \flat(\ref{run2-ADP-1})$ - $\flat(\ref{run2-ADP-5}), \flat(\ref{run2-ADP-7})\}$.

\begin{example}\label{example:dup-problem}    
    If we used GVRFs that can
    duplicate annotations, then 
    the dependency graph processor would not be sound.
    The reason is that  $\Proc_{\mathtt{DG}}$
    maps ADP problems without annotations to
    the empty set. However, this would be unsound 
    if we had GVRFs, because then the ADP problem with
    $\ta \to \{1:\tb\}^{\ttrue}$ and $\td(x) \to \{1:\tc(x,\td(x))\}^{\ttrue}$ would not be 
    $\mathtt{AST}$.
    Here, the use of GVRFs would lead to the following CT
    with an infinite number of
    $(\mathbf{at})$ steps that rewrite $\tA$ to $\tb$.
     
    \vspace*{-0.05cm}
    \begin{center}
        \scriptsize
        \begin{tikzpicture}
            \tikzstyle{adam}=[rectangle,thick,draw=black!100,fill=white!100,minimum size=4mm]
            \tikzstyle{empty}=[rectangle,thick,minimum size=4mm]
            
            \node[adam] at (0, 0)  (a) {$1:\td(\tA)$};
            \node[adam] at (2.1, 0)  (b) {$1:\tc(\tA,\td(\tA))$};
            \node[adam] at (4.6, 0)  (c) {$1:\tc(\tb,\td(\tA))$};
            \node[empty] at (6.5, 0)  (d) {$\ldots$};
            
            \draw (a) edge[->] (b);
            \draw (b) edge[->] (c);
            \draw (c) edge[->] (d);
            \end{tikzpicture}
    \end{center}
 \end{example}

\vspace*{-.2cm}

\subsubsection{Adaption for $\mathtt{bAST}$:}

Here, ADPs that are not in the considered SCC $\PP_i$
may still be necessary for the initial
steps from the basic start term to the SCC. Thus, while we remove the annotations of ADPs
outside the SCC $\PP_i$ in the second component $\PP$ of a basic ADP problem $(\InI,\PP)$,
we add (the original versions of) those ADPs to $\InI$
that reach the SCC $\PP_i$
in the $(\InI \cup \PP)$-dependency graph. Let
  $\PP_i\!\!\uparrow$ be the set of all $\JJ \subseteq (\InI \cup
\PP)\setminus \PP_i$ such that  all ADPs of $\JJ$
reach  $\PP_i$
in the $(\InI \cup \PP)$-dependency graph,
and for all pairs of ADPs $\alpha,\beta \in \JJ$ with $\alpha \neq \beta$,
 $\alpha$ reaches $\beta$ or
$\beta$ reaches $\alpha$ in the $(\InI \cup \PP)$-dependency graph.
Furthermore, $\JJ$ must be maximal w.r.t.\ these properties, i.e.,  if $\alpha \not\in \JJ$
then $\alpha$ does not reach
$\PP_i$ or there exists a $\beta \in \JJ$ such that $\alpha$ does not reach
$\beta$
and $\beta$ does not reach $\alpha$.

\begin{restatable}[Dependency Graph Processor for $\mathtt{bAST}$]{theorem}{ProbDepGraphProcBast}\label{theorem:prob-DGP-bAST}
    For the SCCs $\PP_1,\linebreak \ldots, \PP_n$ of the
    $\PP$-dependency graph, the processor
    $\Proc_{\mathtt{DG}}(\InI,\PP)=
    \{(\mathcal{J} \cup
    \flat(\InI \setminus \mathcal{J}),\linebreak \PP_i \cup \flat(\PP \setminus
    \PP_i)) \mid 1 \leq i \leq n, \, \mathcal{J} \in \PP_i\!\!\uparrow\}$
    is sound and complete for $\mathtt{bAST}$.
\end{restatable}

As remarked in \Cref{ADPs and Chains for AST}, every basic ADP problem 
$(\InI,\PP)$ can be replaced by $(\InI\setminus\PP,\PP)$. Thus,
this should be done after every
application of a processor.

\begin{example}\label{Ex:DepGraphAlg-bAST}
    To prove
    $\mathtt{bAST}$ of $\R_{\mathsf{alg}}$,
    we start with
    $(\emptyset,\DPair{\R_{\mathsf{alg}}})$.
    The SCC $\{\eqref{run2-ADP-6}\}$ is only reachable from $\eqref{run2-ADP-1}$ and
    $\eqref{run2-ADP-2}$,
    leading to the basic ADP problem $(\{\eqref{run2-ADP-1},
    \eqref{run2-ADP-2}\},\paper{ \;}\report{\linebreak} \{\eqref{run2-ADP-6}, \flat(\ref{run2-ADP-1})$ - $\flat(\ref{run2-ADP-5}), \flat(\ref{run2-ADP-7})\})$.
    The SCC $\{\eqref{run2-ADP-1}, \eqref{run2-ADP-2}\}$ is not reachable from other
    ADPs and thus, here we obtain
    $(\emptyset, \;
    \{\eqref{run2-ADP-1}, \eqref{run2-ADP-2}, \flat(\ref{run2-ADP-3})$ -
    $\flat(\ref{run2-ADP-7})\})$, etc.
\end{example}

\begin{example}\label{example:usable-rules-ce-rules-full}
  The next ADP problem $\PP_{\tg}$ illustrates the reachability component.

    \vspace*{-.5cm}

    \hspace*{-.7cm}\begin{minipage}[t]{4cm}
        \begin{align}    
        \tinit &\to \{1:\tF(\tg)\}^{\ttrue} \label{usable-rule-counter-rule-0}
        \end{align}
    \end{minipage}
    \hspace*{-1.8cm}
    \begin{minipage}[t]{10cm}
        \begin{align}
            \tg &\to \{\nicefrac{1}{2}:\tc(\tg,\tg,\tg,\tg),
            \nicefrac{1}{2}:\tz\}^{\ttrue} \label{usable-rule-counter-rule-2} \\
            \tf(\tc(x_1,x_2,x_3,x_4)) &\to
            \{1:\tc(\tF(x_1),\tF(x_2),\tF(x_3),\tF(x_4))\}^{\ttrue} \label{usable-rule-counter-rule-1}
        \end{align}
    \end{minipage}

    \vspace*{.2cm}
    
    \noindent
    Although \eqref{usable-rule-counter-rule-2} has no annotations, the basic
    ADP problem $(\emptyset, \PP_{\tg})$ is not $\mathtt{bAST}$:
    
    \vspace*{-0.05cm}
    \begin{center}
        \scriptsize
        \hspace*{.05cm}
        \begin{tikzpicture}
            \tikzstyle{adam}=[rectangle,thick,draw=black!100,fill=white!100,minimum size=4mm]
            \tikzstyle{empty}=[rectangle,thick,minimum size=4mm]
            
            \node[adam] at (-2, 0)  (z) {$1:\tInit$};
            \node[adam] at (-0.2, 0)  (a) {$1:\tF(\tg)$};
            \node[adam] at (2.5, 0.55)  (b2) {$\nicefrac{1}{2}:\tF(\tz)$};
            \node[adam] at (2.5, 0)  (b) {$\nicefrac{1}{2}:\tF(\tc(\tg,\tg,\tg,\tg))$};
            \node[adam] at (6.4, 0)  (c) {$\nicefrac{1}{2}:\tc(\tF(\tg),\tF(\tg),\tF(\tg),
              \tF(\tg))$};
            \node[empty] at (9.3, 0)  (d) {$\ldots$};
            
            \draw (z) edge[->] (a);
            \draw (a) edge[->] (b);
            \draw (a) edge[->] (b2);
            \draw (b) edge[->] (c);
            \draw (c) edge[->] (d);
            \end{tikzpicture}
    \end{center}
    
    \vspace*{-0.05cm}
    \noindent
    This is a random walk biased  towards non-termination, where the number of $\tF(\tg)$
    subterms increases by $3$ or decreases by $1$, both
    with probability $\nicefrac{1}{2}$. \hspace*{2cm}\\
\end{example}

\vspace*{-.55cm}

    \begin{wrapfigure}[3]{r}{1.6cm}
\paper{\vspace*{-.9cm}}\report{\vspace*{-.1cm}}
      \scriptsize
                    \hspace*{-.17cm}\begin{tikzpicture}
                \node[shape=rectangle,draw=black!100, minimum size=3mm] (A) at (0,0)
                     {\eqref{usable-rule-counter-rule-0}};
                \node[shape=rectangle,draw=black!100, minimum size=3mm] (B) at (0,-0.7)
                     {\eqref{usable-rule-counter-rule-1}};
               \node[shape=rectangle,draw=black!100, minimum size=3mm] (C) at (1,0)
                    {\eqref{usable-rule-counter-rule-2}};
            
                \path [->] (A) edge (B);
                \path [->, loop right] (B) edge (B);            
            \end{tikzpicture}      
    \end{wrapfigure}
    Since the only SCC of the $\PP_{\tg}$-dependency graph on the right
    is $\{\eqref{usable-rule-counter-rule-1}\}$,  $\Proc_{\mathtt{DG}}$
    replaces $\tF(\tg)$ by
    $\tf(\tg)$ in \eqref{usable-rule-counter-rule-0}
    and obtains $\PP_{\tg}' = \{
    \flat(\ref{usable-rule-counter-rule-0}), \eqref{usable-rule-counter-rule-2},
    \eqref{usable-rule-counter-rule-1} \}$. However,
    $(\emptyset, \PP_{\tg}')$  would be
    $\mathtt{bAST}$. So for the soundness of the dependency
    graph processor, we have to add the original ADP \eqref{usable-rule-counter-rule-0} to
    the reachability component and obtain  $(\{ \eqref{usable-rule-counter-rule-0} \},
    \PP_{\tg}')$ which is again not $\mathtt{bAST}$.

\subsection{Usable Terms Processor}

The dependency graph processor removes either
all annotations from an ADP or none.
But an ADP can still contain terms $t$ with annotated root where no instance
$t \sigma_1$ rewrites to an instance $\ell^\#
\sigma_2$ of a
left-hand side $\ell$ of an ADP
with annotations.
The \emph{usable terms processor} removes the annotation from the root 
of such \emph{non-usable}
terms like $\tD(\ldots)$ in 
$\DPair{\R_2} = \{\eqref{run1-ADP-1}, \eqref{run1-ADP-2} \}$.
So instead of whole ADPs, here we consider the subterms in the right-hand sides of an
ADP individually.

\begin{restatable}[Usable Terms Processor for $\mathtt{iAST}$]{theorem}{IUsableTermsProc}\label{theorem:prob-iUPP}
    Let $\ell_1 \in \TT$ and $\PP$ be an ADP problem.
    We call $t \in \TT^{\#}$ with
    $\rootsym(t) \in \SignatureD^\#$ \defemph{innermost usable} w.r.t.\ $\ell_1$ and
    $\PP$  if there
    are substitutions $\sigma_1, \sigma_2$ and
    an $\ell_2 \ruleArr{}{}{} \mu_2 \in \PP$ where $\mu_2$ contains an
    annotated symbol,
    such that $\anno_{\{\varepsilon\}}(t) \sigma_1 \ito_{\nonprob(\PP)}^*
    \ell_2^\# \sigma_2$ and both $\ell_1 \sigma_1$ and $\ell_2
    \sigma_2$ are in $\ANF_{\PP}$.
    Let 
    $\Delta_{\ell,\PP}(s) = \{ \pi \in \posT(s) \mid s|_\pi$ is innermost usable
    w.r.t.\ $\ell$ and $\PP\,\}$. 
    The transformation that removes the annotations from the roots of all
    non-usable terms in the
    right-hand sides is $\mathcal{T}_\mathtt{UT}(\PP) \!=\! \{ \ell \!\to\! 
    \{ p_1: \#_{\Delta_{\ell,\PP}(r_1)}(r_1), \ldots, p_k:\#_{\Delta_{\ell,\PP}(r_k)}(r_k)\}^{m}
    \mid  \ell \!\to\! \{ p_1: r_1, \ldots, p_k:r_k\}^{m} \!\in\! \PP \}$.
    Then $\Proc_{\mathtt{UT}}(\PP) = \{\mathcal{T}_\mathtt{UT}(\PP)\}$ is sound and complete for $\mathtt{iAST}$.
\end{restatable} 

\noindent
So for $\DPair{\R_2}$, $\Proc_{\mathtt{UT}}$ replaces $\eqref{run1-ADP-1}$ 
by $\;\tg \to \{\nicefrac{3}{4}:\td(\tG), \nicefrac{1}{4}:\tz\}^{\ttrue} \quad (\ref{run1-ADP-1}')$.

\vspace*{-.2cm}

\subsubsection{Adaption for $\mathtt{AST}$ and $\mathtt{bAST}$:}

Similar to the dependency graph, for full rewriting, 
we remove the $\ANF$ requirement and allow non-innermost steps
to reach the next ADP.
To adapt the processor to $\mathtt{bAST}$,
in the reachability component we consider usability w.r.t.\ $\InI \cup
\PP$, since one may use both $\InI$ and $\PP$ in the initial steps.

\begin{restatable}[Usable Terms Processor for $\mathtt{AST}$ and $\mathtt{bAST}$]{theorem}{UsableTermsProc}\label{theorem:prob-UPP}
    We call $t \in \TT^{\#}$ with
    $\rootsym(t) \in \SignatureD^\#$ \defemph{usable} w.r.t.\ an ADP problem $\PP$ if there
    are substitutions $\sigma_1, \sigma_2$ and
    an $\ell_2 \ruleArr{}{}{} \mu_2 \in \PP$ where $\mu_2$ contains an
    annotated symbol,
    such that $\anno_{\{\varepsilon\}}(t) \sigma_1 \to_{\nonprob(\PP)}^* \ell_2^\# \sigma_2$.
    Let $\Delta_{\PP}(s) = \{ \pi \in \posT(s) \mid s|_\pi$ is usable
    w.r.t.\ $\PP\,\}$ 
    and $\mathcal{T}_\mathtt{UT}(\PP) \!=\! \{ \ell \!\to\! 
    \{ p_1: \#_{\Delta_{\PP}(r_1)}(r_1), \ldots, p_k:\#_{\Delta_{\PP}(r_k)}(r_k)\}^{m}
    \mid  \ell \!\to\! \{ p_1: r_1, \ldots, p_k:r_k\}^{m} \!\in\! \PP \}$.
    Then $\Proc_{\mathtt{UT}}(\PP) = \{\mathcal{T}_\mathtt{UT}(\PP)\}$ is sound and complete for $\mathtt{AST}$ 
    and $\Proc_{\mathtt{UT}}(\InI, \PP) =
    \{(\mathcal{T}_\mathtt{UT}(\InI\cup \PP),\mathcal{T}_\mathtt{UT}(\PP))\}$ is sound and
    complete for $\mathtt{bAST}$.
\end{restatable} 

\begin{example}\label{ex:UsableTermsAlg}
    For $\mathtt{AST}$, $\Proc_{\mathtt{UT}}$ transforms 
    $\{\eqref{run2-ADP-1}, \eqref{run2-ADP-2}, \flat(\ref{run2-ADP-3})$ -
    $\flat(\ref{run2-ADP-7})\}$
    from \Cref{Ex:DepGraphAlg} into
    $\{(\ref{run2-ADP-1}'), (\ref{run2-ADP-2}'), \flat(\ref{run2-ADP-3})$ -
    $\flat(\ref{run2-ADP-7})\}$ with
    \begin{align*}
        \tloopOne(y) &\to \{\nicefrac{1}{2}:\tLoopOne(\tdouble(y)), \;
        \nicefrac{1}{2}:\tloopTwo(\tdouble(y))\}^{\ttrue}
        \tag{$\ref{run2-ADP-1}'$}\\
        \tloopOne(y) &\to \{\nicefrac{1}{3}:\tLoopOne(\ttriple(y)), \;
        \nicefrac{2}{3}:\tloopTwo(\ttriple(y))\}^{\ttrue} \tag{$\ref{run2-ADP-2}'$} 
    \end{align*} 
    The reason is that the left-hand sides of the only ADPs with annotations in the ADP
    problem have the root $\tloopOne$. Thus,
    $\tLoopTwo$-,
    $\tD$-, or $\tTriple$-terms are not usable.
    
    For $\mathtt{bAST}$, applying $\Proc_{\mathtt{UT}}$ to
    $(\{\eqref{run2-ADP-1}, \eqref{run2-ADP-2}\}, \{\eqref{run2-ADP-6},
    \flat(\ref{run2-ADP-1})$ - $\flat(\ref{run2-ADP-5}), \flat(\ref{run2-ADP-7})\})$
    and afterwards removing those ADPs from the reachability component that also occur in the second
    component yields
    $(\{(\ref{run2-ADP-1}'), (\ref{run2-ADP-2}'')\}, \{\eqref{run2-ADP-6},
    \flat(\ref{run2-ADP-1})$ - $\flat(\ref{run2-ADP-5}), \flat(\ref{run2-ADP-7})\})$ with
    \begin{align*}
        \tloopOne(y) &\to \{\nicefrac{1}{3}:\tLoopOne(\tTriple(y)), \; \nicefrac{2}{3}:\tloopTwo(\tTriple(y))\}^{\ttrue} \tag{$\ref{run2-ADP-2}''$}
    \end{align*}
    The reason is that the left-hand sides of ADPs with annotations in their right-hand sides
    have the root symbols $\tloopOne$ (in 
    $\eqref{run2-ADP-1}$ and $\eqref{run2-ADP-2}$) or $\ttriple$ (in $\eqref{run2-ADP-6}$).
\end{example}

\subsection{Usable Rules Processor}

In an innermost rewrite step,  all variables 
of the used rule are instantiated with normal forms.
The \emph{usable rules processor} detects rules that cannot be 
used below annotations in right-hand sides of ADPs when their variables 
are instantiated with normal forms.
For these rules we can set their flag to $\tfalse$,
indicating that the annotated subterms on their right-hand sides may still lead to a
non-$\mathtt{iAST}$ sequence, but the context of these annotations 
is irrelevant.

\begin{restatable}[Usable Rules Processor for $\mathtt{iAST}$]{theorem}{ProbUsRulesProc}\label{def:prob-usable-rules}
    Let $\PP$ be an ADP problem\linebreak
    and  for  $f\!\in\!\SignatureADC$,  let $\rules_{\PP}(f) =
    \{\ell\!\to\!\mu^{\true}\!\in\!\PP \mid \rootsym(\ell)\!=\!f\}$. 
    For $t\!\in\!\TT^{\#}$,\linebreak
    its \defemph{usable rules} $\urules_{\PP}(t)$ are the smallest set 
    with $\urules_{\PP}(x)=\emptyset$ for all $x\!\in\!\VSet$
    and $\urules_{\PP}(f(t_1, ..., t_n))\!=\!\rules_{\PP}(f) \cup \bigcup_{i = 1}^n\!\urules_{\PP}(t_i)  \cup \bigcup_{\ell \to \mu^{\true} \in \rules_{\PP}(f), r \in
      \Supp(\mu)} \urules_{\PP}(\flat(r))$,  otherwise.
    The \defemph{usable rules} of $\PP$ are
    $\urules(\PP) = \bigcup_{\ell \to \mu^{m} \in \PP, r \in \Supp(\mu), t \trianglelefteq_{\#} r}
    \urules_{\PP}(t^\#)$.
    Then $\Proc_{\mathtt{UR}}(\PP) = \{
    \urules(\PP) \cup
    \{\ell \to \mu^{\tfalse} \mid \ell
    \to \mu^{m} \in \PP \setminus \urules(\PP)\} \}$ is sound and complete,
    i.e., we turn the  flag of all non-usable rules to $\tfalse$.
\end{restatable}

\begin{example}\label{ex:R2Usable}
    The ADP problem $\{(\ref{run1-ADP-1}'), \eqref{run1-ADP-2}\}$
    has  no subterms below annotations.
    So\linebreak both rules are not
    usable and we set their flags to $\tfalse$ which leads to

  \vspace*{-.6cm}
    \begin{minipage}[t]{5cm}
        \begin{align*}
            \tg &\to \{\nicefrac{3}{4}:\td(\tG), \; \nicefrac{1}{4}:\tz\}^{\tfalse}
            \tag{$\ref{run1-ADP-1}''$}
        \end{align*}
    \end{minipage}
    \hspace*{.5cm}
    \begin{minipage}[t]{5cm}
        \begin{align*}
            \td(x) &\to \{1:\tc(x,x)\}^{\tfalse}
            \tag{$\ref{run1-ADP-2}'$}
        \end{align*}
    \end{minipage}
\end{example}

\vspace*{-.3cm}

\subsubsection{Adaption for $\mathtt{AST}$:}

For full rewriting and arbitrary start terms, the usable rules processor is unsound.
This is already the case for non-probabilistic rewriting, but
in the classical DP framework 
there nevertheless
exist processors for full rewriting based on
usable rules which rely on taking the $C_{\varepsilon}$-rules $\th(x,y) \to x$ and 
$\th(x,y) \to y$ for a fresh function symbol $\th$ into account, see, e.g.,
\cite{giesl2006mechanizing,gieslLPAR04dpframework,DBLP:journals/iandc/HirokawaM07,DBLP:journals/jar/Urbain04}.
However, the following example shows that this is not possible for $\mathtt{AST}$.

\begin{example}\label{example:usable-rules-ce-rules-full2}
   The ADP problem $\PP_{\tg}'$ from \Cref{example:usable-rules-ce-rules-full} 
is not $\mathtt{AST}$. It 
has no usable rules and thus,
$\Proc_{\mathtt{UR}}$ would transform
$\PP_{\tg}'$ into $\PP_{\tg}''$ where the flag of  all ADPs 
       is $\tfalse$.
    However, then
    we can no longer rewrite the argument $\tg$ of $\tF(\tg)$. Similarly, if we
    start with $\tF(\tG)$,  rewriting $\tG$ 
    would remove the annotation of $\tF$ above, i.e., 
    $\tF(\tG) \tored{}{}{\PP_{\tg}''} \{\nicefrac{1}{2}: \tf(\tc(\tg,\tg,\tg,\tg)), \nicefrac{1}{2}:\tf(\tz)\}$.
    Hence, then all CTs are finite. This also holds when
    adding the $C_{\varepsilon}$-ADPs  $\th(x,y) \to \{1:x\}^\ttrue$ and $\th(x,y) \to \{1:y\}^\ttrue$.
\end{example}

Thus, even integrating the $C_{\varepsilon}$-rules to represent
non-determinism would not 
allow a usable rule processor for $\mathtt{AST}$ with
arbitrary start terms. 
Moreover, the corresponding proofs in the non-probabilistic setting rely
on the minimality property, which does not hold in the
probabilistic setting, see \Cref{remark:minimality}.

%\vspace*{-.2cm}

\subsubsection{Adaption for $\mathtt{bAST}$:}

For $\mathtt{bAST}$, we can apply the usable rules processor
as for innermost rewriting. 
Since the start term is basic,  in the first application of an ADP
all variables are instantiated with normal forms.
Hence, the only rules that can be applied for rewrite steps below annotated symbols are
the ones that are introduced in right-hand sides of ADPs.
Therefore, we can use the same definitions as in \Cref{def:prob-usable-rules}
to over-approximate the set of ADPs that can be used below an annotated symbol
in a CT that starts with a basic term.
Here, we have to consider the reachability component as well for the usable rules,
as these ADPs can also be used in the initial rewrite steps.

\begin{restatable}[Usable Rules Processor for
    $\mathtt{bAST}$]{theorem}{ProbUsRulesProc}\label{def:prob-usable-rules-basic-AST}
  The following
  processor is sound and complete for $\mathtt{bAST}$:
  \[ \begin{array}{rcll@{\,}ll}
   \Proc_{\mathtt{UR}}(\InI,\PP) &=& \{
    \bigl(&\bigl( \InI &\cap \; \urules(\InI \cup \PP)\bigr)\;&\cup \;
    \{\ell \to \mu^{\tfalse} \mid \ell
    \to \mu^{m} \in \InI \setminus
    \urules(\InI \cup \PP)\},\\
&&&\bigl( \PP &\cap \;\urules(\InI \cup \PP) \bigr)&\cup \;
    \{\ell \to \mu^{\tfalse} \mid \ell
    \to \mu^{m} \in \PP \setminus
    \urules(\InI \cup \PP)\} \;\bigr) \}.
    \end{array}\]
\end{restatable}

\begin{example}\label{ex:UsableRulesAlg}
    For the basic ADP problem
    $(\{(\ref{run2-ADP-1}'), (\ref{run2-ADP-2}'')\}, \{\eqref{run2-ADP-6},
    \flat(\ref{run2-ADP-1})$ - $\flat(\ref{run2-ADP-5}),\linebreak[3] \flat(\ref{run2-ADP-7})\})$
    from \Cref{ex:UsableTermsAlg}, only the $\tdouble$- and $\ttriple$-ADPs 
    $\flat(\ref{run2-ADP-4}), \flat(\ref{run2-ADP-5}), 
    \eqref{run2-ADP-6},  \flat(\ref{run2-ADP-7})$ are usable.
    So we can set the flag of all other ADPs in this problem
    to $\tfalse$. 
    The same holds for the other  basic ADPs resulting  from the dependency graph and the
    usable terms processor in this example,   i.e., here the usable rules processor also sets the flags of all
    ADPs except the $\tdouble$- and $\ttriple$-ADPs 
    to $\tfalse$. 
\end{example}

\vspace*{-.2cm}

\begin{example}
    To see why we  use 
    $\PP \cap \urules(\InI \cup \PP)$ instead of
    $\urules(\PP)$  in
    \Cref{def:prob-usable-rules-basic-AST}
    (whereas\linebreak $\mathcal{T}_\mathtt{UT}(\PP)$ instead of $\mathcal{T}_\mathtt{UT}(\InI \cup \PP)$
    suffices for the second component in \Cref{theorem:prob-UPP}),
    consider the
    basic ADP problem $(\{ \eqref{usable-rule-counter-rule-0} \},
    \PP_{\tg}')$ from
    \cref{example:usable-rules-ce-rules-full}
    which is not $\mathtt{bAST}$.
    As noted in \Cref{example:usable-rules-ce-rules-full2},
    $\urules(\PP_{\tg}') = \emptyset$, but if one sets the flags 
    of all ADPs in $\PP_{\tg}'$ to $\tfalse$, then all CTs are finite (i.e., then
    $\Proc_{\mathtt{UR}}$
    would be unsound). In contrast, 
    for $\InI = \{
    \eqref{usable-rule-counter-rule-0} \}$, we have 
    $\urules(\InI \cup \PP_{\tg}') = \{ \eqref{usable-rule-counter-rule-2} \}$,
    because $\tg$ occurs below the annotated symbol $\tF$ in
    $\eqref{usable-rule-counter-rule-0}$. Hence,
    $\Proc_{\mathtt{UR}}(\{ \eqref{usable-rule-counter-rule-0} \},
    \PP_{\tg}')$ only sets the flags of all ADPs except $\eqref{usable-rule-counter-rule-2}$
    to $\tfalse$ and thus, the resulting basic ADP problem is still not
    $\mathtt{bAST}$.
\end{example}

\vspace*{-.2cm}

\begin{example}\label{UsRuleUnsoundNonSpare}
    Note that if one used GVRFs,
    then the usable rules processor would be unsound on ADP problems that are not weakly spare.
    For instance, it would transform the ADP problem 
    $(\emptyset, \{(\ref{run1-ADP-1}'), \eqref{run1-ADP-2}\})$ 
    (which is not $\mathtt{bAST}$ when using GVRFs)
    into $(\emptyset, \{(\ref{run1-ADP-1}''),
    (\ref{run1-ADP-2}')\})$ (see \Cref{ex:R2Usable}).
    However, as $(\ref{run1-ADP-2}')$ has the flag $\tfalse$, it cannot be applied at the
    position of the non-annotated symbol $\td$, since \Cref{def:ADPs-and-Rewriting} does not have a 
    case of the form $(\mathbf{nf})$. Hence, $(\emptyset, \{(\ref{run1-ADP-1}''),
    (\ref{run1-ADP-2}')\})$ is $\mathtt{bAST}$.
\end{example}

\vspace*{-.4cm}

\subsection{Reduction Pair Processor}\label{sec:ReductionPairProcessor}

Next we adapt the reduction pair processor
which lifts the direct use of orderings\linebreak from PTRSs to ADP problems.
This processor is the same \pagebreak[3] for $\mathtt{iAST}$ and $\mathtt{AST}$.

To handle expected values, as in
\cite{kassinggiesl2023iAST,FLOPS2024} we only consider
orderings based on po\-lynomial interpretations \cite{lankford1979proving}.
A \emph{polynomial interpretation} $\Pol$ is a $\SignatureADC$-algebra
which maps every function  $f \in \SignatureADC$ to a polynomial $f_{\Pol} \in \IN[\VSet]$.
It is \emph{monotonic} if\linebreak $x > y$ implies $f_{\Pol}(\ldots, x, \ldots) > f_{\Pol}(\ldots,
y, \ldots)$ for all $f \in \SignatureADC$.
$\Pol(t)$ denotes the \emph{interpretation} of
a term $t \in \TT^{\#}$ by $\Pol$.
An arithmetic inequation $\Pol(t_1) >\linebreak \Pol(t_2)$ \emph{holds} if it is true for
all instantiations of its variables by natural numbers.

The constraints (1) - (3) in \Cref{theorem:prob-RPP}
are based on the
conditions of a ranking function for $\mathtt{AST}$ as in \cite{mciver2017new}.
If we prove $\mathtt{AST}$
by considering the rules
$\ell \to \{ p_1:r_{1}, \ldots, p_k: r_k\}$
of a PTRS directly, then we need
a monotonic polynomial interpre\-tation $\Pol$ and require
a weak decrease when comparing
$\Pol(\ell)$ to the expected\linebreak value
$\sum_{1 \leq j \leq k} \, p_j \cdot \Pol(r_j)$
of the right-hand side, and additionally, at least one
$\Pol(r_j)$
must be strictly smaller than 
$\Pol(\ell)$ \cite{kassinggiesl2023iAST}.
For ADPs, we adapt these con\-straints
by comparing
the value $\Pol(\ell^\#)$ of
the annotated left-hand side with the\linebreak $\#$-\emph{sum} of the right-hand sides $r_j$, i.e.,
the sum of the polynomial values of their annotated subterms $\val(r_j) = \sum_{t
\trianglelefteq_{\#} r_j} \Pol(t^\#)$.
This allows us to
remove the requirement of (strong) monotonicity (every polynomial $f_{\Pol}$ with natural
coefficients is weakly monotonic, i.e.,
$x \geq y$ implies
$f_{\Pol}(\ldots, x, \ldots) \geq f_{\Pol}(\ldots, y, \ldots)$).

Here, (1) we require a weak decrease when comparing the annotated left-hand side 
with the expected value of $\#$-\emph{sums} in the right-hand side.
The processor then removes the annotations from those ADPs where (2) in addition there is at
least one right-hand side $r_j$ whose $\#$-\emph{sum} is
strictly decreasing.\footnote{In addition, the corresponding non-annotated
right-hand side $\flat(r_j)$ must be at least weakly decreasing.
This ensures that nested annotations behave ``monotonically''.
So we have to ensure that $\Pol(A) > \Pol(B)$
also implies that the $\#$-\emph{sum} of $F(A)$ is greater than $F(B)$,
i.e., $\Pol(A) > \Pol(B)$ must imply that
$\val(F(A)) = \Pol(F(a)) + \Pol(A) > \Pol(F(b)) + \Pol(B) = \val(F(B))$, which is ensured by $\Pol(a) \geq
\Pol(b)$.} 
Finally, (3) for  every rule with the flag $\ttrue$ (which can therefore be used for steps
below annotations),  
the expected value must be weakly decreasing when removing the annotations.
As in \cite{avanzini2020probabilistic,kassinggiesl2023iAST,FLOPS2024}, to ensure 
``monotonicity'' w.r.t.\ expected values, we restrict
ourselves to interpretations with multilinear polynomials, i.e.,
all monomials must have the
form $c \cdot x_1^{e_1} \cdot \ldots \cdot x_n^{e_n}$ with $c \in \NN$ and $e_1,\ldots,e_n \in \{0,1\}$.

\begin{restatable}[{\small Reduction Pair Processor for $\mathtt{iAST}$ \& $\mathtt{AST}$}]{theorem}{ProbRPP}\label{theorem:prob-RPP}
   $\!\!$ Let $\Pol\!:\!\TT^{\#} \to \IN[\VSet]\!$ be $\!\!\!$ a $\!\!\!$
    multilinear $\!\!\!$ polynomial $\!\!$ interpretation. $\!\!\!$ 
    Let $\PP\!=\!\PP_{\geq}\!\uplus\!\PP_{>}\!$ with $\PP_{>}\!\neq\!\emptyset\!$ where:
	\begin{itemize}
		\item[(1)] $\forall \ell \ruleArr{}{}{} \{ p_1:r_{1}, \ldots, p_k:
                  r_k\}^{m} \in \PP : \Pol(\ell^\#) \geq \sum_{1 \leq j \leq k} \, p_j \cdot \val(r_j)$. 
		\item[(2)] $\forall \ell \ruleArr{}{}{} \{ p_1:r_{1}, \ldots, p_k:
                  r_k\}^{m} \in \PP_{>} : \exists j \in \{1,\ldots,k\} : \Pol(\ell^\#) > \val(r_j)$.\\
		    If $m = \ttrue$, then we additionally have $\Pol(\ell) \geq
                    \Pol(\flat(r_j))$.
	\item[(3)] $\forall \ell \ruleArr{}{}{} \{ p_1:r_{1}, \ldots, p_k: r_k\}^{\ttrue}
          \in \PP : \Pol(\ell) \geq \sum_{1 \leq j \leq k} \, p_j \cdot \Pol(\flat(r_j))$.                   
		\end{itemize}
	Then $\Proc_{\mathtt{RP}}(\PP) = \{\PP_{\geq} \cup \flat(\PP_{>})\}$ is sound and
        complete for $\mathtt{iAST}$
        and $\mathtt{AST}$.
\end{restatable}

\begin{example}\label{RPPexample-iAST}
    To conclude $\mathtt{iAST}$ for $\R_2$ we have to remove all remaining
    annotations in the ADP problem $\{(\ref{run1-ADP-1}''), (\ref{run1-ADP-2}')\}$ from
    \Cref{ex:R2Usable} (then another application of the dependency graph processor
    yields the empty set of ADP problems).
    Here, we can use the reduction pair processor with the polynomial interpretation that
    maps $\tG$ to $1$,  
    and all other symbols to $0$. Then $(\ref{run1-ADP-2}')$ is weakly decreasing,
    and
    $(\ref{run1-ADP-1}'')$ is strictly decreasing, since  (1)
    $\Pol(\tG) = 1 \geq \nicefrac{3}{4} \cdot \val(\td(\tG)) +  \nicefrac{1}{4} \cdot
    \val(\tz) = \nicefrac{3}{4} \cdot \Pol(\tG) = \nicefrac{3}{4}$
    and (2) $\Pol(\tG) = 1 > \val(\tz) = 0$. Thus, the annotation
    of $\tG$ in $(\ref{run1-ADP-1}'')$ is deleted.

    Note that 
    this polynomial interpretation would also satisfy the constraints for $\DPair{\R_2} =
    \{ \eqref{run1-ADP-1}, \eqref{run1-ADP-2} \}$ from \Cref{example:running-1-ADPs},
    i.e., it would allow 
    us to remove the annotations from the canonical ADP directly. Hence, if we 
    extended our approach for $\mathtt{AST}$
    to GVRFs
    that can duplicate annotations, then the reduction pair processor would be
    unsound, as it would allow us to falsely ``prove'' $\mathtt{AST}$ of $\DPair{\R_2}$.
    The problem is that
    we compare terms with annotations via their  $\#$-sum, but for duplicating ADPs like
    \eqref{run1-ADP-2},
   $\Pol(\td(x)) \geq \Pol(\tc(x,x))$ does
    not imply
    $\val(\td(\tG)) \geq \val(\tc(\tG,\tG))$ since 
    $\val(\td(\tG))  = \Pol(\tG)$
    and $\val(\tc(\tG,\tG)) = \Pol(\tG) + \Pol(\tG)$.
\end{example}

\begin{example}\label{RPPexample-AST}
    To prove $\mathtt{AST}$ for $\R_{\mathsf{alg}}$, we also have to remove all
    annotations from all remaining sub-problems.
    For instance, for the sub-problem $\{(\ref{run2-ADP-1}'),
    (\ref{run2-ADP-2}'), \flat(\ref{run2-ADP-3})$ - $\flat(\ref{run2-ADP-7})\}$
    from \Cref{ex:UsableTermsAlg},
 we can use the reduction pair processor with the polynomial interpretation that maps $\ts(x)$ to $x+1$,
    $\tdouble(x)$ to $2x$, $\ttriple(x)$ to $3x$, $\tLoopOne(x)$ to $1$, and all other symbols to $0$.
    Then $(\ref{run2-ADP-2}')$ is strictly decreasing, since (1)
    $\Pol(\tLoopOne(y)) = 1 \geq \nicefrac{1}{3} \cdot \val(\tLoopOne(\ttriple(y))) +
    \nicefrac{2}{3} \cdot \val(\tloopTwo(\ttriple(y))) = \nicefrac{1}{3}$ and (2)
    $\Pol(\tLoopOne(y)) = 1 > \val(\tloopTwo(\ttriple(y))) = 0$.
    Similarly, $(\ref{run2-ADP-1}')$ is also strictly decreasing and 
    we can remove all annotations from this ADP problem.
    One can find similar interpretations to delete the remaining annotations
    also from the other remaining sub-problems.
    This proves $\mathtt{AST}$ for $\DPair{\R_{\mathsf{alg}}}$, and hence for $\R_{\mathsf{alg}}$.
\end{example}

\vspace*{-.4cm}

\subsubsection{Adaption for $\mathtt{bAST}$:}

To adapt the reduction pair processor to $\mathtt{bAST}$,  we only have to require
 the  conditions of
 \Cref{theorem:prob-RPP}
 for the second component $\PP$ of a basic ADP problem $(\InI, \PP)$.
 So the reachability component
$\InI$ is needed to determine which  rules are usable
in the usable rules processor,  but it does not result in  any additional
constraints for the reduction pair processor. Thus, proving $\mathtt{bAST}$ is never
harder than proving
$\mathtt{AST}$, since the second component changes in the same way for
$\mathtt{AST}$ and $\mathtt{bAST}$
in all processors
except for the usable rules processor, which is not applicable for $\mathtt{AST}$.
The conditions of
\Cref{theorem:prob-RPP}
ensure that to prove $\mathtt{AST}$,
infinitely many $(\mathbf{at})$ or $(\mathbf{af})$ steps
with
ADPs from $\PP_{>}$ do not have to be regarded
 anymore and thus, we can remove their
 annotations in $\PP$. However,  these ADPs may still be
 applied in finitely many initial $(\mathbf{at})$ or $(\mathbf{af})$ steps. Thus,
similar to the dependency graph processor, we have to keep the original annotated ADPs
from $\PP_{>}$ in the reachability
component  $\InI$.

\begin{restatable}[Reduction Pair Processor for $\mathtt{bAST}$]{theorem}{ProbRPPBast}\label{theorem:prob-RPP-bAST}
    Let $\Pol: \TT^{\#} \to \IN[\VSet]$ be a
    multilinear polynomial interpretation
and let $\PP = \PP_{\geq} \uplus \PP_{>}$ with $\PP_{>}\neq\emptyset$ satisfy
    the conditions of \Cref{theorem:prob-RPP}.
    Then $\Proc_{\mathtt{RP}}(\InI, \PP) =    \{(\InI \cup \PP_{>}, \PP_{\geq} \cup \flat(\PP_{>}))\}$ is sound and complete for $\mathtt{bAST}$.
\end{restatable}

\begin{example}
    If we only want to prove $\mathtt{bAST}$ of $\R_{\mathsf{alg}}$, 
    then the application of the reduction pair processor is easier than in \Cref{RPPexample-AST},
    as we have less constraints.
    For instance, consider the basic ADP problem from \Cref{ex:UsableRulesAlg}
    which results from $(\{(\ref{run2-ADP-1}'), (\ref{run2-ADP-2}'')\}, \{\eqref{run2-ADP-6},
    \flat(\ref{run2-ADP-1})$ - $\flat(\ref{run2-ADP-5}), \flat(\ref{run2-ADP-7})\})$
    by setting the flags of all ADPs except
    the $\tdouble$- and $\ttriple$-ADPs  $\flat(\ref{run2-ADP-4}), \flat(\ref{run2-ADP-5}), 
    \eqref{run2-ADP-6},  \flat(\ref{run2-ADP-7})$ to $\tfalse$. When using the polynomial
    interpretation
    $\Pol(\tTriple(x)) = x$, $\Pol(\ts(x)) = x+1$, $\Pol(\tdouble(x)) = 2x$, and
    $\Pol(\ttriple(x)) = 3x$, the ADP $\eqref{run2-ADP-6}$ is strictly decreasing and
    $\flat(\ref{run2-ADP-4})$ - $\flat(\ref{run2-ADP-7})$ are weakly decreasing. Thus, we can
    remove all annotations without having to regard any of the other (probabilistic) ADPs.
    In contrast, when proving $\mathtt{AST}$ instead of $\mathtt{bAST}$, all ADPs in the 
    corresponding ADP problem
    $\{\eqref{run2-ADP-6}, \flat(\ref{run2-ADP-1})$ - $\flat(\ref{run2-ADP-5}),
    \flat(\ref{run2-ADP-7})\}$
    have the flag $\ttrue$ and thus, here we have to find a polynomial interpretation which
    also makes the  ADPs $\flat(\ref{run2-ADP-1})$ - $\flat(\ref{run2-ADP-3})$ weakly decreasing. 
\end{example}

\subsection{Probability Removal Processor}

Finally, in proofs with the ADP framework, one may obtain
ADP problems $\PP$ with a non-probabilistic structure, 
i.e., every ADP has the form $\ell \to \{1:r\}^{m}$.
Then the \emph{probability removal processor} allows us to switch
to ordinary (non-probabilistic) DPs.
Ordinary DP problems for termination of TRSs have two components $(\mathcal{D}, \R)$: 
a set of dependency pairs $\mathcal{D}$, i.e., rules with annotations only at the roots of both sides,
and a TRS $\R$ containing rules that can be used below the annotations.
Such a DP problem is considered to be \emph{(innermost) non-terminating} if there exists
an
\emph{infinite chain} $t_0, t_1, t_2, \ldots$
with $t_i \to_{\mathcal{D}} \circ \to_{\R}^* t_{i+1}$ ($t_i \itodr \circ \itorstar t_{i+1}$)
for all $i \in \IN$.
Here, ``$\circ$'' denotes composition and $\itodr$ is the restriction of $\to_{\mathcal{D}}$ to rewrite steps
where the used redex is in $\NF_{\R}$.
This definition corresponds to an infinite chain tree consisting of only a single path.

\begin{restatable}[Probability Removal Processor for $\mathtt{iAST}$]{theorem}{INPP}\label{theorem:prob-NPP}
	Let $\PP$ be an ADP problem where every ADP in $\PP$ has the
    form $\ell \to \{1:r\}^{m}$.
    Let $\nonprobDP(\PP) = \{\ell^\# \to t^\# \mid \ell \to \{1\!:r\}^{m} \in \PP, t \trianglelefteq_{\#} r\}$.
    Then $\PP$ is $\mathtt{iAST}$ iff the
    non-probabilistic DP problem $(\nonprobDP(\PP),\nonprob(\PP))$ is innermost terminating.  
       So the processor $\Proc_{\mathtt{PR}}(\PP) = \emptyset$
    is sound and complete for $\mathtt{iAST}$ iff $(\nonprobDP(\PP), \nonprob(\PP))$ is
    innermost  terminating.
\end{restatable}

\subsubsection{Adaption for $\mathtt{AST}$ and $\mathtt{bAST}$:}

$\Proc_{\mathtt{PR}}$ works in an analogous way
for $\mathtt{(b)AST}$, i.e.,
for both $\mathtt{AST}$ and $\mathtt{bAST}$, 
we can switch to ordinary DPs for full rewriting.
Of course, here the ``only if'' direction does not hold for $\mathtt{bAST}$ because the
non-probabilistic DP framework considers arbitrary (possibly non-basic) start terms.

\begin{restatable}[Probability Removal Processor for $\mathtt{bAST}$ and $\mathtt{AST}$]{theorem}{NPP}\label{theorem:prob-NPP-AST-bAST}
    Let $\PP$ be an ADP problem where every ADP in $\PP$ has the
    form $\ell \to \{1:r\}^{m}$.
    Then $\PP$ is $\mathtt{AST}$ iff the
    non-probabilistic DP problem $(\nonprobDP(\PP),\nonprob(\PP))$ is
    terminating.
    So the processor $\Proc_{\mathtt{PR}}(\PP) = \emptyset$
    is sound and complete for $\mathtt{AST}$ iff $(\nonprobDP(\PP), \nonprob(\PP))$ is
    terminating.
    Similarly, $(\InI, \PP)$ is $\mathtt{bAST}$ if $(\nonprobDP(\PP),\nonprob(\PP))$ is
    terminating. So $\Proc_{\mathtt{PR}}(\InI, \PP)\linebreak = \emptyset$ is sound and complete for $\mathtt{bAST}$ if 
    $(\nonprobDP(\PP), \nonprob(\PP))$ is terminating.
\end{restatable}

\subsection{Switching From Full to Innermost AST}

In the non-probabilistic DP framework for analyzing termination of TRSs,
there is a processor to switch from full to innermost rewriting
if the DP problem satisfies certain conditions \cite[Thm.\ 32]{gieslLPAR04dpframework}.
This is useful as the DP framework for innermost termination is more
powerful
than the one for full termination and in this way, one can switch to the innermost case
for certain sub-problems, even if the whole TRS does not belong to any
class where innermost termination implies termination.
However, the soundness of this processor relies on the minimality property, which
does not hold in the probabilistic setting, see 
\Cref{remark:minimality}.
Indeed, the  following example which corresponds to \cite[Ex.\ 3.15]{thiemanndiss2007}
shows that a similar processor in the ADP framework would be unsound.

\begin{example}
    The ADP problem
    with $\tf(x) \to \{1:\tF(\ta)\}^{\ttrue}$ and $\ta \to \{1:\ta\}^{\ttrue}$ is not 
    $\mathtt{AST}$ as we can rewrite $\tF(\ta)$ to itself with the $\tf$-ADP.
    However, it is $\mathtt{iAST}$ as in innermost evaluations, we have to rewrite the inner
    $\ta$, which does not terminate but does not use any annotations,  
    i.e., any $(\mathbf{at})$ or $(\mathbf{af})$ steps.
    The ADPs are non-overlapping, and  left- and right-linear. Thus,
    \Cref{properties-eq-AST-iAST-1} to switch from full to innermost $\mathtt{AST}$ cannot be
    applied on the level of ADP problems.
\end{example}

Hence, for
$\mathtt{AST}$ of PTRSs that satisfy the conditions of \Cref{properties-eq-AST-iAST-1}
or \ref{properties-eq-AST-iAST-2},
one should apply the ADP framework for $\mathtt{iAST}$ \cite{FLOPS2024}, because
its processors are more powerful. But otherwise, one has to use our novel ADP framework
for full $\mathtt{AST}$.

\section{Conclusion and Evaluation}\label{Evaluation}

In this paper, we introduced the first DP framework for 
$\mathtt{AST}$ and $\mathtt{bAST}$ of PTRSs, which is based on the 
existing ADP framework from \cite{FLOPS2024} for $\mathtt{iAST}$.
It is particularly useful when analyzing $\mathtt{(b)AST}$
of overlapping PTRSs,
as for such PTRSs we cannot use the criteria of \cite{FoSSaCS2024}
for classes of PTRSs where $\mathtt{iAST}$  implies $\mathtt{(b)AST}$.

Compared to the non-probabilistic DP framework for  termination of TRSs\linebreak \cite{arts2000termination,gieslLPAR04dpframework,giesl2006mechanizing,hirokawa2005automating,DBLP:journals/iandc/HirokawaM07},
analyzing $\mathtt{AST}$ automatically is significantly more difficult due to\linebreak 
the lack of  a ``minimality property'' in the probabilistic setting,
which would allow\linebreak several further processors.
Moreover, the ADP framework for PTRSs is
restricted to multilinear reduction pairs.
The following table compares the ADP frameworks for $\mathtt{AST}$, $\mathtt{bAST}$, and $\mathtt{iAST}$.
The parts in \emph{italics} show the differences to the non-probabilistic
DP framework.
Here, 
``S'' and ``C'' stand  for ``sound'' and ``complete''.

\smallskip

\begin{center}
    \scriptsize
    \begin{tabular}{|c|c|c|c|}
        \hline
        Processor & ADP for $\mathtt{AST}$ & ADP for $\mathtt{bAST}$ & ADP for $\mathtt{iAST}$\\
        \hline
        \hline
        Chain Crit.\ & \textcolor{green}{S} \& \textcolor{blue}{C} \emph{for
          non-duplicating} & \textcolor{green}{S} \& \textcolor{blue}{C} \emph{for
          weakly spare} & \textcolor{green}{S} \& \textcolor{blue}{C} \\[1px]
        \hline
        Dep.\ Graph & \textcolor{green}{S} \& \textcolor{blue}{C} & \textcolor{green}{S} \& \textcolor{blue}{C} & \textcolor{green}{S} \& \textcolor{blue}{C} \\
        \hline
        \emph{Usable Terms} & \textcolor{green}{S} \& \textcolor{blue}{C} & \textcolor{green}{S} \& \textcolor{blue}{C} & \textcolor{green}{S} \& \textcolor{blue}{C} \\
        \hline
        Usable Rules & \textcolor{red}{$\lnot$ S} \emph{(even with $C_\varepsilon$-Rules)}& \textcolor{green}{S} \& \textcolor{blue}{C} & \textcolor{green}{S} \& \textcolor{blue}{C} \\[1px]
        \hline
        Reduction Pairs & \textcolor{green}{S} \& \textcolor{blue}{C} \emph{(multilinearity)}& \textcolor{green}{S} \& \textcolor{blue}{C} \emph{(multilinearity)} & \textcolor{green}{S} \& \textcolor{blue}{C} \emph{(multilinearity)} \\
        \hline
        \emph{Probability Removal} & \textcolor{green}{S} \& \textcolor{blue}{C} & \textcolor{green}{S} \& \textcolor{blue}{C} & \textcolor{green}{S} \& \textcolor{blue}{C} \\[1px]
        \hline
    \end{tabular}
\end{center}

\smallskip

For our experimental evaluation, we compared all existing approaches
to prove $\mathtt{(b)AST}$ of PTRSs.
More precisely, we compared our implementation of the novel ADP framework for
 $\mathtt{(b)AST}$ in
a new version of \aprove{} \cite{JAR-AProVE2017}
with the old version of \aprove{} that only implements the techniques from \cite{FoSSaCS2024,kassinggiesl2023iAST,FLOPS2024},
and  with the direct application of polynomial interpretations from
\cite{kassinggiesl2023iAST}.\footnote{In addition,  an alternative technique to analyze
PTRSs via a direct application of 
interpretations 
was presented in \cite{avanzini2020probabilistic}.
However, \cite{avanzini2020probabilistic} analyzes $\mathtt{PAST}$
(or rather \emph{strong}  $\mathtt{AST}$), and a
comparison with their technique can be found in \cite{kassinggiesl2023iAST}.}

To this end, we extended the existing benchmark set of 118 PTRSs from
\cite{FoSSaCS2024} by
  12 new examples including all PTRSs
  presented in this paper and PTRSs for typical 
probabilistic algorithms on lists and trees.
Of these 130 examples, the direct application of polynomials can find 37 (1) $\mathtt{AST}$ proofs,
\emph{old} \aprove{} shows $\mathtt{AST}$ for
50 (1) PTRSs, 
and our \emph{new} \aprove{} version proves $\mathtt{AST}$ for 
58 (6) examples.
In brackets we indicate the number of $\mathtt{AST}$ proofs when only regarding the 
12 new examples.
The 118 benchmarks from \cite{FoSSaCS2024} lack non-determinism by overlapping rules and
thus,
here we are only able to prove $\mathtt{AST}$ for three more examples
than \emph{old}\linebreak \aprove{}.
In contrast, our new 12 examples contain non-determinism and  create random
data objects, which are accessed or modified afterwards  (see \Cref{Examples}).\linebreak
Our experiments show that our novel ADP framework can for the first time
prove $\mathtt{AST}$ of such PTRSs.
If we consider basic start terms, the numbers rise to 62 (1) for \emph{old} \aprove{} and
74 (8) for \emph{new} \aprove{}.
For details on our experiments and for instructions on how to run our implementation in
\aprove{} via its \emph{web
interface} or locally, see \url{https://aprove-developers.github.io/ADPFrameworkFullAST}.

Reduction pairs were also adapted to disprove reachability \cite{reach2022Akihisa}, 
and thus, 
in the\linebreak future
we will also
integrate
reachability analysis into the ADP framework for $\mathtt{bAST}$.\linebreak
Moreover, we  aim to 
analyze stronger properties like $\mathtt{PAST}$
via DPs.
Here, we will again start with innermost evaluation,
which is easier to analyze.
Fur\-thermore, we want to develop methods to automatically
disprove $\mathtt{(P)AST}$ of PTRSs.

\vspace*{-.1cm}

\paragraph{Acknowledgments.}
This paper is dedicated to Joost-Pieter Katoen whose
groundbreaking work on verification of probabilistic programs
laid the foundations for this whole research area. His scientific excellence,
his enthusiasm in developing outstanding new research results, and his 
energy and commitment in the establishment of new research projects
(like, e.g., the DFG research training group \textsf{UnRAVeL})
are outstanding. While originally we only analyzed ``classical'' (non-probabilistic)
programs, it is due to 
Joost-Pieter and this research training group
that we extended the focus of our research
towards probabilistic programs.  Joost-Pieter
is not only a major inspiration for our work and a fantastic chair of the research
training group \textsf{UnRAVeL}, but he is a great and close colleague, and we
look
forward to many more joint years together in Aachen at the Chair i2.

\paper{\printbibliography}
\report{\bibliographystyle{splncs04}
 %\bibliography{biblioReport}
  \providecommand{\noopsort}[1]{}

}

\appendix

\report{\clearpage

\section*{Appendix}

\vspace*{-.1cm}

In \Cref{Examples}, we present three examples to demonstrate how our novel
ADP
framework can be used for full rewriting on
data structures like lists or trees. 
\Cref{appendix} contains all proofs for our new contributions and observations.

\section{Examples}\label{Examples}

In this section, we show that
in contrast to most other techniques for analyzing $\mathtt{AST}$, due to
probabilistic term rewriting, our approach is also suitable for the analysis
of algorithms on algebraic data structures other than numbers.}

\paper{

\section{Appendix}\label{Examples}
In this appendix, we present three examples to demonstrate how our novel
ADP
framework can be used for full rewriting on
data structures like lists or trees.  They
show that
in contrast to most other techniques for analyzing $\mathtt{AST}$, due to
probabilistic term rewriting, our approach is also suitable for the analysis
of algorithms on algebraic data structures other than numbers.}

\subsection{Lists}\label{Examples-List}

We start with algorithms on lists.
Similar to \Cref{alg1}, the following algorithm first creates a random list,
filled with random numbers, and afterwards  
uses the list for further computation.
In general, algorithms that access or modify randomly generated lists
can be analyzed by our new ADP framework.

The algorithm below computes the sum of all numbers in the generated list.\linebreak
Here, natural numbers are again represented via the constructors $\tz$ and $\ts$, and\linebreak
lists are represented via $\tnil$ (for the empty list) and $\tcons$, where, e.g.,
$\tcons(\ts(\tz),\linebreak \tcons(\ts(\tz), \tcons(\tz, \tnil)))$ represents the list $[1,1,0]$. The function
$\tcreateL(\xs)$ adds a prefix of arbitrary length filled with arbitrary natural numbers
in front of the list $\xs$. Moreover, $\tapp(\xs,\ys)$ concatenates the two lists $\xs$
and $\ys$. Finally, for a non-empty list $\xs$ of numbers, $\tsum(\xs)$ computes a
singleton list 
whose only element is the sum of all numbers in $\xs$. So
$\tsum(\tcons(\ts(\tz), \tcons(\ts(\tz), \tcons(\tz, \tnil))))$ evaluates to
$\tsum(\ts(\ts(\tz)), \tnil)$.

\vspace*{-.2cm}

{\scriptsize
\begin{align*}
    \tinit & \to \{ 1 : \tsum(\tcreateL(\tnil))\} \\
    \taddNum(x, \xs) & \to \{ \nicefrac{1}{2} : \tcons(x, \xs) ,\nicefrac{1}{2} : \taddNum(\ts(x), \xs)\} \\
    \tcreateL(\xs) & \to \{ \nicefrac{1}{2} : \taddNum(\tz, \xs) ,\nicefrac{1}{2} : \tcreateL(\taddNum(\tz, \xs))\} \\
    \tplus(\tz, y) & \to \{ 1 : y\} \\
    \tplus(\ts(x), y) & \to \{ 1 : \ts(\tplus(x, y))\} \\
    \tsum(\tcons(x, \tnil)) & \to \{ 1 : \tcons(x, \tnil)\} \\
    \tsum(\tcons(x, \tcons(y, \ys))) & \to \{ 1 : \tsum(\tcons(\tplus(x, y), \ys))\} \\
    \tsum(\tapp(\xs, \tcons(x, \tcons(y, \ys)))) & \to \{ 1
    : \tsum(\tapp(\xs, \tsum(\tcons(x, \tcons(y, \ys)))))\} \\ 
    \tapp(\tcons(x, \xs), \ys) & \to \{ 1 : \tcons(x, \tapp(\xs, \ys))\} \\
    \tapp(\tnil, \ys) & \to \{ 1 : \ys\} \\
    \tapp(\xs, \tnil) & \to \{ 1 : \xs \}\!
\end{align*}
}

Note that the left-hand sides of the two rules $\tapp(\tnil, \ys) \to \{ 1 : \ys \}$ and $\tapp(\xs, \tnil) \to \{ 1 : \xs \}$
overlap
and moreover, the last $\tsum$-rule overlaps with the first $\tapp$-rule.
Hence, we cannot use the techniques from \cite{FoSSaCS2024} to analyze full
$\mathtt{AST}$ of
this PTRS.
Furthermore, there exists no polynomial ordering that proves $\mathtt{AST}$ for this
example directly (i.e., without the use of DPs), because the left-hand side of the last
$\tsum$-rule is embedded in its right-hand side.
With our new ADP framework, \aprove{}
can now prove $\mathtt{AST}$ of this example automatically.

Next, consider the following adaption of this example.
Here, we only create lists of even numbers.

\vspace*{-.2cm}

{\scriptsize
\begin{align*}
    \tinit & \to \{ 1: \tsum(\tcreateL(\tnil))\} \\
    \taddNum(x, \xs) & \to \{ \nicefrac{1}{2} : \tcons(\tplus(x, x), \xs) ,\nicefrac{1}{2} : \taddNum(\ts(x), \xs)\} \\
    \tcreateL(\xs) & \to \{ \nicefrac{1}{2} : \taddNum(\tz, \xs) ,\nicefrac{1}{2} : \tcreateL(\taddNum(\tz, \xs)) \}\\
    \tplus(\tz, y) & \to \{ 1 : y\} \\
    \tplus(\ts(x), y) & \to \{ 1 : \ts(\tplus(x, y))\} \\
    \tsum(\tcons(x, \tnil)) & \to \{ 1: \tcons(x, \tnil)\} \\
    \tsum(\tcons(x, \tcons(y, \xs))) & \to \{ 1: \tsum(\tcons(\tplus(x, y), \xs))\} \\
    \tsum(\tapp(\xs, \tcons(x, \tcons(y, \ys)))) & \to \{ 1 : \tsum(\tapp(\xs, \tsum(\tcons(x, \tcons(y, \ys)))))\} \\
    \tapp(\tcons(x, \xs), \ys) & \to \{ 1 : \tcons(x, \tapp(\xs, \ys))\} \\
    \tapp(\tnil, \ys) & \to \{ 1 : \ys\} \\
    \tapp(\xs, \tnil) & \to \{ 1 : \xs\} \!
\end{align*}
}

Due to the subterm $\tplus(x, x)$ in the right-hand side, the $\taddNum$-rule is duplicating.
Hence, we cannot use the ADP framework for $\mathtt{AST}$.
However, the PTRS is weakly spare, as the arguments of $\tplus$ cannot contain defined
function symbols if we start with a basic term.
Hence, \aprove{} can use the ADP framework for $\mathtt{bAST}$ and successfully prove
$\mathtt{bAST}$ of this example.

\subsection{Trees}\label{Examples-Tree}

As another example, our new ADP framework
can also deal with trees.
In the following algorithm (adapted from \cite{AG01}), we consider binary trees represented
via $\tleaf$ and $\ttree(x,y)$,
where $\tconcat(x,y)$ replaces the rightmost leaf of the tree $x$ by $y$.
The algorithm first creates two random trees and then checks
whether the first tree has less leaves than the second one.

\vspace*{-.2cm}

{\scriptsize
\begin{align*}
    \tinit & \to \{ 1 : \tlessleaves(\tcreateT(\tleaf), \tcreateT(\tleaf)) \} \\
    \tconcat(\tleaf, y) & \to \{ 1 : y \} \\
    \tconcat(\ttree(u, v), y) & \to \{ 1 : \ttree(u, \tconcat(v, y)) \} \\
    \tlessleaves(x, \tleaf) & \to \{ 1 : \tfalse \} \\
    \tlessleaves(\tleaf, \ttree(x, y)) & \to \{ 1 : \ttrue \} \\
    \tlessleaves(\ttree(u, v), \ttree(x, y)) & \to \{ 1 : \tlessleaves(\tconcat(u, v), \tconcat(x, y)) \} \\
    \tcreateT(\xs) & \to \{ 1 : \xs \} \\
    \tcreateT(\xs) & \to \{ \nicefrac{1}{3} : \xs ,\nicefrac{1}{3} : \tcreateT(\ttree(\xs, \tleaf)) ,\nicefrac{1}{3} : \tcreateT(\ttree(\tleaf, \xs)) \} \!
\end{align*}
}

Note that the last two rules are overlapping. 
Again, our new ADP framework is able to prove $\mathtt{AST}$ for this example,
while both \cite{FoSSaCS2024} and the direct application of polynomial interpretations fail.

\report{\section{Proofs}\label{appendix}

In this section, we give all proofs for our new results and observations.
In the following, let $L(x) = (p_x^{\F{T}}, t_x^{\F{T}})$ denote the labeling of the node
$x$ in the chain tree $\F{T}$. 
We say that a CT (or RST) $\F{T}$ \emph{converges (or terminates) with probability}
$p \in \IR$ if we have $|\F{T}|_{\ctleaf} = p$.
Moreover, we often write
$\anno_{\varepsilon}(t)$ instead of $\anno_{\{\varepsilon\}}(t)$ and
$\annoD(t)$ instead of $\anno_{\pos_{\SignatureD}(t)}(t)$ to annotate all defined symbols
in a term $t$.
We can now start with the proof of the sound and complete chain criterion for $\mathtt{AST}$.

\ProbChainCriterionFull*

\begin{myproof}
    \smallskip
    
    \noindent
    \underline{\emph{Soundness:}} Assume that $\R$ is not $\mathtt{AST}$.
    Then, there exists an $\R$-RST $\F{T}=(V,E,L)$
    whose root is labeled with $(1:t)$ for some term $t \in \TT$
    that converges with probability $<1$.
    We will construct a $\DPair{\R}$-CT $\F{T}' = (V,E,L',V \setminus \ctleaf^{\F{T}})$
    with the same underlying tree structure and an adjusted labeling such that
    $p_x^{\F{T}} = p_x^{\F{T}'}$ for all $x \in V$, where all the inner nodes are in $A$. 
    Since the tree structure and the probabilities are the same, we then get $|\F{T}|_{\ctleaf} = |\F{T}'|_{\ctleaf}$.
    To be precise, the set of leaves in $\F{T}$ is equal to the set of leaves in $\F{T}'$, and they have the same probabilities.
    Since $|\F{T}|_{\ctleaf} < 1$, we thus have $|\F{T}'|_{\ctleaf} < 1$.
    Hence, there exists a $\DPair{\R}$-CT $\F{T}'$ that converges with probability $<1$
    and $\DPair{\R}$ is not $\mathtt{AST}$ either.
    \vspace*{-10px}
    \begin{center}
        \scriptsize
        \begin{tikzpicture}
            \tikzstyle{adam}=[thick,draw=black!100,fill=white!100,minimum size=4mm, shape=rectangle split, rectangle split parts=2,rectangle split
            horizontal]
            \tikzstyle{empty}=[rectangle,thick,minimum size=4mm]
            
            \node[adam] at (-3.5, 0)  (a) {$1$ \nodepart{two} $t$};
            \node[adam] at (-5, -0.8)  (b) {$p_1$ \nodepart{two} $t_{1}$};
            \node[adam] at (-2, -0.8)  (c) {$p_2$ \nodepart{two} $t_{2}$};
            \node[adam] at (-6, -1.6)  (d) {$p_3$ \nodepart{two} $t_3$};
            \node[adam] at (-4, -1.6)  (e) {$p_4$ \nodepart{two} $t_4$};
            \node[adam] at (-2, -1.6)  (f) {$p_5$ \nodepart{two} $t_5$};
            \node[empty] at (-6, -2.4)  (g) {$\ldots$};
            \node[empty] at (-4, -2.4)  (h) {$\ldots$};
            \node[empty] at (-2, -2.4)  (i) {$\ldots$};

            \node[empty] at (-0.5, -1.2)  (arrow) {\Huge $\leadsto$};
            
            \node[adam,pin={[pin distance=0.1cm, pin edge={,-}] 140:\tiny \textcolor{blue}{$A$}}] at (3.5, 0)  (a2) {$1$ \nodepart{two} $\annoD(t)$};
            \node[adam,pin={[pin distance=0.1cm, pin edge={,-}] 140:\tiny \textcolor{blue}{$A$}}] at (2, -0.8)  (b2) {$p_1$ \nodepart{two} $\annoD(t_1)$};
            \node[adam,pin={[pin distance=0.1cm, pin edge={,-}] 45:\tiny \textcolor{blue}{$A$}}] at (5, -0.8)  (c2) {$p_2$ \nodepart{two} $\annoD(t_2)$};
            \node[adam,pin={[pin distance=0.1cm, pin edge={,-}] 140:\tiny \textcolor{blue}{$A$}}] at (1, -1.6)  (d2) {$p_3$ \nodepart{two} $\annoD(t_3)$};
            \node[adam,pin={[pin distance=0.1cm, pin edge={,-}] 45:\tiny \textcolor{blue}{$A$}}] at (3, -1.6)  (e2) {$p_4$ \nodepart{two} $\annoD(t_4)$};
            \node[adam,pin={[pin distance=0.1cm, pin edge={,-}] 45:\tiny \textcolor{blue}{$A$}}] at (5, -1.6)  (f2) {$p_5$ \nodepart{two} $\annoD(t_5)$};
            \node[empty] at (1, -2.4)  (g2) {$\ldots$};
            \node[empty] at (3, -2.4)  (h2) {$\ldots$};
            \node[empty] at (5, -2.4)  (i2) {$\ldots$};
        
            \draw (a) edge[->] (b);
            \draw (a) edge[->] (c);
            \draw (b) edge[->] (d);
            \draw (b) edge[->] (e);
            \draw (c) edge[->] (f);
            \draw (d) edge[->] (g);
            \draw (e) edge[->] (h);
            \draw (f) edge[->] (i);

            \draw (a2) edge[->] (b2);
            \draw (a2) edge[->] (c2);
            \draw (b2) edge[->] (d2);
            \draw (b2) edge[->] (e2);
            \draw (c2) edge[->] (f2);
            \draw (d2) edge[->] (g2);
            \draw (e2) edge[->] (h2);
            \draw (f2) edge[->] (i2);
        \end{tikzpicture}
    \end{center}
    \vspace*{-10px}
    We label all nodes $x \in V$ in $\F{T}'$ with $\annoD(t_x)$, where $t_x$ is the term for the node $x$ in $\F{T}$.
    The annotations ensure that we rewrite with Case $(\mathbf{at})$ of \Cref{def:ADPs-and-Rewriting-full} so that the node $x$ is contained in $A$. 
    We only have to show that $\F{T}'$ is indeed a valid CT, i.e., that the edge relation represents valid rewrite steps with $\tored{}{}{\DPair{\R}}$.
    Let $x \in V \setminus \ctleaf$ and $xE = \{y_1, \ldots, y_k\}$ be the set of its successors.
    Since $x$ is not a leaf, we have $t_x \to_{\R} \{\tfrac{p_{y_1}}{p_x}:t_{y_1}, \ldots, \tfrac{p_{y_k}}{p_x}:t_{y_k}\}$.
    This means that there is a rule $\ell \to \{p_1:r_1, \ldots, p_k:r_k\} \in \R$, a position $\pi$, and a substitution $\sigma$ such that ${t_x}|_\pi = \ell\sigma$.
    Furthermore, we have $t_{y_j} = t_x[r_j \sigma]_{\pi}$ for all $1 \leq j \leq k$.
    
    The corresponding ADP for the rule is $\ell \to \{ p_1 : \annoD(r_1), \ldots, p_k : \annoD(r_k) \}^{\ttrue}$.
    Furthermore, $\pi \in \posT(\annoD(t_x))$ as all defined symbols are annotated in $\annoD(t_x)$.
    Hence, we can rewrite $\annoD(t_x)$ with $\ell \to \{ p_1 : \annoD(r_1), \ldots, p_k : \annoD(r_k) \}^{\ttrue}$, 
    using the position $\pi$, the substitution $\sigma$, and Case $(\mathbf{at})$ of \Cref{def:ADPs-and-Rewriting-full} applies.
    Furthermore, we take some VRF $(\varphi_j)_{1 \leq j \leq k}$ that is surjective on
    the positions of the variables in the right-hand side, i.e., for all $1 \leq j \leq k$
    and all positions $\tau \in \pos_{\VSet}(r_j)$ there exists a $\tau' \in
    \pos_{\VSet}(\ell)$ such that $\varphi_j(\tau') = \tau$. 
    Such a VRF must exist, since $\R$ is non-duplicating.
    We have $\annoD(t_x) \tored{}{}{\DPair{\R}} \{p_1: \annoD(t_{y_1}), \ldots, p_k: \annoD(t_{y_k})\}$ since
    by rewriting with $(\mathbf{at})$ we get $\annoD(t_x)[\anno_{\Phi_j}(r_j \sigma)]_{\pi} = \annoD(t_{y_j})$
    with $\Phi_j$ defined as in \Cref{def:ADPs-and-Rewriting-full}.
    Note that since the used VRF is surjective on the variable positions of the right-hand side, we do not remove any annotation in the substitution.
    Furthermore, we annotated all defined symbols in $r_j$.
    Thus, we result in $\annoD(t_{y_j})$ where all defined symbols are annotated again.
    \smallskip
   
    \noindent
    \underline{\emph{Completeness:}} Assume that $\DPair{\R}$ is not $\mathtt{AST}$.
    Then, there exists a $\DPair{\R}$-CT $\F{T} = (V,E,L,A)$ whose root is labeled with $(1:t)$ for some annotated term $t \in \TT^{\#}$
    that converges with probability $<1$.
    We will construct an $\R$-RST $\F{T}' = (V,E,L')$ with the same underlying tree structure and an adjusted labeling such that $p_x^{\F{T}} = p_x^{\F{T}'}$ for all $x \in V$.
    Since the tree structure and the probabilities are the same, we then get $|\F{T}'|_{\ctleaf} = |\F{T}|_{\ctleaf} < 1$.
    Therefore, there exists an $\R$-RST $\F{T}'$ that converges with probability $<1$.
    Hence, $\R$ is not $\mathtt{AST}$ either.

    \begin{center}
        \scriptsize
        \begin{tikzpicture}
            \tikzstyle{adam}=[thick,draw=black!100,fill=white!100,minimum size=4mm, shape=rectangle split, rectangle split parts=2,rectangle split
            horizontal]
            \tikzstyle{empty}=[rectangle,thick,minimum size=4mm]
            
            \node[adam] at (-3.5, 0)  (a) {$1$ \nodepart{two} $t$};
            \node[adam] at (-5, -0.8)  (b) {$p_1$ \nodepart{two} $t_{1}$};
            \node[adam] at (-2, -0.8)  (c) {$p_2$ \nodepart{two} $t_{2}$};
            \node[adam] at (-6, -1.6)  (d) {$p_3$ \nodepart{two} $t_3$};
            \node[adam] at (-4, -1.6)  (e) {$p_4$ \nodepart{two} $t_4$};
            \node[adam] at (-2, -1.6)  (f) {$p_5$ \nodepart{two} $t_5$};
            \node[empty] at (-6, -2.4)  (g) {$\ldots$};
            \node[empty] at (-4, -2.4)  (h) {$\ldots$};
            \node[empty] at (-2, -2.4)  (i) {$\ldots$};

            \node[empty] at (-0.5, -1.2)  (arrow) {\Huge $\leadsto$};
            
            \node[adam] at (3.5, 0)  (a2) {$1$ \nodepart{two} $\flat(t)$};
            \node[adam] at (2, -0.8)  (b2) {$p_1$ \nodepart{two} $\flat(t_1)$};
            \node[adam] at (5, -0.8)  (c2) {$p_2$ \nodepart{two} $\flat(t_2)$};
            \node[adam] at (1, -1.6)  (d2) {$p_3$ \nodepart{two} $\flat(t_3)$};
            \node[adam] at (3, -1.6)  (e2) {$p_4$ \nodepart{two} $\flat(t_4)$};
            \node[adam] at (5, -1.6)  (f2) {$p_5$ \nodepart{two} $\flat(t_5)$};
            \node[empty] at (1, -2.4)  (g2) {$\ldots$};
            \node[empty] at (3, -2.4)  (h2) {$\ldots$};
            \node[empty] at (5, -2.4)  (i2) {$\ldots$};
        
            \draw (a) edge[->] (b);
            \draw (a) edge[->] (c);
            \draw (b) edge[->] (d);
            \draw (b) edge[->] (e);
            \draw (c) edge[->] (f);
            \draw (d) edge[->] (g);
            \draw (e) edge[->] (h);
            \draw (f) edge[->] (i);

            \draw (a2) edge[->] (b2);
            \draw (a2) edge[->] (c2);
            \draw (b2) edge[->] (d2);
            \draw (b2) edge[->] (e2);
            \draw (c2) edge[->] (f2);
            \draw (d2) edge[->] (g2);
            \draw (e2) edge[->] (h2);
            \draw (f2) edge[->] (i2);
        \end{tikzpicture}
    \end{center}
    \vspace*{-10px}
    We label all nodes $x \in V$ in $\F{T}'$ with $\flat(t_x)$, where $t_x$ is the term for the node $x$ in $\F{T}$, 
    i.e., we remove all annotations.
    We only have to show that $\F{T}'$ is indeed a valid RST, 
    i.e., that the edge relation represents valid rewrite steps with $\to_{\R}$, 
    but this follows directly from the fact that if we remove all annotations in \Cref{def:ADPs-and-Rewriting-full},
    then we get the ordinary probabilistic term rewriting relation again.
\end{myproof}

Next, we prove the sound and complete chain criterion for $\mathtt{bAST}$.
In the following,
for two (possibly annotated) terms $s,t$ we define $s \doteq t$ if $\flat(s) = \flat(t)$.

\ProbChainCriterionFullBasic*

\begin{myproof}
  Since the reachability component of the canonical basic ADP problem is empty, we just
  consider $\DPair{\R}$-CTs.

    \smallskip 
    
    \noindent
    \underline{\emph{Soundness:}} 
    We use the same construction as for $\mathtt{AST}$ but the definition of the terms 
    in the new $\DPair{\R}$-CT is slightly different.
    Previously, all defined symbols were annotated.
    Now, we do not annotate all of them, but we may remove annotations from defined symbols
    if the corresponding subterm at this position is in normal form.
    Furthermore, the initial term $t$ is a basic term and thus, $\annoD(t) = t^\#$.

    \vspace*{-10px}
    \begin{center}
        \scriptsize
        \begin{tikzpicture}
            \tikzstyle{adam}=[thick,draw=black!100,fill=white!100,minimum size=4mm, shape=rectangle split, rectangle split parts=2,rectangle split
            horizontal]
            \tikzstyle{empty}=[rectangle,thick,minimum size=4mm]
            
            \node[adam] at (-3.5, 0)  (a) {$1$ \nodepart{two} $t$};
            \node[adam] at (-5, -0.8)  (b) {$p_1$ \nodepart{two} $t_{1}$};
            \node[adam] at (-2, -0.8)  (c) {$p_2$ \nodepart{two} $t_{2}$};
            \node[adam] at (-6, -1.6)  (d) {$p_3$ \nodepart{two} $t_3$};
            \node[adam] at (-4, -1.6)  (e) {$p_4$ \nodepart{two} $t_4$};
            \node[adam] at (-2, -1.6)  (f) {$p_5$ \nodepart{two} $t_5$};
            \node[empty] at (-6, -2.4)  (g) {$\ldots$};
            \node[empty] at (-4, -2.4)  (h) {$\ldots$};
            \node[empty] at (-2, -2.4)  (i) {$\ldots$};

            \node[empty] at (-0.5, -1.2)  (arrow) {\Huge $\leadsto$};
            
            \node[adam,pin={[pin distance=0.1cm, pin edge={,-}] 140:\tiny \textcolor{blue}{$A$}}] at (3.5, 0)  (a2) {$1$ \nodepart{two} $t^\#$};
            \node[adam,pin={[pin distance=0.1cm, pin edge={,-}] 140:\tiny \textcolor{blue}{$A$}}] at (2, -0.8)  (b2) {$p_1$ \nodepart{two} $t'_1$};
            \node[adam,pin={[pin distance=0.1cm, pin edge={,-}] 45:\tiny \textcolor{blue}{$A$}}] at (5, -0.8)  (c2) {$p_2$ \nodepart{two} $t'_2$};
            \node[adam,pin={[pin distance=0.1cm, pin edge={,-}] 140:\tiny \textcolor{blue}{$A$}}] at (1, -1.6)  (d2) {$p_3$ \nodepart{two} $t'_3$};
            \node[adam,pin={[pin distance=0.1cm, pin edge={,-}] 45:\tiny \textcolor{blue}{$A$}}] at (3, -1.6)  (e2) {$p_4$ \nodepart{two} $t'_4$};
            \node[adam,pin={[pin distance=0.1cm, pin edge={,-}] 45:\tiny \textcolor{blue}{$A$}}] at (5, -1.6)  (f2) {$p_5$ \nodepart{two} $t'_5$};
            \node[empty] at (1, -2.4)  (g2) {$\ldots$};
            \node[empty] at (3, -2.4)  (h2) {$\ldots$};
            \node[empty] at (5, -2.4)  (i2) {$\ldots$};
        
            \draw (a) edge[->] (b);
            \draw (a) edge[->] (c);
            \draw (b) edge[->] (d);
            \draw (b) edge[->] (e);
            \draw (c) edge[->] (f);
            \draw (d) edge[->] (g);
            \draw (e) edge[->] (h);
            \draw (f) edge[->] (i);

            \draw (a2) edge[->] (b2);
            \draw (a2) edge[->] (c2);
            \draw (b2) edge[->] (d2);
            \draw (b2) edge[->] (e2);
            \draw (c2) edge[->] (f2);
            \draw (d2) edge[->] (g2);
            \draw (e2) edge[->] (h2);
            \draw (f2) edge[->] (i2);
        \end{tikzpicture}
    \end{center}
    \vspace*{-10px}
    We construct the new labeling $L'$ for the $\DPair{\R}$-CT inductively such that for all inner nodes $x \in V \setminus \ctleaf$ with children nodes $xE = \{y_1,\ldots,y_k\}$ we have $t_x' \tored{}{}{\DPair{\R}} \{\tfrac{p_{y_1}}{p_x}:t_{y_1}', \ldots, \tfrac{p_{y_k}}{p_x}:t_{y_k}'\}$.
    Let $X \subseteq V$ be the set of nodes $x$ where we have already defined the labeling $L'(x)$.
    Furthermore, for any term $t \in \TT$ let
    $\normalfont{\PosDPoss}(t,\R) = \{\pi \mid \pi \in \posD(t), t|_\pi \notin \NF_{\R}\}$.
    During our construction, we ensure that the following property holds:
    \begin{equation} \label{chain-crit-1-soundness-induction-hypothesis}
        \parbox{.9\textwidth}{For every node $x \in X$ we have $t_x \doteq t_x'$ and $\PosDPoss(t_x,\R) \subseteq \posT(t_x')$.}
    \end{equation}
    This means that the corresponding term $t_x$ for the node $x$ in $\F{T}$ 
    has the same structure as the term $t_x'$ in $\F{T}'$,
    and additionally, at least all possible redexes in $t_x$ are annotated in $t_x'$.
    The annotations ensure that we rewrite with Case $(\mathbf{at})$ of \Cref{def:ADPs-and-Rewriting-full} so that the node $x$ is contained in $A$. 
    We label the root of $\F{T}'$ with $t^\#$.
    Here, we have $t \doteq t^\#$ and $\PosDPoss(t,\R) = \{\varepsilon\} = \posT(t^\#)$, since $t$ is a basic term.
    As long as there is still an inner node $x \in X$ such that its successors are not contained in $X$, we do the following.
    Let $xE = \{y_1, \ldots, y_k\}$ be the set of its successors.
    We need to define the corresponding terms $t_{y_1}', \ldots, t_{y_k}'$ for the nodes $y_1, \ldots, y_k$.
    Since $x$ is not a leaf, we have $t_x \to_{\R} \{\tfrac{p_{y_1}}{p_x}:t_{y_1}, \ldots, \tfrac{p_{y_k}}{p_x}:t_{y_k}\}$, i.e., there is a rule $\ell \to \{p_1:r_1, \ldots, p_k:r_k\} \in \R$, a position $\pi$, and a substitution $\sigma$ such that ${t_x}|_\pi = \ell\sigma$.
    Furthermore, we have $t_{y_j} = t_x[r_j \sigma]_{\pi}$ for all $1 \leq j \leq k$.
    
    The corresponding ADP for the rule is $\ell \to \{ p_1 : \annoD(r_1), \ldots, p_k : \annoD(r_k) \}^{\ttrue}$.
    Furthermore, $\pi \in \PosDPoss(t_x,\R) \subseteq_{(IH)} \posT(t_x')$ and $t_x \doteq_{(IH)} t_x'$.
    Hence, we can rewrite $t_x'$ with $\ell \to \{ p_1 : \annoD(r_1), \ldots, p_k : \annoD(r_k) \}^{\ttrue}$, 
    using the position $\pi$
    and the substitution $\sigma$, and Case $(\mathbf{at})$ of \Cref{def:ADPs-and-Rewriting-full} applies.
    Additionally, we use a VRF $(\varphi_j)_{1 \leq j \leq k}$ that is surjective on the
    positions of those variables 
    that occur at least as often in the left-hand side as in the right-hand sides. The
    positions of all other variables (i.e., all variables that are duplicated) are
    mapped to $\bot$.
    Note that $\R$ is weakly spare, hence such variables can only be instantiated by normal forms.
    We get $t_x' \tored{}{}{\DPair{\R}} \{p_1: t_{y_1}', \ldots, p_k: t_{y_k}'\}$ with
    $t_{y_j}' = t_x'[\anno_{\Phi_j}(r_j\sigma)]_{\pi}$ by $(\mathbf{at})$ 
    with $\Phi_j$ defined as in \Cref{def:ADPs-and-Rewriting-full}.
    This means that we have $t_{y_j} \doteq t_{y_j}'$.
    It remains to prove $\PosDPoss(t_{y_j},\R) \subseteq \posT(t_{y_j}')$ for all $1 \leq j \leq k$.
    For all positions $\tau \in \PosDPoss(t_{y_j},\R) = \PosDPoss(t_x[r_j \sigma]_{\pi},\R)$ that are orthogonal or above $\pi$, we have $\tau \in \PosDPoss(t_{x},\R) \subseteq_{(IH)} \posT(t_{x}')$, and all annotations orthogonal or above $\pi$ remain in $t_{y_j}'$ as they were in $t_{x}'$.
    For all positions $\tau \in \PosDPoss(t_{y_j},\R) = \PosDPoss(t_x[r_j \sigma]_{\pi},\R)$ that are below $\pi$, we have two possibilities: (1) at least the defined root symbol of $t_{y_j}|_{\tau}$ is inside $r_j$, 
    and thus $\tau \in \posT(t_{y_j}')$, as all defined symbols of $r_j$ are annotated in 
    $t_{y_j}' = t_x'[\anno_{\Phi_j}(r_j) \sigma]_{\pi}$,
      or (2) $\tau$ is below a non-duplicated variable 
    (otherwise the subterm $t_{y_j}|_{\tau}$ has to be a normal form and thus,
    $\tau \notin \PosDPoss(t_{y_j},\R)$),
    and hence, it is still annotated in $t_{y_j}'$ due to the used VRF.
    This ends the induction proof for this direction.
    \smallskip
    
    \noindent
    \underline{\emph{Completeness:}} Completely the same as in the proof of \Cref{theorem:prob-chain-criterion-full}.
\end{myproof}

Next, we consider the theorems regarding the processors 
that we adapted from \cite{FLOPS2024} to our new framework for $\mathtt{AST}$ and
$\mathtt{bAST}$.
We first recapitulate the notion of a \emph{sub-chain tree} from \cite{reportkg2023iAST}.

\begin{definition}[Subtree, Sub-CT] \label{def:chain-tree-induced-sub}
    Let $\PP$ be an ADP problem and let $\F{T} = (V,E,L,A)$ be a tree that satisfies Conditions (1)-(5) of a $\PP$-CT.
	Let $W \subseteq V$ be non-empty, weakly connected, and for all $x \in W$ we have $xE \cap W = \emptyset$ or $xE \cap W = xE$.
	Then, we define the \defemph{subtree} (or \defemph{sub-CT} if it satisfies Condition (6) as well) $\F{T}[W]$ by $\F{T}[W] = (W,E \cap (W \times W),L^W,A \cap (W \setminus W_{\ctleaf}))$.
	Here, $W_{\ctleaf}$ denotes the leaves of the tree $G^{\F{T}[W]} = (W, E \cap (W
        \times W))$ so that the new set $A  \cap (W \setminus W_{\ctleaf})$ only contains inner nodes.
	Let $w \in W$ be the root of $G^{\F{T}[W]}$.
	To ensure that the root of our subtree has the probability $1$ again,
	we use the labeling $L^W(x) = (\frac{p_{x}^{\F{T}}}{p_w^{\F{T}}}: t_{x}^{\F{T}})$
	for all nodes $x \in W$.
    If $W$ contains the root of $(V, E)$, then we call the sub-chain tree \defemph{grounded}.
\end{definition}

\begin{example}
    Reconsider the PTRS $\R_1$ containing the only rule $\tg \to \{\nicefrac{3}{4}:\td(\tg), \, \nicefrac{1}{4}:\tz\}$.
    Below one can see the $\R_1$-RST from \Cref{Probabilistic Term Rewriting} (on the left), and the subtree that starts at the node of the term $\td(\tg)$ (on the right).
    Note that the probabilities are normalized such that the root has the probability $1$ again.
    \vspace*{-5px}
    \begin{center}
        \scriptsize
        \begin{tikzpicture}
            \tikzstyle{adam}=[thick,draw=black!100,fill=white!100,minimum size=4mm, shape=rectangle split, rectangle split parts=2,rectangle split
            horizontal]
            \tikzstyle{empty}=[rectangle,thick,minimum size=4mm]
            
            \node[adam] at (-3, 0)  (a) {$1$\nodepart{two}$\tg$};
            \node[adam] at (-4, -0.7)  (b) {$\nicefrac{3}{4}$\nodepart{two}$\td(\tg)$};
            \node[adam,label=below:{$\quad \mathtt{NF}_{\R_{1}}$}] at (-2, -0.7)  (c) {$\nicefrac{1}{4}$\nodepart{two}$\tz$};
            \node[adam] at (-5, -1.4)  (d) {$\nicefrac{9}{16}$\nodepart{two}$\td(\td(\tg))$};
            \node[adam,label=below:{$\mathtt{NF}_{\R_{1}}$}] at (-3, -1.4)  (e) {$\nicefrac{3}{16}$\nodepart{two}$\td(\tz)$};
            \node[empty] at (-5.5, -2)  (f) {$\ldots$};
            \node[empty] at (-4.5, -2)  (g) {$\ldots$};
            
            \draw (a) edge[->] (b);
            \draw (a) edge[->] (c);
            \draw (b) edge[->] (d);
            \draw (b) edge[->] (e);
            \draw (d) edge[->] (f);
            \draw (d) edge[->] (g);

            \node[adam] at (2, -0.7)  (b2) {$1$\nodepart{two}$\td(\tg)$};
            \node[adam] at (1, -1.4)  (d2) {$\nicefrac{3}{4}$\nodepart{two}$\td(\td(\tg))$};
            \node[adam,label=below:{$\mathtt{NF}_{\R_{1}}$}] at (3, -1.4)  (e2) {$\nicefrac{1}{4}$\nodepart{two}$\td(\tz)$};
            \node[empty] at (0.5, -2)  (f2) {$\ldots$};
            \node[empty] at (1.5, -2)  (g2) {$\ldots$};

            \draw (b2) edge[->] (d2);
            \draw (b2) edge[->] (e2);
            \draw (d2) edge[->] (f2);
            \draw (d2) edge[->] (g2);
        \end{tikzpicture}
    \end{center}
    \vspace*{-10px}
\end{example}

The property of being non-empty and weakly connected ensures that the resulting 
graph $G^{\F{T}[W]}$ is a tree again.
The property that we either have $xE \cap W = \emptyset$ or $xE \cap W = xE$ ensures that the sum of 
probabilities for the successors of a node $x$ is equal to the probability for the node $x$ itself.

Next, we recapitulate a lemma
and adapt another important lemma from \cite{reportkg2023iAST}.
Afterwards, we prove the theorems on the processors.
We start with the \emph{A-partition lemma}.
This lemma was proven in \cite{reportkg2023iAST}
(where it was called ``P-partition lemma'')
and still applies to our new ADP problems, since the structure of our
CTs is the same as in \cite{kassinggiesl2023iAST,reportkg2023iAST}.

\begin{lemma}[A-Partition Lemma]\label{lemma:p-partition}
    Let $\PP$ be an ADP problem and let $\F{T} = (V,E,L,A)$ be a $\PP$-CT that converges with probability $<1$.
    Assume that we can partition $A = A_1 \uplus A_2$ such that every sub-CT that only contains $A$-nodes from $A_1$ converges with probability $1$.
    Then there is a grounded sub-CT $\F{T}'$ that converges with probability $<1$ such that every infinite path has an infinite number of nodes from $A_2$.
\end{lemma}

\begin{myproof}
    See \cite{reportkg2023iAST}, as this proof does not depend on the used rewrite strategy but just on the structure of a chain tree.
\end{myproof}
  
Next, we adapt the proof of the \emph{starting lemma} from
\cite{reportkg2023iAST,FLOPSreport2024}.
It shows that w.l.o.g., we can assume 
that we label the root of our CT with $(1:t)$
for an annotated term $t$ such that $\posT(t) = \{\varepsilon\}$, i.e., only the root is annotated.
This obviously holds for $\mathtt{bAST}$, but
the starting lemma shows that it can also be assumed for
$\mathtt{AST}$.
For the following proofs,
we extend \Cref{def:ADPs-and-Rewriting-full} by the missing $(\mathbf{nf})$ case.
As discussed in \Cref{Probabilistic Annotated Dependency Pairs}, this does not change the definition of $\mathtt{AST}$ for chain
trees, but it allows us to still perform the same rewrite steps in a CT if annotations are
removed.

\begin{definition}[Rewriting with ADPs Including $(\mathbf{nf})$]\label{def:ADPs-and-Rewriting-full-nf}
    Let $\PP$ be an ADP problem.
    A term $s \in \TT^{\#}$ \defemph{rewrites} with $\PP$ to
    $\mu = \{p_1:t_1,\ldots,p_k:t_k\}$ (denoted $s \tored{}{}{\PP} \mu$)
    if there are an ADP $\ell \ruleArr{}{}{} \{ p_1:r_{1}, \ldots, p_k: r_k\}^{m} \in
    \PP$, a VRF $(\varphi_j)_{1 \leq j \leq k}$ for this ADP, a substitution $\sigma$,
    and a position $\pi \in \pos_{\SignatureD \cup 
      \SignatureA}(s)$ such that $\flat(s|_\pi)=\ell\sigma$,
    and for all $1 \leq j \leq k$ we have
    \begin{equation*}
        \begin{array}{rllllll@{\hspace*{1cm}}l}
        t_j &=                  &s[\anno_{\Phi_j}(r_j\sigma)]_{\pi}         & \text{if} & \pi \in \posT(s)    & \text{and} & m = \ttrue   & (\mathbf{at})\\
        t_j &= \disannoPos{\pi}(&s[\anno_{\Phi_j}(r_j\sigma)]_{\pi})        & \text{if} & \pi \in \posT(s)    & \text{and} & m = \tfalse  & (\mathbf{af})\\
        t_j &=                  &s[\anno_{\Psi_j}(r_j\sigma)]_{\pi}  & \text{if} & \pi \not\in\posT(s) & \text{and} & m = \ttrue   & (\mathbf{nt})\\
        t_j &= \disannoPos{\pi}(&s[\anno_{\Psi_j}(r_j\sigma)]_{\pi})  & \text{if} & \pi \not\in\posT(s) & \text{and} & m = \tfalse   & (\mathbf{nf})\!
        \end{array}
    \end{equation*}
    {\small Here, $\Psi_j\!=\!\{\varphi_j(\rho).\tau \mid \rho \!\in\! \pos_{\VSet}(\ell), \,
    \varphi_j(\rho)\!\neq\!\bot,  \, \rho.\tau \!\in\! \posT(s|_{\pi}) \}$
    and $\Phi_j\!=\!\posT(r_j) \cup \Psi_j$.}
\end{definition}

\begin{lemma}[Starting Lemma]\label{lemma:starting}
    If an ADP problem $\PP$ is not $\mathtt{AST}$, then there exists a $\PP$-CT
    $\F{T}$ with $|\F{T}|_{\ctleaf} < 1$ that starts with $(1:t)$ where $\posT(t) = \{\varepsilon\}$. 
\end{lemma}

\begin{myproof}
    We prove the contraposition.
    Assume that every $\PP$-CT $\F{T}$
    converges with probability $1$ if it 
    starts with $(1:t)$ and $\posT(t) = \{\varepsilon\}$.
    We now prove that then also every $\PP$-CT $\F{T}$ 
    that starts with $(1:t)$ for some arbitrary term $t$ converges with probability $1$, 
    and thus $\PP$ is $\mathtt{AST}$\@.
    We prove the claim by induction on the number of annotations in the initial term $t$.

    If $t$ contains no annotation, then the CT
    starting with $(1:t)$ is trivially finite (it cannot contain an infinite path, since there are no nodes in $A$) and hence, it converges with probability $1$.
    Next, if $t$ contains exactly one annotation at position $\pi$, then we can ignore everything above the annotation, as we will never use an $A$-step above the annotated position, and we cannot duplicate or change annotations by rewriting above them, since we use VRFs and not GVRFs.
    However, for $t|_{\pi}$ with $\posT(t|_{\pi}) = \{\varepsilon\}$,
    we know by our assumption that such a CT converges with probability $1$.

    Now we regard the induction step, and assume for a contradiction that for a term $t$ with $n > 1$ annotations, there is a CT $\F{T}$ that
    converges with probability $< 1$.
    Here, our induction hypothesis is that every $\PP$-CT $\F{T}$ that starts with $(1:t')$, where $t'$ contains $m$ annotations for some $1 \leq m < n$ converges with probability $1$.\linebreak
    Let $\Pi_1 = \{\tau\}$ and $\Pi_2 = \{\chi \in \posT(t) \mid \chi \neq \tau\}$ for some $\tau \in \posT(t)$ and consider the two terms $\anno_{\Pi_1}(t)$ and $\anno_{\Pi_2}(t)$, which contain both strictly less than $n$ annotations.
    By our induction hypothesis, we know that every $\PP$-CT that starts with $(1:\anno_{\Pi_1}(t))$ or $(1:\anno_{\Pi_2}(t))$ converges with probability $1$.
    Let $\F{T}_1 = (V, E, L_1, A_1)$ be the tree that starts with $(1:\anno_{\Pi_1}(t))$ 
    and uses the same rules as we did in $\F{T}$. (Here, the new definition of the rewrite
    relation $\tored{}{}{\PP}$ including the case
    $(\mathbf{nf})$ from \Cref{def:ADPs-and-Rewriting-full-nf} is needed in order to ensure that one can still use the same rules as in $\F{T}$ although we now may have less annotations.)
    
    We can partition $A$ into the sets $A_1$ and $A_2 = A \setminus A_1$.
    Note that $\F{T}_1$ itself may not be a $\PP$-CT again, since there might exist paths without 
    an infinite number of $A_1$-nodes, 
    but obviously every subtree $\F{T}'_1$ of $\F{T}_1$ such that every infinite path has an infinite number 
    of $A_1$-nodes is a $\PP$-CT again.
    Moreover, by extending such a subtree to be grounded, i.e., adding the initial path from the root of
    $\F{T}_1$ to $\F{T}'_1$,
    we created a $\PP$-CT that starts with $\anno_{\Pi_1}(t)$, 
    and hence by our induction hypothesis, converges with probability $1$. Thus, this also
    holds for $\F{T}'_1$.
    
    We want to use the A-partition lemma (\Cref{lemma:p-partition}) for the tree $\F{T}$.
    For this, we have to show that every sub-CT $\F{T}'_1$ of $\F{T}$ 
    that only contains $A$-nodes from $A_1$ converges with probability $1$.
    But since $\F{T}'_1$ only contains $A$-nodes from $A_1$
    it must either contain infinitely many $A_1$-nodes, 
    and by the previous paragraph it converges with probability $1$,
    or it contains only finitely many $A_1$-nodes, hence must be finite itself,
    and also converges with probability $1$.

    Now, we have shown that the conditions for the A-partition lemma (\cref{lemma:p-partition}) are satisfied.
    Thus, we can apply the A-partition lemma to obtain a grounded sub-CT
    $\F{T}'$ of $\F{T}$ with $|\F{T}'|_{\ctleaf} < 1$ such that on every infinite path, we have an infinite number of $A_2$ nodes.
    Let $\F{T}_2$ be the tree that starts with $\anno_{\Pi_2}(t)$
    and uses the same rules as we did in $\F{T}'$.
    Again, all local properties for a $\PP$-CT are satisfied.
    Additionally, this time we know that every infinite path has an infinite number of $A_2$-nodes in $\F{T}'$, hence we also know that the global property for $\F{T}_2$ is satisfied.
    This means that $\F{T}_2$ is a $\PP$-CT that starts with
    $\anno_{\Pi_2}(t)$ and with $|\F{T}_2|_{\ctleaf} < 1$.
    This is our desired contradiction, which proves the induction step.
\end{myproof}

As only the root position of the term at the root of the CT is annotated, we can 
assume that
the step from the root of the CT to its children corresponds to a 
rewrite step at the root of this term.
The reason is that we have to eventually rewrite at this root position of the term in a CT that converges with 
probability $<1$. Hence, w.l.o.g.\ we can start with this root rewrite step.

Next, we adapt the soundness and completeness proofs for the processors from \cite{FLOPSreport2024} from
innermost to full rewriting.

\ProbDepGraphProc*

\begin{myproof}
    Let $\overline{X} = X \cup \flat(\PP \setminus X)$ for $X \subseteq \PP$.
    \smallskip
   
    \noindent
    \underline{\emph{Completeness:}} Every $\overline{\PP_i}$-CT is also a $\PP$-CT with fewer annotations in the terms.
    So if some $\overline{\PP_i}$ is not $\mathtt{AST}$, then there exists a $\overline{\PP_i}$-CT $\F{T}$ that converges with probability $<1$. 
    By adding annotations to the terms of the tree, we result in a $\PP$-CT that converges with probability $<1$ as well.
    Hence, if $\overline{\PP_i}$ is not $\mathtt{AST}$, then $\PP$ is not $\mathtt{AST}$ either.
     \medskip
   
    \noindent
    \underline{\emph{Soundness:}} Let $\F{G}$ be the $\PP$-dependency graph.
    Suppose that every $\overline{\PP_i}$-CT converges with probability $1$ for all $1 \leq i \leq n$.
    We prove that then also every $\PP$-CT converges with probability 1.
    Let $W = \{\PP_1, \ldots, \PP_n\} \cup \{\{v\} \subseteq \PP \mid v$ is not in an SCC  of $\F{G}\}$ be the set of all SCCs and all singleton sets of nodes that do not belong to any SCC\@.
    The core steps of this proof are the following:
    \begin{enumerate}
      \item[1.] We show that every ADP problem $\overline{X}$ with $X \in W$ is $\mathtt{AST}$\@.
      \item[2.] We show that composing SCCs maintains the $\mathtt{AST}$ property.
       \item[3.] We show that for every $X \in W$, the ADP problem $\overline{\bigcup_{X >_{\F{G}}^* Y}Y}$ is $\mathtt{AST}$ by induction on $>_{\F{G}}$.
      \item[4.] We conclude that $\PP$ must be $\mathtt{AST}$\@.
    \end{enumerate}
    Here, for two $X_1,X_2 \in W$ we say that $X_2$ is a \emph{direct successor} of $X_1$ (denoted
    $X_1 >_{\F{G}} X_2$) if there exist nodes $v \in X_1$ and $w \in X_2$ such that there is
    an edge from $v$ to $w$ in $\F{G}$.
  
    \medskip
  
    \noindent
    \textbf{\underline{1. Every ADP problem $\overline{X}$ with $X \in W$ is $\mathtt{AST}$\@.}}
  
    \noindent
    We start by proving the following:
    \begin{equation}
      \label{W is AST}
      \mbox{Every ADP problem $\overline{X}$ with $X \in W$ is $\mathtt{AST}$\@.}
    \end{equation}
    To prove~\eqref{W is AST}, note that if $X$ is an SCC, then it follows from our assumption that $\overline{X}$ is $\mathtt{AST}$\@.
    If $X$ is a singleton set of a node that does not belong to any SCC,
    then assume for a contradiction that $\overline{X}$ is not $\mathtt{AST}$\@.
    By \cref{lemma:starting} there exists an $\overline{X}$-CT $\F{T} = (V,E,L,A)$ that converges with probability
    $< 1$ and starts with $(1:t)$ where $\posT(t) = \{\varepsilon\}$
    and $\flat(t) = s\theta$ for a substitution $\theta$ and some ADP $s \to \{ p_1:t_1,
    \ldots, p_k:r_k \}^m \in \overline{X}$.
    If $s \to \ldots \notin X$, then the resulting terms after the first rewrite step contain no annotations anymore and this cannot start a CT that converges with probability $<1$.
    Hence, we have $s \to \ldots \in X$ and thus,
    $X = \{s \to \ldots \}$, since $X$ is a singleton set.
    Assume for a contradiction that there exists a node $x \in A$ in $\F{T}$ that is not the root and introduces new annotations.
    W.l.o.G., let $x$ be reachable from the root without traversing any other node that introduces new annotations.
    This means that for the corresponding term $t_x$ for
    node $x$ there is a $t' \trianglelefteq_{\#} t_x$ at position $\tau$ such that $t' = s \sigma'$ for
    some substitution $\sigma'$ and the only ADP $s \to \ldots \in X$ (since this
    is the only ADP in $\overline{X}$ that contains any annotations in the right-hand side).
    Let $(z_0, \ldots, z_m)$ with $z_m = x$ be the path from the root to $x$ in $\F{T}$.
    The first rewrite step at the root must be $s \theta \tored{}{}{\overline{X}} \{p_1:r_1 \theta, \ldots, p_k:r_k \theta\}$.
    After that, we only use ADPs with the flag $\ttrue$ below the annotated position that
    will be used for the rewrite step at node $x$,
    as otherwise, the position $\tau$ would not be annotated in $t_x$.
    Therefore, we must have an $1 \leq j \leq k$ and a $t'' \trianglelefteq_{\#} r_j$ such that 
    $t''^\# \theta \to_{\nonprob(\PP)}^* s^\# \sigma'$, 
    which means that there must be a self-loop for the only
    ADP in $X$, which is a contradiction to our assumption that $X$ is a
    singleton consisting of an ADP that is not in any SCC of $\F{G}$.\linebreak
    \indent Now, we have proven that the $\overline{X}$-CT $\F{T}$ does not introduce new annotations.
    By definition of a $\PP$-CT, every infinite path must contain an infinite number of nodes in $A$, i.e., nodes where we rewrite at an annotation.
    Thus, every path in $\F{T}$ must be finite, which means that $\F{T}$ is finite itself,
    as the tree is finitely branching.
    But every finite CT converges with probability $1$, which is a contradiction to
    our assumption that $\F{T}$
    converges with probability $<1$. 
  
    \medskip
  
    \noindent
    \textbf{\underline{2. Composing SCCs maintains the $\mathtt{AST}$ property.}}
  
    \noindent
    Next, we show that composing SCCs maintains the $\mathtt{AST}$ property. More precisely, we prove
    the following:
    \begin{equation}
    \label{Composing AST}
      \parbox{.9\textwidth}{Let $\hat{X} \subseteq W$ and $\hat{Y} \subseteq W$
        such that there are no $X_1,X_2 \!\in\! \hat{X}$ and $Y \!\in\! \hat{Y}$ which
        satisfy both $X_1 >_{\F{G}}^* Y >_{\F{G}}^* X_2$ and $Y \not\in \hat{X}$, and such that there are no $Y_1,Y_2 \!\in\! \hat{Y}$ and $X \!\in\! \hat{X}$ which
        satisfy both $Y_1 >_{\F{G}}^* X >_{\F{G}}^* Y_2$ and $X \not\in \hat{Y}$.
      If both $ \overline{\bigcup_{X \in \hat{X}} X} $ and $ \overline{\bigcup_{Y \in \hat{Y}} Y} $ are $\mathtt{AST}$, then $ \overline{\bigcup_{X \in \hat{X}} X \cup \bigcup_{Y \in \hat{Y}} Y} $ is $\mathtt{AST}$.}
    \end{equation}
    To show~\eqref{Composing AST}, we assume that both $\overline{\bigcup_{X \in \hat{X}} X}$ and $\overline{\bigcup_{Y \in \hat{Y}} Y}$ are $\mathtt{AST}$\@.
    Let $\overline{Z} = \overline{\bigcup_{X \in \hat{X}} X \cup \bigcup_{Y \in \hat{Y}} Y}$.
    The property in~\eqref{Composing AST} for $\hat{X}$ and $\hat{Y}$ says that a
    path between two nodes from $\bigcup_{X \in \hat{X}} X$ that only traverses nodes
    from $Z$ must also be a path that only traverses
    nodes from $\bigcup_{X \in \hat{X}} X$, so that $\bigcup_{Y \in \hat{Y}} Y$ cannot be used to ``create'' new paths between two nodes from $\bigcup_{X \in \hat{X}} X$, and vice versa.
    Assume for a contradiction that $\overline{Z}$ is not $\mathtt{AST}$\@.
    By \cref{lemma:starting} there exists a $\overline{Z}$-CT $\F{T} = (V,E,L,A)$ that converges with probability
    $< 1$ and starts with $(1:t)$ where $\posT(t) = \{\varepsilon\}$ and $\flat(t) = s
    \theta$ for a substitution $\theta$ and an ADP $s \to \ldots \in \overline{Z}$.
    
    If $s \to \ldots \notin \bigcup_{X \in \hat{X}} X \cup \bigcup_{Y \in \hat{Y}} Y$, then the resulting terms contain no annotations anymore and this cannot start a CT that converges with probability $<1$.
    W.l.o.g., we may assume that the ADP that is used for the rewrite step at
    the root is in $\bigcup_{X \in \hat{X}} X$.
    Otherwise, we simply swap $\bigcup_{X \in \hat{X}} X$ with $\bigcup_{Y \in \hat{Y}} Y$ in the following.
  
    We can partition the set $A$ of our $\overline{Z}$-CT $\F{T}$ into the sets
    \begin{itemize}
      \item[$\bullet$] $A_1 := \{x \in A \mid x$ together with the labeling and its successors represents a step with an ADP from $\bigcup_{X \in \hat{X}} X\}$
      \item[$\bullet$] $A_2 := A \setminus A_1$
    \end{itemize}
    Note that in the case of $x \in A_2$, we know that $x$ together with its successors and the labeling represents a step with an ADP from $\PP \setminus \bigcup_{X \in \hat{X}} X$.
    We know that every $\overline{\bigcup_{Y \in \hat{Y}} Y}$-CT converges with probability $1$, since $\overline{\bigcup_{Y \in \hat{Y}} Y}$ is $\mathtt{AST}$\@.
    Thus, also every $\overline{\bigcup_{Y \in \hat{Y}} Y \setminus \bigcup_{X \in \hat{X}} X}$-CT converges with probability $1$ (as it contains fewer annotations than $\overline{\bigcup_{Y \in \hat{Y}} Y}$).
    Furthermore, we have $|\F{T}|_{\ctleaf} < 1$ by our assumption.
    By the A-partition lemma (\cref{lemma:p-partition}) we can find a grounded sub $\overline{Z}$-CT $\F{T}' = (V',E',L',A')$ with $|\F{T}'|_{\ctleaf} < 1$ such that every infinite path has an infinite number of $A_1$-edges.
    Since $\F{T}'$ is a grounded sub-CT of $\F{T}$ it must also start with $(1:t)$.
  
    We now construct a $\overline{\bigcup_{X \in \hat{X}} X}$-CT $\F{T}'' = (V',E',L'',A'')$ 
    with $A_1 \cap A' \subseteq A''$
    that has the same underlying tree structure and adjusted labeling
    such that all nodes get the same probabilities as in $\F{T}'$.
    Since the tree structure and the probabilities are the same, we then obtain $|\F{T}'|_{\ctleaf} = |\F{T}''|_{\ctleaf}$.
    To be precise, the set of leaves in $\F{T}'$ is equal to the set of leaves in $\F{T}''$, and every leaf has the same probability.
    Since $|\F{T}'|_{\ctleaf} < 1$ we thus have $|\F{T}''|_{\ctleaf} < 1$, which is a contradiction to our
    assumption that $\overline{\bigcup_{X \in \hat{X}} X}$ is $\mathtt{AST}$\@.
  
    \vspace*{-10px}
    \begin{center}
      \small
      \begin{tikzpicture}
        \tikzstyle{adam}=[thick,draw=black!100,fill=white!100,minimum size=4mm, shape=rectangle split, rectangle split parts=2,rectangle split
        horizontal]
        \tikzstyle{empty}=[rectangle,thick,minimum size=4mm]
        
        \node[adam,pin={[pin distance=0.1cm, pin edge={,-}] 145:\tiny \textcolor{blue}{$A_1$}}] at (-3.5, 0)  (a) {$1$ \nodepart{two} $t$};
        \node[adam] at (-5, -0.8)  (b) {$p_1$ \nodepart{two} $t_{1}'$};
        \node[adam,pin={[pin distance=0.1cm, pin edge={,-}] 35:\tiny \textcolor{blue}{$A_2$}}] at (-2, -0.8)  (c) {$p_2$ \nodepart{two} $t_{2}'$};
        \node[adam,pin={[pin distance=0.1cm, pin edge={,-}] 145:\tiny \textcolor{blue}{$A_1$}}] at (-6, -1.6)  (d) {$p_3$ \nodepart{two} $t_3'$};
        \node[adam,pin={[pin distance=0.1cm, pin edge={,-}] 35:\tiny \textcolor{blue}{$A_2$}}] at (-4, -1.6)  (e) {$p_4$ \nodepart{two} $t_4'$};
        \node[adam,pin={[pin distance=0.1cm, pin edge={,-}] 145:\tiny \textcolor{blue}{$A_1$}}] at (-2, -1.6)  (f) {$p_5$ \nodepart{two} $t_5'$};
        \node[empty] at (-6, -2.4)  (g) {$\ldots$};
        \node[empty] at (-4, -2.4)  (h) {$\ldots$};
        \node[empty] at (-2, -2.4)  (i) {$\ldots$};
  
        \node[empty] at (-0.5, -1)  (arrow) {\Huge $\leadsto$};
        
        \node[adam,pin={[pin distance=0.1cm, pin edge={,-}] 145:\tiny \textcolor{blue}{$A_1$}}] at (3.5, 0)  (a2) {$1$ \nodepart{two} $t$};
        \node[adam] at (2, -0.8)  (b2) {$p_1$ \nodepart{two} $t''_{1}$};
        \node[adam] at (5, -0.8)  (c2) {$p_2$ \nodepart{two} $t''_{2}$};
        \node[adam,pin={[pin distance=0.1cm, pin edge={,-}] 145:\tiny \textcolor{blue}{$A_1$}}] at (1, -1.6)  (d2) {$p_3$ \nodepart{two} $t''_3$};
        \node[adam] at (3, -1.6)  (e2) {$p_4$ \nodepart{two} $t''_4$};
        \node[adam,pin={[pin distance=0.1cm, pin edge={,-}] 145:\tiny \textcolor{blue}{$A_1$}}] at (5, -1.6)  (f2) {$p_5$ \nodepart{two} $t''_5$};
        \node[empty] at (1, -2.4)  (g2) {$\ldots$};
        \node[empty] at (3, -2.4)  (h2) {$\ldots$};
        \node[empty] at (5, -2.4)  (i2) {$\ldots$};
      
        \draw (a) edge[->] (b);
        \draw (a) edge[->] (c);
        \draw (b) edge[->] (d);
        \draw (b) edge[->] (e);
        \draw (c) edge[->] (f);
        \draw (d) edge[->] (g);
        \draw (e) edge[->] (h);
        \draw (f) edge[->] (i);
  
        \draw (a2) edge[->] (b2);
        \draw (a2) edge[->] (c2);
        \draw (b2) edge[->] (d2);
        \draw (b2) edge[->] (e2);
        \draw (c2) edge[->] (f2);
        \draw (d2) edge[->] (g2);
        \draw (e2) edge[->] (h2);
        \draw (f2) edge[->] (i2);
      \end{tikzpicture}
    \end{center}
    \vspace*{-10px}
    The core idea of this construction is that annotations introduced by rewrite steps at a node $x \in A_2$ are not
    important for our computation.
    The reason is that if annotations are introduced using an ADP from 
    $\bigcup_{Y \in \hat{Y}} Y$ that is not contained in $\bigcup_{X \in \hat{X}} X$,
    then by the prerequisite of~\eqref{Composing AST}, we know that such an ADP 
    has no path in the dependency graph to an ADP in $\bigcup_{X \in \hat{X}} X$.
    Hence, by definition of the dependency graph, we are never able to use these terms 
    for a rewrite step with an ADP from $\bigcup_{X \in \hat{X}} X$ to introduce new annotations.
    We can therefore apply the non-annotated ADP from $\bigcup_{Y \in \hat{Y}} Y$ to perform the rewrite step.
  
    We now construct the new labeling $L''$ for the $\overline{\bigcup_{X \in \hat{X}} X}$-CT $\F{T}''$ recursively.
    Let $Q \subseteq V$ be the set of nodes where we have already defined the labeling $L''$.
    Furthermore, for any term $t'_x$, let $\Junk_{\hat{X}}(t'_x)$ denote the positions of all annotated subterms $s \trianglelefteq_{\#} t'_x$ 
    that can never be used for a rewrite step with an ADP from $\hat{X}$, as indicated by the dependency graph.
    To be precise, we define $\pi \in \Junk_{\hat{X}}(t'_x)$:$\Leftrightarrow$ there
    is no $A \in W$ with $A >_{\F{G}}^* X$
    for some $X \in \hat{X}$ such that there is
    an ADP $\ell \to \{p_1: r_1, \ldots, p_k: r_k\}^{m} \in A$, and a substitution $\sigma$ with $\annoEps(t_x'|_{\pi}) \to_{\nonprob(\PP)}^* \ell^\# \sigma$.
    During our construction, we ensure that the following property holds:
    \begin{equation}\label{dep-graph-construction-induction-hypothesis}
      \parbox{.9\textwidth}{For every $x \in Q$ we have $t'_x \doteq t''_x$ and 
      $\posT(t'_x) \setminus \Junk_{\hat{X}}(t'_x) \subseteq \posT(t''_{x})$.}
    \end{equation}
  
    We start by setting $t''_v = t'_v$ for the root $v$ of $\F{T}'$.
    Here, our property~\eqref{dep-graph-construction-induction-hypothesis} is clearly satisfied.
    As long as there is still an inner node $x \in Q$ such that its successors are not contained in $Q$, we do the following.
    Let $xE = \{y_1, \ldots, y_k\}$ be the set of its successors.
    We need to define the corresponding terms for the nodes $y_1, \ldots, y_k$ in $\F{T}''$.
    Since $x$ is not a leaf and $\F{T}'$ is a $\overline{Z}$-CT, we have $t'_x \tored{}{}{\overline{Z}} \{\tfrac{p_{y_1}}{p_x}:t'_{y_1}, \ldots, \tfrac{p_{y_k}}{p_x}:t'_{y_k}\}$,
    and hence, we have to deal with the following two cases:\linebreak
    \vspace*{-0.4cm}
    \begin{enumerate}
        \item If we use an ADP from $\bigcup_{X \in \hat{X}} X$ in $\F{T}'$, then we perform
        the rewrite step with the same ADP, the same VRF $(\varphi_j)_{1 \leq j \leq k}$, the same position $\pi$, and the same substitution in $\F{T}''$.
        Since we have $t'_x \doteq_{(IH)} t''_x$, we also get $t'_{y_j} \doteq t''_{y_j}$ for all $1 \leq j \leq k$.
        Furthermore, since we rewrite at position $\pi$ it cannot be in $\Junk_{\hat{X}}(t'_x)$, and hence, 
        if $\pi \in \posT(t'_x)$, then also $\pi \in \posT(t''_{x})$ by \eqref{dep-graph-construction-induction-hypothesis}.
        Thus, whenever we create annotations in the rewrite step in $\F{T}'$ (a step with
        $(\mathbf{af})$ or $(\mathbf{at})$), then we do the same in $\F{T}''$ (the step is also an $(\mathbf{af})$ or $(\mathbf{at})$ step, respectively), and whenever we remove annotations in the rewrite step in $\F{T}''$ (a step with $(\mathbf{af})$ or $(\mathbf{nf})$), then the same happened in $\F{T}'$ (the step is also an $(\mathbf{af})$ or $(\mathbf{nf})$ step).
        Therefore, we also get $\posT(t'_{y_j}) \setminus \Junk_{\hat{X}}(t'_{y_j}) \subseteq \posT(t''_{y_j})$ for all $1 \leq j \leq k$ and \eqref{dep-graph-construction-induction-hypothesis} is again satisfied.
        \item If we use an ADP from $\PP \setminus \bigcup_{X \in \hat{X}} X$ in $\F{T}'$, and we use the ADP $\ell \to \{p_1:r_1, \ldots, p_k:r_k\}^{m}$, 
        then we can use $\ell \to \{p_1:\flat(r_1), \ldots, p_k:\flat(r_k)\}^{m}$ instead, with the same VRF $(\varphi_j)_{1 \leq j \leq k}$, the same position $\pi$, and the same substitution.
        Note that if $\pi \in \posT(t'_x)$, then all the annotations introduced by the ADP are in $\Junk_{\hat{X}}(t'_{y_j})$ for all $1 \leq j \leq k$, since the used ADP is not in $\bigcup_{X \in \hat{X}} X$ and by \eqref{Composing AST} we cannot use another ADP to create a path in the dependency graph to a node in $\bigcup_{X \in \hat{X}} X$ again.
        Otherwise, we remove the annotations during the application of the rule anyway.
        Again, \eqref{dep-graph-construction-induction-hypothesis} is satisfied.
    \end{enumerate}
    We have now shown that~\eqref{Composing AST} holds.
  
    \medskip
  
    \noindent
    \textbf{\underline{3. For every $X \in W$, the ADP problem $\overline{\bigcup_{X >_{\F{G}}^* Y}Y}$ is $\mathtt{AST}$\@.}}
  
    \noindent
    Using~\eqref{W is AST} and~\eqref{Composing AST}, by induction on $>_{\F{G}}$ we now prove that
    \begin{equation}
      \label{SCC induction} \mbox{for every $X \in W$, the ADP problem $\overline{\bigcup_{X >_{\F{G}}^* Y}Y}$ is $\mathtt{AST}$\@.}
    \end{equation}
    Note that $>_{\F{G}}$ is well founded, \pagebreak[2] since $\F{G}$ is finite.
    
    For the base case, we consider an $X \in W$ that is minimal w.r.t.\ $>_{\F{G}}$.
    Hence, we have $\bigcup_{X >_{\F{G}}^* Y} Y = X$.
    By~\eqref{W is AST}, $\overline{X}$ is $\mathtt{AST}$\@.
  
    For the induction step, we consider an $X \in W$ and assume that
    $\overline{\bigcup_{Y >_{\F{G}}^* Z} Z}$ is $\mathtt{AST}$ for every $Y \in W$ with $X >_{\F{G}}^+ Y$.
    Let $\mathtt{Succ}(X) = \{Y \in W \mid X >_{\F{G}} Y\} = \{Y_1, \ldots Y_m\}$ be the set of all direct successors of $X$.
    The induction hypothesis states that $\overline{\bigcup_{Y_u >_{\F{G}}^* Z} Z}$ is $\mathtt{AST}$ for all $1 \leq u \leq m$.
    We first prove by induction that for all $1 \leq u \leq m$,
    $\overline{\bigcup_{1 \leq i \leq u} \bigcup_{Y_i >_{\F{G}}^* Z} Z}$ is $\mathtt{AST}$\@.
    
    In the inner induction base, we have $u = 1$ and hence $\overline{\bigcup_{1 \leq i \leq u} \bigcup_{Y_i >_{\F{G}}^* Z} Z} = \overline{\bigcup_{Y_1 >_{\F{G}}^* Z} Z}$.
    By our outer induction hypothesis we know that $\overline{\bigcup_{Y_1 >_{\F{G}}^* Z} Z}$ is $\mathtt{AST}$\@.
    
    In the inner induction step, assume that the claim holds for some $1 \leq u < m$.
    Then $\overline{\bigcup_{Y_{u+1} >_{\F{G}}^* Z} Z}$ is $\mathtt{AST}$ by our outer induction hypothesis and\linebreak $\overline{\bigcup_{1 \leq i \leq u} \bigcup_{Y_{i} >_{\F{G}}^* Z} Z}$ is $\mathtt{AST}$ by our inner induction hypothesis.
    By~\eqref{Composing AST}, we know that then $\overline{\bigcup_{1 \leq i \leq u+1} \bigcup_{Y_{i} >_{\F{G}}^* Z} Z}$ is $\mathtt{AST}$ as well.
    The conditions for~\eqref{Composing AST} are clearly satisfied, as we use the
    reflexive, transitive closure $>_{\F{G}}^*$ of the direct successor relation 
    in both $\bigcup_{1 \leq i \leq u} \bigcup_{Y_{i} >_{\F{G}}^* Z} Z$ and $\bigcup_{Y_{u+1} >_{\F{G}}^* Z} Z$.
    
    Now we have shown that $\overline{\bigcup_{1 \leq i \leq m} \bigcup_{Y_i >_{\F{G}}^* Z} Z}$ is $\mathtt{AST}$\@.
    We know that $\overline{X}$ is $\mathtt{AST}$ by our assumption and that $\overline{\bigcup_{1 \leq i \leq m} \bigcup_{Y_i >_{\F{G}}^* Z} Z}$ is $\mathtt{AST}$\@.
    Hence, by~\eqref{Composing AST} we obtain that $\overline{\bigcup_{X >_{\F{G}}^* Y} Y}$ $\mathtt{AST}$\@.
    Again, the conditions of~\eqref{Composing AST} are satisfied, since $X$ is strictly
    greater w.r.t.\ $>_{\F{G}}^+$ than all $Z$ with $Y_i >_{\F{G}}^* Z$ for some $1 \leq i \leq m$.

    \medskip
  
    \noindent
    \textbf{\underline{4. $\PP$ is $\mathtt{AST}$\@.}}
  
    \noindent
    In~\eqref{SCC induction} we have shown that $\overline{\bigcup_{X >_{\F{G}}^* Y} Y}$ for every $X \in W$ is $\mathtt{AST}$\@.
    Let $X_1, \ldots, X_m\linebreak \in W$ be the maximal elements of $W$ w.r.t.\ $>_{\F{G}}$.
    By induction, one can prove that $\overline{\bigcup_{1 \leq i \leq u} \bigcup_{X_i >_{\F{G}}^* Y}
    Y}$ is $\mathtt{AST}$ for all $1 \leq u \leq m$ by~\eqref{Composing AST}, analogous to the previous induction.
    Again, the conditions of~\eqref{Composing AST} are satisfied as we use
    the reflexive, transitive closure of $>_{\F{G}}$.
    In the end, we know that $\overline{\bigcup_{1 \leq i \leq m} \bigcup_{X_i >_{\F{G}}^* Y} Y} = \PP$ is $\mathtt{AST}$ and this ends the proof.
\end{myproof}

\ProbDepGraphProcBast*

\begin{myproof}
    Let $\overline{X}^{\PP} = X \cup \flat(\PP \setminus X)$ for $X \subseteq \PP$
    and $\overline{X}^{\InI} = X \cup \flat(\InI \setminus X)$ for $X \subseteq \InI \cup \PP$.
    \smallskip
    
    \noindent
    \underline{\emph{Completeness:}} Every $(\overline{\JJ}^{\mathcal{I}} \cup \overline{\PP_i}^{\PP})$-CT is also a $(\mathcal{I} \cup \PP)$-CT with fewer annotations in the terms.
    So if some $(\overline{\mathcal{J}}^{\mathcal{I}}, \overline{\PP_i}^{\PP})$ is not
    $\mathtt{bAST}$, then there exists a $(\overline{\mathcal{J}}^{\mathcal{I}} \cup
    \overline{\PP_i}^{\PP})$-CT $\F{T}$ that converges with probability $<1$ and uses
    $\overline{\mathcal{J}}^{\mathcal{I}}\setminus\overline{\PP_i}^{\PP}$
    only finitely often. 
    By adding annotations to the terms of the tree, we result in an $(\mathcal{I} \cup
    \PP)$-CT that converges with probability $<1$ as well and uses ADPs from
    $\InI\setminus\PP$ only finitely often.
    Hence, if $(\overline{\mathcal{J}}^{\mathcal{I}}, \overline{\PP_i}^{\PP})$ is not
    $\mathtt{bAST}$, then $(\mathcal{I},\PP)$ is not $\mathtt{bAST}$ either. 
    \medskip

    \noindent
    \underline{\emph{Soundness:}}
    Let $(\mathcal{I},\PP)$ be not $\mathtt{bAST}$.
    Then there exists an $(\InI,\PP)$-CT $\F{T}$ that converges with probability
    $< 1$, whose root is labeled with $(1: t^\#)$
    for a basic term $t$, and $\InI\setminus\PP$ is used only finitely often.
    So there exists a depth $H \in \IN$, such that we only use ADPs from $\PP$ and no ADPs from $\InI\setminus\PP$ anymore.
    All the subtrees that start at depth $H$ are $\PP$-CTs and one of them needs to 
    converge with probability $< 1$, since $\F{T}$ converges with probability $< 1$.
    W.l.o.G., let $x$ be the root node of such a subtree that converges with probability $< 1$ (with $x$  at depth $H$).
    We can use the previous proof of \Cref{theorem:prob-DGP}
    to show that then there exists an SCC $\PP_i \subseteq \PP$ and a CT $\F{T}'$ such that
    $\F{T}'$ starts with $(1:t_x)$, where $t_x$ is the term of the node $x$ in $\F{T}$,
    $\F{T}'$ converges with probability $< 1$, and
    $\F{T}'$ is a $\overline{\PP_i}^{\PP}$-CT.
    It remains to show that we can reach the root term $t_x$ of $\F{T}'$
    from a basic term in 
    a $(\overline{\mathcal{J}}^{\mathcal{I}} \cup \overline{\PP_i}^{\PP})$-CT with
    $\mathcal{J} \in \PP_i\!\!\uparrow$.

    Let $\ell \to \mu$ be the ADP used at the root of $\F{T}$.
    Since the root of  $\F{T}$ is $t^\#$ for a basic term $t$, the ADP from $\PP_i$ applied to
    $t_x$ must be reachable from $\ell \to \mu$ in the $(\InI \cup \PP)$-dependency graph.  
    Hence, there exists a $\mathcal{J} \in \PP_i\!\!\uparrow$ such that one can reach $t_x$
    from $t^\#$ by applying only ADPs from $\overline{\mathcal{J}}^{\mathcal{I}} \cup
    \overline{\PP_i}^{\PP}$. Therefore,
    $\F{T}''$ together with the prefix tree that includes the
    path from $t^\#$ to $t_x$ is a 
    $(\overline{\mathcal{J}}^{\mathcal{I}},\overline{\PP_i}^{\PP})$-CT that converges with probability $< 1$.
    Hence, $(\overline{\mathcal{J}}^{\mathcal{I}},\overline{\PP_i}^{\PP})$
    is not $\mathtt{bAST}$.
\end{myproof}

\UsableTermsProc*

\begin{myproof}
   
    \noindent
    \underline{\emph{Completeness:}} 
    We only prove this direction for $\mathtt{AST}$.
    The proof for $\mathtt{bAST}$ is completely analogous.
    Every $\mathcal{T}_\mathtt{UT}(\PP)$-CT is also a $\PP$-CT with fewer annotations in the terms.
    So if $\mathcal{T}_\mathtt{UT}(\PP)$ is not $\mathtt{AST}$, then there exists a $\mathcal{T}_\mathtt{UT}(\PP)$-CT $\F{T}$ that converges with probability $<1$. 
    By adding annotations to the terms of the tree, we result in a $\PP$-CT that converges with probability $<1$ as well.
    Hence, if $\mathcal{T}_\mathtt{UT}(\PP)$ is not $\mathtt{AST}$, then $\PP$ is not $\mathtt{AST}$ either.
    \medskip
   
    \noindent
    \underline{\emph{Soundness:}} 
    Here, the proofs for $\mathtt{AST}$ and $\mathtt{bAST}$ differ slightly, similar to the proofs of \Cref{theorem:prob-DGP} and \Cref{theorem:prob-DGP-bAST}.
    We start with the proof of $\mathtt{AST}$.
    
    Let $\PP$ be not $\mathtt{AST}$.
    Then by \Cref{lemma:starting} there exists a $\PP$-CT $\F{T} = (V,E,L,A)$ that converges with probability
    $< 1$ whose root is labeled with $(1: t)$ and $\posT(t) = \{\varepsilon\}$. 
    We will now create a $\mathcal{T}_\mathtt{UT}(\PP)$-CT $\F{T}' = (V,E,L',A)$, with the same underlying tree structure, and an adjusted labeling such that $p_x^{\F{T}} = p_x^{\F{T}'}$ for all $x \in V$.
    Since the tree structure and the probabilities are the same, we then get $|\F{T}'|_{\ctleaf} = |\F{T}|_{\ctleaf} < 1$, and hence $\mathcal{T}_\mathtt{UT}(\PP)$ is not $\mathtt{AST}$ either.
    
    We construct the new labeling $L'$ for the $\mathcal{T}_\mathtt{UT}(\PP)$-CT $\F{T}'$ recursively.
    Let $X \subseteq V$ be the set of nodes where we have already defined the labeling $L'$.
    During our construction, we ensure that the following property holds for every node $x \in X$:
    \begin{equation}\label{usable-terms-soundness-induction-hypothesis}
        \parbox{.9\textwidth}{For every $x \in X$ we have $t_x \doteq t'_x$ and $\posT(t_x) \setminus \Junk(t_x) \subseteq \posT(t'_x)$.}
    \end{equation}
    Here, for any term $t_x$, let $\Junk(t_x)$ be the set of positions 
    that can never be used for a rewrite step with an ADP that contains annotations.
    To be precise, we define $\pi \in \Junk(t_x)$:$\Leftrightarrow$ there
    is no ADP $\ell \to \{p_1: r_1, \ldots, p_k: r_k\}^{m} \in \PP$ with annotations
    and no substitution $\sigma$ such that $\annoEps(t_x|_{\pi}) \to_{\nonprob(\PP)}^* \ell^\# \sigma$.
  
    We start with the same term $t$ at the root.
    Here, our property~\eqref{usable-terms-soundness-induction-hypothesis} is clearly satisfied.
    As long as there is still an inner node $x \in X$ such that its successors are not contained in $X$, we do the following.
    Let $xE = \{y_1, \ldots, y_k\}$ be the set of its successors.
    We need to define the terms for the nodes $y_1, \ldots, y_k$ in $\F{T}'$.
    Since $x$ is not a leaf and $\F{T}$ is a $\PP$-CT, we have $t_x \tored{}{}{\PP} \{\tfrac{p_{y_1}}{p_x}:t_{y_1}, \ldots, \tfrac{p_{y_k}}{p_x}:t_{y_k}\}$.
    If we performed a step with $\tored{}{}{\PP}$ using the ADP $\ell \to \{ p_1: r_1, \ldots, p_k:r_k\}^{m}$, the VRF $(\varphi_j)_{1 \leq j \leq k}$,
    the position $\pi$, and the substitution $\sigma$ in $\F{T}$,
    then we can use the ADP $\ell \!\to\! 
    \{ p_1: \#_{\Delta_{\PP}(r_1)}(r_1), \ldots, p_k:\#_{\Delta_{\PP}(r_k)}(r_k)\}^{m}$ with the same VRF $(\varphi_j)_{1 \leq j \leq k}$, the same position $\pi$, and the same substitution $\sigma$.
    Now, we directly get $t_{y_j} \doteq t'_{y_j}$ for all $1 \leq j \leq k$.
    To prove $\posT(t_{y_j}) \setminus \Junk(t_{y_j}) \subseteq \posT(t'_{y_j})$,
    note that if $\pi \in \posT(t_x) \cap \Junk(t_x)$,
    then $\ell \to \{ p_1: r_1, \ldots, p_k:r_k\}^{m}$ contains no annotations by definition of $\Junk(t_x)$.
    Therefore, it does not matter whether we rewrite with case $(\mathbf{at})$ or $(\mathbf{nt})$ ($(\mathbf{af})$ or $(\mathbf{nf})$).
    Otherwise, if $\pi \in \posT(t_x) \setminus \Junk(t_x)$, then
    the original rule contains the same terms with possibly more annotations, 
    but all missing annotations are in $\Junk(t_x)$ 
    by definition of $\#_{\Delta_{\PP}(r_j)}(r_j)$.
    Thus, we get $\posT(t_{y_j}) \setminus \Junk(t_{y_j}) \subseteq \posT(t'_{y_j})$ for all $1 \leq j \leq k$.

    Next, we consider $\mathtt{bAST}$.
    Let $(\InI,\PP)$ be not $\mathtt{bAST}$.
    Then there exists an $(\InI,\PP)$-CT $\F{T}$ that converges with probability
    $< 1$, whose root is labeled with $(1: t^\#)$
    for a basic term $t$, and $\InI\setminus\PP$ is used only finitely often.
    So there exists a depth $H \in \IN$, such that we only use ADPs from $\PP$ and no ADPs from $\InI\setminus\PP$ anymore.
    All the subtrees that start at depth $H$ are $\PP$-CTs and one of them needs to 
    converge with probability $< 1$, since $\F{T}$ converges with probability $< 1$.
    W.l.o.G., let $x$ be the root node of such a subtree that converges with probability $< 1$ (with $x$ at depth $H$).
    We can use the previous proof for $\mathtt{AST}$
    to show that then there exists a $\mathcal{T}_\mathtt{UT}(\PP)$-CT $\F{T}'$ such that
    $\F{T}'$ starts with $(1:t_x)$, where $t_x$ is the term of the node $x$ in $\F{T}$,
    and $\F{T}'$ converges with probability $< 1$.
    It remains to show that we can reach the root term $t_x$ of $\F{T}'$
    from a basic term in 
    a $(\mathcal{T}_\mathtt{UT}(\InI \cup \PP),\mathcal{T}_\mathtt{UT}(\PP))$-CT.
    But since we consider both $\InI$ and $\PP$ in $\mathcal{T}_\mathtt{UT}(\InI \cup \PP)$,
    we can use the same construction as before to show that we can reach a term $t_x'$ with
    $t_x \doteq t'_x$ and $\posT(t_x) \setminus \Junk(t_x) \subseteq \posT(t'_x)$, which
    is sufficient.
\end{myproof}

\ProbUsRulesProc*

\begin{myproof}
  Let $\overline{X} =
(X \cap \urules(\mathcal{I} \cup \PP))
  \cup \{\ell \to \mu^{\tfalse} \mid \ell \to \mu^{m} \in X \setminus
\urules(\mathcal{I} \cup \PP)\}$.
    \smallskip
   
    \noindent
    \underline{\emph{Completeness:}} Every $(\overline{\mathcal{I}} \cup \overline{\PP})$-CT is also an $(\mathcal{I} \cup \PP)$-CT with fewer annotations in the terms.
   If $(\overline{\mathcal{I}}, \overline{\PP})$ is not $\mathtt{bAST}$, then there exists
   an $(\overline{\mathcal{I}} \cup \overline{\PP})$-CT $\F{T}$ that converges with
   probability $<1$ and uses ADPs from
   $\overline{\mathcal{I}}\setminus \overline{\PP}$ only finitely often. 
   By adding annotations to the terms of the tree, we result in an
   $(\mathcal{I} \cup \PP)$-CT that converges with probability $<1$ as well
and uses ADPs from $\InI \setminus \PP$ only finitely often.   
    Hence, if $(\overline{\mathcal{I}}, \overline{\PP})$ is not $\mathtt{bAST}$, then $(\mathcal{I}, \PP)$ is not $\mathtt{bAST}$ either.

    \smallskip
   
    \noindent
    \underline{\emph{Soundness:}} Assume that $(\mathcal{I}, \PP)$ is not $\mathtt{bAST}$\@.
    Then there exists an $(\mathcal{I} \cup \PP)$-CT  that converges with probability
    $< 1$, whose root is labeled with $(1: t^\#)$ for a basic term $t$,
        and $\InI\setminus\PP$ is used only finitely often. 
    As $t$ is basic,
    in the first rewrite step at the root of the tree, the substitution only instantiates
    variables of the ADP by normal forms. 
 
    By the definition of usable rules, as in the non-probabilistic case, 
    rules $\ell \to \mu^m \in \mathcal{I} \cup \PP$ that are not usable 
    (i.e., $\ell \to \mu^m \not\in \urules(\mathcal{I} \cup \PP)$) 
    will never be used below an annotated symbol in such an $(\mathcal{I} \cup \PP)$-CT.
    Hence, we can also view $\F{T}$ as an $(\overline{\mathcal{I}} \cup \overline{\PP})$-CT that
    converges with probability $<1$ and thus $(\overline{\mathcal{I}}, \overline{\PP})$ is not $\mathtt{bAST}$\@.
\end{myproof}

In the following, we use the \emph{prefix ordering}
($\pi \leq \tau \Leftrightarrow \text{ there exists } \chi \in \IN^* \text{ such that } \pi.\chi = \tau$)
to compare positions.

\ProbRPP*

\begin{myproof}
    For the proof for $\mathtt{iAST}$, see \cite{FLOPSreport2024}.
    Let $\overline{\PP} = \PP_{\geq} \cup \flat(\PP_{>})$.
    \smallskip
   
    \noindent
    \underline{\emph{Completeness:}} Every $\overline{\PP}$-CT is also a $\PP$-CT with fewer annotations in the terms.
    So if $\overline{\PP}$ is not $\mathtt{AST}$, then there exists a $\overline{\PP}$-CT $\F{T}$ that converges with probability $<1$. 
    By adding annotations to the terms of the tree, we result in a $\PP$-CT that converges with probability $<1$ as well.
    Hence, if $\overline{\PP}$ is not $\mathtt{AST}$, then $\PP$ is not $\mathtt{AST}$ either.

    \smallskip
   
    \noindent
    \underline{\emph{Soundness:}} This proof uses the proof idea for $\mathtt{AST}$ from~\cite{mciver2017new}.
    The core steps of the proof are the following:
	\begin{enumerate}
		\item[(I)] We extend the conditions (1), (2), and (3) to rewrite steps instead of just rules (and thus, to edges of a CT).
		\item[(II)] We create a CT $\F{T}^{\leq N}$ for any $N \in \IN$.
		\item[(III)] We prove that $|\F{T}^{\leq N}|_{\ctleaf} \geq p_{min}^{N}$ for any $N \in \IN$.
		\item[(IV)] We prove that $|\F{T}^{\leq N}|_{\ctleaf}=1$ for any $N \in \IN$.
		\item[(V)] Finally, we prove that $|\F{T}|_{\ctleaf}=1$.
	\end{enumerate}
        Here, $p_{min}$ is the minimal probability occurring in $\PP$.
	Parts (II) to (V) remain completely the same as in \cite{reportkg2023iAST}.
    We only show that we can adjust part (I) to our new rewrite relation for $\mathtt{AST}$.

    \smallskip

    \noindent
    \textbf{\underline{(I) We extend the conditions to rewrite steps instead of just rules}}

    \noindent
    We show that the conditions (1), (2), and (3) of the lemma extend to rewrite steps instead of just rules:
	\begin{enumerate}
		\item[(a)] If $s \to \{ p_1:t_1, \ldots, p_k:t_k \}$ using a rewrite rule $\ell \to \{ p_1:r_1, \ldots, p_k:r_k \}$\linebreak with $\Pol(\ell) \geq \Pol(r_j)$ for some $1 \leq j \leq k$, then we have $Pol(s) \geq Pol(t_j)$.
		\item[(b)] If $a \tored{}{}{\PP} \{ p_1:b_1, \ldots, p_k:b_k \}$ using the ADP $\ell \to \{ p_1:r_1, \ldots, p_k:r_k \}^{m} \in \PP_>$ at a position $\pi \in \posT(a)$, then $\val(a) > \val(b_j)$ for some $1 \leq j \leq k$.
		\item[(c)] If $s \to \{ p_1:t_1, \ldots, p_k:t_k \}$ using a rewrite rule $\ell \to \{ p_1:r_1, \ldots, p_k:r_k \}$\linebreak with $Pol(\ell) \geq \sum_{1 \leq j \leq k} p_j \cdot \Pol(r_j)$, then $Pol(s) \geq \sum_{1 \leq j \leq k} p_j \cdot \Pol(t_j)$.
		\item[(d)] If $a \tored{}{}{\PP} \{ p_1:b_1, \ldots, p_k:b_k \}$ using the ADP $\ell \to \{ p_1:r_1, \ldots, p_k:r_k \}^{m} \in \PP$, then $\val(a) \geq \sum_{1 \leq j \leq k} p_j \cdot \val(b_j)$.
	\end{enumerate}

	\begin{itemize}
    	\item[(a)] 
        In this case, there exist a rule $\ell \to \{ p_1:r_1, \ldots, p_k:r_k \}$ with
        $\Pol(\ell) \geq \Pol(r_j)$ for some $1 \leq j \leq k$, a substitution $\sigma$,
        and a position $\pi$ of $s$ such that $s|_\pi =\ell\sigma$ and
        $t_h = s[r_h \sigma]_\pi$ for all $1 \leq h \leq k$. 
        
        We perform structural induction on $\pi$.
        So in the induction base, let $\pi = \varepsilon$.
        Hence, we have $s = \ell\sigma \to \{ p_1:r_1 \sigma, \ldots, p_k:r_k \sigma\}$.
        By assumption, we have $\Pol(\ell) \geq \Pol(r_j)$ for some $1 \leq j \leq k$.
        As these inequations hold for all instantiations of the occurring variables, for $t_j = r_j\sigma$ we have
        \[ \Pol(s) = \Pol(\ell\sigma) \geq \Pol(r_j\sigma) = \Pol(t_j). \]
        
        In the induction step, we have $\pi = i.\pi'$, $s = f(s_1,\ldots,s_i,\ldots,s_n)$, 
        $f \in \Sigma$, $s_i \to \{ p_1:t_{i,1}, \ldots, p_k:t_{i,k} \}$, and $t_j =
        f(s_1,\ldots,t_{i,j},\ldots,s_n)$ with $t_{i,j} = s_i[r_j\sigma]_{\pi'}$ for all $1 \leq j \leq k$.
        Then by the induction hypothesis we have $Pol(s_i) \geq Pol(t_{i,j})$.
        For $t_j = f(s_1,\ldots,t_{i,j},\ldots,s_n)$ we obtain
        \[
        \begin{array}{lcl}
            \Pol(s) & = & \Pol(f(s_1,\ldots,s_i,\ldots,s_n)) \\
                & = & f_{\Pol}(\Pol(s_1),\ldots,\Pol(s_i),\ldots,\Pol(s_n)) \\
                & \geq & f_{\Pol}(\Pol(s_1),\ldots,\Pol(t_{i,j}),\ldots,\Pol(s_n)) \\
                &  & \hspace*{1cm} \text{ \textcolor{blue}{(by weak monotonicity of $f_{\Pol}$ and $Pol(s_i) \geq Pol(t_{i,j})$)}} \\
            & = & \Pol(f(s_1,\ldots,t_{i,j},\ldots,s_n)) \\
                & = & \Pol(t_j).
        \end{array}
        \]

	    \item[(b)] In this case, there exist an ADP $\ell \to \{ p_1:r_1, \ldots, p_k:r_k \}^{m} \in \PP_>$, 
	    a VRF $(\varphi_j)_{1 \leq j \leq k}$, a substitution $\sigma$, 
        and position $\pi \in \posT(a)$ with $\flat(a|_{\pi}) = \ell \sigma$ and $b_j \doteq a[r_j \sigma]_{\pi}$.
        First, assume that $m = \ttrue$.
        Let $I_1 = \{\tau \in \posT(a) \mid \tau < \pi\}$ be the set of positions of all
        annotations strictly above $\pi$, $I_2 = \{\tau \in \posT(a) \mid \gamma \in
        \pos_{\VSet}(\ell), \pi < \tau \leq \pi.\gamma\}$ be the set of positions of all
        annotations inside the left-hand side $\ell$ of the used redex $\ell \sigma$ (but
        not on the root of $\ell$),
        $I_3 = \{\tau \in \posT(a) \mid \gamma \in \pos_{\VSet}(\ell), \pi.\gamma < \tau\}$ be the set of positions of all annotations inside the substitution, and let $I_4 = \{\tau \in \posT(a) \mid \tau \bot \pi\}$ be the set of positions of all annotations orthogonal to $\pi$.
        Furthermore, for each
        $\tau \in I_1$ let $\kappa_\tau$ be the position such that $\tau.\kappa_\tau = \pi$,
        and for each $\tau \in I_3$ let $\gamma_\tau$ and $\rho_\tau$ be the positions such that 
        $\gamma_\tau \in \pos_{\VSet}(\ell)$ and $\pi.\gamma_\tau.\rho_\tau = \tau$.
        By Requirement (2), there exists a $1 \leq j \leq k$ with $\Pol(\ell^\#) > \val(r_j) = \sum_{s
        \trianglelefteq_{\#} r_j} \Pol(s^\#)$ and, additionally, $\Pol(\ell) \geq \Pol(\flat(r_j))$ since $m = \ttrue$.
        As these inequations hold for all instantiations of the occurring variables, we
        have

{\scriptsize\[
            \begin{array}{lcl}
            \val(a) & = & \sum_{s \trianglelefteq_{\#} a} \Pol(s^\#) \\
                & = & \Pol(\annoEps(s|_{\pi})) + \sum_{\tau \in I_1}
            \Pol(\annoEps(a|_{\tau})) +
            \sum_{\tau \in I_2} \Pol(\annoEps(a|_{\tau})) + \sum_{\tau \in I_3} \Pol(\annoEps(a|_{\tau}))\\
                && \; + \sum_{\tau \in I_4} \Pol(\annoEps(a|_{\tau}))\\
            & \geq & \Pol(\annoEps(s|_{\pi})) + \sum_{\tau \in I_1}
            \Pol(\annoEps(a|_{\tau})) +  \sum_{\tau \in I_3} \Pol(\annoEps(a|_{\tau}))\\
                && \; + \sum_{\tau \in I_4} \Pol(\annoEps(a|_{\tau}))\\
                    && \hspace*{.5cm} \text{\textcolor{blue}{(removed $I_2$)}}\\
            & = & \Pol(\annoEps(\ell) \sigma) + \sum_{\tau \in I_1}
            \Pol(\annoEps(a|_{\tau})) +  \sum_{\tau \in I_3} \Pol(\annoEps(a|_{\tau}))\\
                && \; + \sum_{\tau \in I_4} \Pol(\annoEps(a|_{\tau}))\\
                    && \hspace*{.5cm} \text{\textcolor{blue}{(as $\annoEps(s|_{\pi}) =
                \annoEps(\ell) \sigma$)}} \\ 
            & > & \sum_{s \trianglelefteq_{\#} r_j} \Pol(\annoEps(s)\sigma) + \sum_{\tau \in I_1}
            \Pol(\annoEps(a|_{\tau})) +  \sum_{\tau \in I_3} \Pol(\annoEps(a|_{\tau}))\\
            && \; + \sum_{\tau \in I_4} \Pol(\annoEps(a|_{\tau}))\\            
                    && \hspace*{.5cm} \text{\textcolor{blue}{(as $\Pol(\annoEps(\ell)) >
                \sum_{s \trianglelefteq_{\#} r_j} \Pol(\annoEps(s))$, hence
                $\Pol(\annoEps(\ell)\sigma) > \sum_{s \trianglelefteq_{\#} r_j}
                \Pol(\annoEps(s)\sigma)$)}}\\
             & \geq & \sum_{s \trianglelefteq_{\#} r_j\sigma}
            \Pol(\annoEps(s))
            + \sum_{\tau \in I_1} \Pol(\annoEps(a|_{\tau}[r_j \sigma]_{\kappa_{\tau}}))
             \\
                && \; + \sum_{\tau \in I_3} \Pol(\annoEps(a|_{\tau})) + \sum_{\tau \in I_4} \Pol(\annoEps(a|_{\tau})) \\
                                && \hspace*{.5cm} \text{\textcolor{blue}{(by $\Pol(\ell)
                \geq \Pol(r_j)$ and (a))}}\\
           &\geq& \sum_{s \trianglelefteq_{\#} r_j\sigma}
            \Pol(\annoEps(s))
            + \sum_{\tau \in I_1} \Pol(\annoEps(a|_{\tau}[r_j \sigma]_{\kappa_{\tau}}))   \\
                && \;   + \sum_{\tau \in I_3, \varphi_j(\gamma_\tau) \neq \bot}
   \Pol(\annoEps(b_j|_{\pi.\varphi_j(\gamma_\tau).\rho_\tau})) + \sum_{\tau \in I_4} \Pol(\annoEps(a|_{\tau})) \\
                                && \hspace*{.5cm} \text{\textcolor{blue}{(moving $\tau = \pi.\gamma_\tau.\rho_\tau \in I_3$ via the VRF)}}\\
                & = & \sum_{s \trianglelefteq_{\#} b_j} \Pol(s^\#) \\
            & = & \val(b_j)\!    
            \end{array}
        \]}

        \noindent
        In the case $m = \tfalse$, we remove $\sum_{\tau \in I_1}
        \Pol(\annoEps(a|_{\tau}[r_j \sigma]_{\kappa_\tau}))$, so that the inequation remains
        correct. 

	    \item[(c)] In this case, there exist a rule $\ell \to \{ p_1:r_1, \ldots, p_k:r_k \}$ with $\Pol(\ell) \geq \sum_{1 \leq j \leq k} p_j \cdot \Pol(r_j)$, a substitution $\sigma$, and a position $\pi$ of $s$ such that $s|_\pi =\ell\sigma$, and $t_j = s[r_j \sigma]_\pi$ for all $1 \leq j \leq k$.
        
        We perform structural induction on $\pi$.
        So in the induction base $\pi = \varepsilon$ we have $s = \ell\sigma \to \{ p_1:r_1\sigma, \ldots, p_k:r_k\sigma \}$.
        As $\Pol(\ell) \geq \sum_{1 \leq j \leq k} p_j \cdot \Pol(r_j)$ holds for all instantiations of the occurring variables, for $t_j = r_j\sigma$ we obtain
        \[
            \Pol(s) \;=\; \Pol(\ell\sigma) \;\geq\;\sum_{1 \leq j \leq k} p_j \cdot \Pol(r_j\sigma) \;=\; \sum_{1 \leq j \leq k} p_j \cdot \Pol(t_j). 
        \]
        
        In the induction step, we have $\pi = i.\pi'$, $s = f(s_1,\ldots,s_i,\ldots,s_n)$,
        $s_i \to \{ p_1:t_{i,1}, \ldots, p_k:t_{i,k} \}$, and $t_j =
        f(s_1,\ldots,t_{i,j},\ldots,s_n)$
        with $t_{i,j} = s_i[r_j\sigma]_{\pi'}$ for all $1 \leq j \leq k$.
        Then by the induction hypothesis we have $\Pol(s_i) \geq \sum_{1 \leq j \leq k} p_j \cdot \Pol(t_{i,j})$.
        Thus, we have
        \[
            \begin{array}{lcl}
            \Pol(s) & = & \Pol(f(s_1,\ldots,s_i,\ldots,s_n)) \\
                & = & f_{\Pol}(\Pol(s_1),\ldots,\Pol(s_i),\ldots,\Pol(s_n)) \\
                & \geq & f_{\Pol}(\Pol(s_1),\ldots,\sum_{1 \leq j \leq k} p_j \cdot \Pol(t_{i,j}),\ldots,\Pol(s_n)) \\
                &   & \; \text{\textcolor{blue}{(by weak monotonicity of $f_{\Pol}$ and $\Pol(s_i) \geq \sum_{1 \leq j \leq k} p_j \cdot \Pol(t_{i,j})$)}} \\
                & = & \sum_{1 \leq j \leq k} p_j \cdot f_{\Pol}(\Pol(s_1),\ldots,\Pol(t_{i,j}),\ldots,\Pol(s_n)) \\
                &   & \; \text{\textcolor{blue}{(as $f_{\Pol}$ is multilinear)}} \\
                & = & \sum_{1 \leq j \leq k} p_j \cdot \Pol(f(s_1,\ldots,t_{i,j},\ldots,s_n))\\
                & = & \sum_{1 \leq j \leq k} p_j \cdot
                \Pol(t_j).
            \end{array}
        \]

	    \item[(d)] In this case, there exist an ADP $\ell \to \{ p_1:r_1, \ldots, p_k:r_k \}^{m} \in \PP$, 
        a substitution $\sigma$, and position $\pi$ with $\flat(a|_{\pi}) = \ell \sigma$ and $b_j \doteq a[r_j \sigma]_{\pi}$.
        First, assume that $m = \ttrue$ and $\pi \in \posT(a)$.
        Let $I_1 = \{\tau \in \posT(a) \mid \tau < \pi\}$ be the set of positions of all
        annotations strictly above $\pi$, $I_2 = \{\tau \in \posT(a) \mid \gamma \in
        \pos_{\VSet}(\ell), \pi < \tau \leq \pi.\gamma\}$ be the set of positions of all
        annotations inside the left-hand side $\ell$ of the used redex $\ell \sigma$ (but
        not on the root of $\ell$), $I_3 = \{\tau \in \posT(a) \mid \gamma \in \pos_{\VSet}(\ell), \pi.\gamma < \tau\}$ be the set of positions of all annotations inside the substitution, and let $I_4 = \{\tau \in \posT(a) \mid \tau \bot \pi\}$ be the set of positions of all annotations orthogonal to $\pi$.
        Furthermore, for each $\tau \in I_1$ let $\kappa_\tau$ be the position such that
        $\tau.\kappa_\tau
        = \pi$,
        and for each $\tau \in I_3$ let $\gamma_\tau$ and $\rho_\tau$ be the positions such that 
        $\gamma_\tau \in \pos_{\VSet}(\ell)$ and $\pi.\gamma_\tau.\rho_\tau = \tau$.
        By Requirement (1), we have $\Pol(\annoEps(\ell)) \geq \sum_{1 \leq j \leq k} p_j \cdot \sum_{t \trianglelefteq_{\#} r_j} \Pol(\annoEps(t))$ and by (3) we have $\Pol(\ell) \geq \sum_{1 \leq j \leq k} p_j \cdot \Pol(\flat(r_j))$.
        As these inequations hold for all instantiations of the occurring variables, we have

{\scriptsize 
        \begin{longtable}{lcl}
            $\val(a)$ & $=$ & $\sum_{t \trianglelefteq_{\#} a} \Pol(t^\#)$\\
            & $=$ & $\Pol(\annoEps(a|_{\pi})) + \sum_{\tau \in I_1}
          \Pol(\annoEps(a|_{\tau}))
          + \sum_{\tau \in I_2} \Pol(\annoEps(a|_{\tau})) + \sum_{\tau \in I_3} \Pol(\annoEps(a|_{\tau}))$ \\
            && $\; + \sum_{\tau \in I_4} \Pol(\annoEps(a|_{\tau}))$ \\
            & $\geq$ & $\Pol(\annoEps(a|_{\pi})) + \sum_{\tau \in I_1} \Pol(\annoEps(a|_{\tau})) + \sum_{\tau \in I_3} \Pol(\annoEps(a|_{\tau})) + \sum_{\tau \in I_4} \Pol(\annoEps(a|_{\tau}))$ \\
                && \hspace*{2cm} \text{\textcolor{blue}{(removed $I_2$)}}\\
            & $=$ & $\Pol(\annoEps(\ell)\sigma) + \sum_{\tau \in I_1} \Pol(\annoEps(a|_{\tau})) + \sum_{\tau \in I_3} \Pol(\annoEps(a|_{\tau})) + \sum_{\tau \in I_4} \Pol(\annoEps(a|_{\tau}))$ \\
                && \hspace*{2cm} \text{\textcolor{blue}{(as $a|_{\pi} = \annoEps(\ell) \sigma$)}} \\
            & $\geq$ & $\sum_{1 \leq j \leq k} p_j \cdot \sum_{t \trianglelefteq_{\#} r_j \sigma} \Pol(\annoEps(t)) + \sum_{\tau \in I_1} \Pol(\annoEps(a|_{\tau})) + \sum_{\tau \in I_3} \Pol(\annoEps(a|_{\tau}))$ \\
            && $\; + \sum_{\tau \in I_4} \Pol(\annoEps(a|_{\tau}))$ \\
                &  &  \hspace*{2cm} \text{\textcolor{blue}{(by $\Pol(\annoEps(\ell)) \geq \sum_{1 \leq j \leq k} p_j \cdot \sum_{t \trianglelefteq_{\#} r_j} \Pol(\annoEps(t))$,}}  \\
                &  &  \hspace*{3cm} \text{\textcolor{blue}{hence $\Pol(\annoEps(\ell)\sigma) \geq \sum_{1 \leq j \leq k} p_j \cdot \sum_{t \trianglelefteq_{\#} r_j \sigma} \Pol(\annoEps(t))$)}}  \\
            & $\geq$ & $\sum_{1 \leq j \leq k} p_j \cdot \sum_{t \trianglelefteq_{\#} r_j \sigma} \Pol(\annoEps(t)) + \sum_{\tau \in I_1} \sum_{1 \leq j \leq k} p_j \cdot \Pol(\annoEps(a|_{\tau}[r_j \sigma]_{\kappa_\tau}))$\\
            &  & $+ \sum_{\tau \in I_3} \Pol(\annoEps(a|_{\tau})) + \sum_{\tau \in I_4} \Pol(\annoEps(a|_{\tau}))$\\
                &  &  \hspace*{2cm} \text{\textcolor{blue}{(by $\Pol(\ell) \geq \sum_{1 \leq j \leq k} p_j \cdot \Pol(r_j)$ and (c))}} \\
            & $=$ & $\sum_{1 \leq j \leq k} p_j \cdot \sum_{t \trianglelefteq_{\#} r_j \sigma} \Pol(\annoEps(t)) + \sum_{1 \leq j \leq k} \sum_{\tau \in I_1} p_j \cdot \Pol(\annoEps(a|_{\tau}[r_j \sigma]_{\kappa_\tau}))$\\
            &  & $+ \sum_{\tau \in I_3} \Pol(\annoEps(a|_{\tau})) + \sum_{\tau \in I_4} \Pol(\annoEps(a|_{\tau}))$\\
            & $=$ & $\sum_{1 \leq j \leq k} p_j \cdot \sum_{t \trianglelefteq_{\#} r_j \sigma} \Pol(\annoEps(t)) + \sum_{1 \leq j \leq k} p_j \cdot \sum_{\tau \in I_1} \Pol(\annoEps(a|_{\tau}[r_j \sigma]_{\kappa_\tau}))$\\
            &  & $+ \sum_{\tau \in I_3} \Pol(\annoEps(a|_{\tau})) + \sum_{\tau \in I_4} \Pol(\annoEps(a|_{\tau}))$\\
            & $=$ & $\sum_{1 \leq j \leq k} p_j \cdot \sum_{t \trianglelefteq_{\#} r_j \sigma} \Pol(\annoEps(t)) + \sum_{1 \leq j \leq k} p_j \cdot \sum_{\tau \in I_1} \Pol(\annoEps(a|_{\tau}[r_j \sigma]_{\kappa_\tau}))$\\
            &  & $+ \sum_{1 \leq j \leq k} p_j \cdot \sum_{\tau \in I_3} \Pol(\annoEps(a|_{\tau})) + \sum_{1 \leq j \leq k} p_j \cdot \sum_{\tau \in I_4} \Pol(\annoEps(a|_{\tau}))$\\
            & $=$ & $\sum_{1 \leq j \leq k} p_j \cdot \big(\sum_{t \trianglelefteq_{\#} r_j \sigma} \Pol(\annoEps(t)) + \sum_{\tau \in I_1} \Pol(\annoEps(a|_{\tau}[r_j \sigma]_{\kappa_\tau}))$ \\
            && $\; + \sum_{\tau \in I_3} \Pol(\annoEps(a|_{\tau})) + \sum_{\tau \in I_4} \Pol(\annoEps(a|_{\tau}))\big)$ \\
            & $\geq$ & $\sum_{1 \leq j \leq k} p_j \cdot \big(\sum_{t \trianglelefteq_{\#} r_j \sigma} \Pol(\annoEps(t)) + \sum_{\tau \in I_1} \Pol(\annoEps(a|_{\tau}[r_j \sigma]_{\kappa_\tau}))$ \\
            && $\; + \sum_{\tau \in I_3, \varphi_j(\gamma_\tau) \neq \bot} \Pol(\annoEps(b_j|_{\pi.\varphi_j(\gamma_\tau).\rho_\tau})) + \sum_{\tau \in I_4} \Pol(\annoEps(a|_{\tau}))\big)$ \\
          &  & \hspace*{2cm} \text{\textcolor{blue}{(moving
              $\tau = \pi.\gamma_\tau.\rho_\tau \in I_3$ via the VRF)}}\\
            & $=$ & $\sum_{1 \leq j \leq k} p_j \cdot \sum_{t \trianglelefteq_{\#} b_j} \Pol(t^\#)$ \\
            & $=$ & $\val(b_j) \!$
        \end{longtable}
        }

        \noindent
        In the case $\pi \notin \posT(a)$, we need to remove
        $\Pol(\annoEps(\ell)\sigma)$ as this annotated subterm does not exist in $a$,
        and therefore also $\sum_{t \trianglelefteq_{\#} r_j \sigma} \Pol(t^\#)$
        in the end, leading to the same result.
        In the case $m = \tfalse$, we additionally remove $\sum_{i \in I_1} \Pol(\annoEps(a|_{\tau}[r_j \sigma]_{\kappa_\tau}))$ in the end.
	\end{itemize}
    The rest is completely analogous to the proof in \cite{kassinggiesl2023iAST}.
\end{myproof}

\ProbRPPBast*

\begin{myproof}
   Analogous to the proof of \Cref{theorem:prob-RPP}.
    We can only use the ADPs from $\InI\setminus\PP$ finitely often,
    hence we can ignore them for the constraints in the theorem.
\end{myproof}

\NPP*

\begin{myproof}
    Let $\PP$ be an ADP problem such that every ADP in
    $\PP$ has the form $\ell \to \{1:r\}^{m}$.
    Note that every $\PP$-chain tree is a single (not necessarily finite) path.
    For such a chain tree $\F{T}$ that is only a single path, we have only two possibilities for $|\F{T}|_{\ctleaf}$.
    If the path is finite, then $|\F{T}|_{\ctleaf} = 1$, since we have a single leaf in this tree with probability $1$.
    Otherwise, we have an infinite path, which means that there is no leaf at all and
    hence $|\F{T}|_{\ctleaf} = 0$. 

\medskip
    
    \noindent
    \underline{\emph{``only if''}}

    \noindent        
    This direction only works for $\mathtt{AST}$.
    Assume that $(\nonprobDP(\PP),\nonprob(\PP))$ is not terminating.
    Then there exists an infinite $(\nonprobDP(\PP),\nonprob(\PP))$-chain
    \[t_0^\# \to_{\nonprobDP(\PP)} \circ \to_{\nonprob(\PP)}^* t_1^\# \to_{\nonprobDP(\PP)} \circ \to_{\nonprob(\PP)}^* t_2^\# \to_{\nonprobDP(\PP)} \circ \to_{\nonprob(\PP)}^* ...\] 
    such that for all $i \in \IN$ we have $t_i^\# = \ell_i^\# \sigma_i$ for
    some dependency pair $\ell_i^\# \to r_i^\# \in \nonprobDP(\PP)$ and some substitution $\sigma_i$.
    From this infinite $(\nonprobDP(\PP),\nonprob(\PP))$-chain, we will now construct an infinite $\PP$-chain tree $\F{T} = (V,E,L,A)$.
    As explained above, we then know that this infinite $\PP$-chain tree must be an
    infinite path, and thus $|\F{T}|_{\ctleaf} = 0$, which means that $\PP$ is not
    $\mathtt{AST}$, and thus, a processor with $\Proc_{\mathtt{PR}}(\PP) = \emptyset$
    would be unsound.
    \begin{center}
        \small
        \begin{tikzpicture}
            \tikzstyle{adam}=[thick,draw=black!100,fill=white!100,minimum size=4mm, shape=rectangle split, rectangle split parts=2,rectangle split
            horizontal]
            \tikzstyle{empty}=[rectangle,thick,minimum size=4mm]
            
            \node[adam,pin={[pin distance=0.1cm, pin edge={,-}] 135:\tiny \textcolor{blue}{$A$}}] at (-2, 0)  (a2) {$1$ \nodepart{two} $t_0^\#$};
            \node[adam] at (0, 0)  (b2) {$1$ \nodepart{two} $a_{1}^{\textcolor{white}{!}}$};
            \node[adam] at (2, 0)  (c2) {$1$ \nodepart{two} $a_{2}^{\textcolor{white}{!}}$};
            \node[empty] at (4, 0)  (d2) {$\ldots$};

            \draw (a2) edge[->] (b2);
            \draw (b2) edge[->] (c2);
            \draw (c2) edge[->] (d2);
        \end{tikzpicture}
    \end{center}
    We start our chain tree with $(1:t_0^\#)$.
    In the non-probabilistic rewrite sequence, we have $t_0^\# \to_{\nonprobDP(\PP)}
    \circ \to_{\nonprob(\PP)}^* t_1^\#$, so there exists a natural number $k \geq
    1$ such that 
    \[\mbox{\small $t_0^\# = \ell_0^\# \sigma_0 \to_{\nonprobDP(\PP)} r_0^\# \sigma_0 = v_1^\#
    \to_{\nonprob(\PP)} v_2^\# \to_{\nonprob(\PP)}
    \ldots \to_{\nonprob(\PP)} v_{k}^\# = t_1^\# = \ell_1^\# \sigma_1$} \]
    Performing the same rewrite steps with $\PP$ yields
    terms $a_1,\ldots,a_k$ such that
    $v_i \trianglelefteq_{\#} a_i$ for all $1 \leq i \leq k$.
    Here, one needs that all ADPs that yield the rules in $\nonprob(\PP)$ have the flag
    $\ttrue$ and thus, the annotations above the redex are not removed.
    For all $1 \leq i \leq k$, we now construct $a_i$ inductively.

    In the induction base ($i = 1$), let 
    $\ell_0 \to \{1: r_0'\}^{m} \in \PP$ be the
    ADP that was used to create the dependency pair $\ell_0^\# \to r_0^\#$ in  $\nonprobDP(\PP)$.
    This means that we have $r_0 \trianglelefteq_{\#} r_0'$.
    Since we have $t_0^\# = \ell_0^\# \sigma_0$, we can also
    rewrite $t_0^\#$ with the ADP $\ell_0 \to \{1:r_0'\}^{m} \in \PP$ and the substitution $\sigma_0$.
    We result in $a_1 = r_0'\sigma_0$ and thus we have $v_1 = r_0
    \sigma_0 \trianglelefteq_{\#} r_0' \sigma_0 = a_1$.

    In the induction step, we assume that we have
    $v_i \trianglelefteq_{\#} a_i$ for some $1 \leq i < k$.
    Let $\pi$ be the annotated position of $a_i$ where $v_i = \flat(a_i)|_\pi$.
    In our non-probabilistic rewrite sequence we have
    $v_i^\# \to_{\nonprob(\PP)} v_{i+1}^\#$
    using a rule $\ell' \to \flat(r') \in \nonprob(\PP)$ and substitution $\delta_i$ at
    a position $\tau \in \IN^+$
    such that $v_i^\#|_{\tau} = \ell' \delta_i$ and $v_{i+1}^\# = v_i^\#[\flat(r') \delta_i]_{\tau}$.
    We can mirror this rewrite step with the ADP $\ell' \to \{1:r'\}^{\ttrue} \in \PP$,
    since by construction we have $v_i \trianglelefteq_{\#} a_i$ and $v_i = \flat(a_i)|_\pi$. 
    We obtain $a_i \tored{}{}{\PP} a_{i+1} = a_i[\anno_{X}(r' \delta_i)]_{\pi.\tau}$ with $X = \Phi_1$ (step with $(\mathbf{at})$) or $X = \Psi_1$ (step with $(\mathbf{nt})$) by rewriting the subterm
    of $a_i$ at position $\pi.\tau$, which implies $v_{i+1} = v_i[\flat(r') \delta_i]_{\tau} \trianglelefteq_{\#}
    a_i[\anno_{X}(r' \delta_i)]_{\pi.\tau} = a_{i+1}$.

    At the end of this induction, we result in $a_{k}$.
    Next, we can mirror the step $t_1^\# \to_{\nonprobDP(\PP)} \circ \to_{\nonprob(\PP)}^*
    t_2^\#$ from our non-probabilistic rewrite sequence with the same construction, etc.
    This results in an infinite $\PP$-chain tree.
    To see that this is indeed a $\PP$-chain tree, note that all the
    local properties are satisfied since every edge represents a rewrite step with $\tored{}{}{\PP}$.
    The global property is also satisfied since,
    in an infinite $(\nonprobDP(\PP),\nonprob(\PP))$-chain, we use
    an infinite number of steps with  $\to_{\nonprobDP(\PP)}$ 
    so that our resulting chain tree has an infinite number of nodes in $A$.
    \medskip

    \noindent
    \underline{\emph{``if''}}

    \noindent	
    This direction works analogously
    for both $\mathtt{AST}$ and $\mathtt{bAST}$, and we prove it only for $\mathtt{AST}$.
    For $\mathtt{bAST}$ we simply ignore the reachability component $\mathcal{I}$.
    Assume that $\PP$ is not $\mathtt{AST}$, i.e., that the processor
    $\Proc_{\mathtt{PR}}(\PP) = \emptyset$ is unsound.
    By \cref{lemma:starting}, there exists a $\PP$-chain tree $\F{T} =
    (V,E,L,A)$ that converges with probability $<1$ and starts with
    $(1:t^\#)$ such that $t^\# = \ell^\# \sigma_0$ for some
    substitution $\sigma_0$ and an ADP $\ell \to \{1:r\}^{m} \in \PP$.
    As explained above, this tree must be an infinite path.
    \begin{center}
        \centering
        \small
            \begin{tikzpicture}
                \tikzstyle{adam}=[thick,draw=black!100,fill=white!100,minimum size=4mm, shape=rectangle split, rectangle split parts=2,rectangle split
                horizontal]
                \tikzstyle{empty}=[rectangle,thick,minimum size=4mm]
    
                \node[adam,pin={[pin distance=0.1cm, pin edge={,-}] 135:\tiny \textcolor{blue}{$A$}}] at (-2, 0)  (a2) {$1$ \nodepart{two} $t^\#$};
                \node[adam] at (0, -0)  (b2) {$1$ \nodepart{two} $a_{1}^{\textcolor{white}{!}}$};
                \node[adam] at (2, 0)  (c2) {$1$ \nodepart{two} $a_{2}^{\textcolor{white}{!}}$};
                \node[empty] at (4, 0)  (d2) {$\ldots$};
    
                \draw (a2) edge[->] (b2);
                \draw (b2) edge[->] (c2);
                \draw (c2) edge[->] (d2);
            \end{tikzpicture}
        \end{center}
    From $\F{T}$, we will now construct an infinite $(\nonprobDP(\PP),\nonprob(\PP))$-chain, which shows that $(\nonprobDP(\PP),\nonprob(\PP))$ is not terminating.
    We start our infinite chain with the term $t_0^\# = t^\#$.
    We have $t^\# \tored{}{}{\PP} \{1:a_1\}$,
    where $a_{1} = \anno_{\Phi_1}(r\sigma_0) = r \sigma_0$.

    There must be a term $r_0 \trianglelefteq_{\#} r$ (i.e.,
      $r_0^\# \sigma_0
    \trianglelefteq
    r \sigma_0$) 
    such that if we replace $a_{1} = r \sigma_0$ with
    $r_0^\# \sigma_0$ and obtain the same $\PP$-chain tree except when we
    would rewrite terms that do not exist anymore (i.e., we ignore these rewrite steps),
    then we still end up in an infinite number of nodes in $A$ (otherwise, $\F{T}$
    would not have an infinite number of nodes in $A$).
    
    Let $\pi$ be the annotated position of $r \sigma_0$ where $r_0 \sigma_0 = \flat(r \sigma_0)|_\pi$.
    We can rewrite the term $t_0^\#$ with the dependency pair $\ell^\# \to r_0^\# \in \nonprobDP(\PP)$, using the substitution $\sigma_0$ since $t_0 = t = \ell \sigma_0$.
    Hence, we result in $t_0^\# \to_{\nonprobDP(\PP)} r_0^\# \sigma_0 = v_1^\#$.
    Next, we mirror the rewrite steps from the $\PP$-chain tree that are performed
    strictly below the root of
    $r_0^\# \sigma_0$ with $\nonprob(\PP)$ until we would rewrite at the root of $r_0^\# \sigma_0$.
    With this construction, we ensure that each the term $v_i^\#$ in our $(\nonprobDP(\PP),\nonprob(\PP))$-chain
    satisfies $v_i \trianglelefteq_{\#} a_i$ and $v_i = \flat(a_i)|_{\pi}$.
    A rewrite step at position $\pi.\tau$ in $a_i$ with $\PP$ is then mirrored with $\nonprob(\PP)$
    in $v_i^\#$ at position $\tau$.
    Note that we only use a finite number of $\nonprob(\PP)$ steps until we rewrite at the root.

 So eventually,  we result in a term $t_1^\# = v_{k}^\#$ with $v_k = \flat(a_k)|_{\pi}$,
    and we rewrite at position $\pi$ in the $\PP$-chain tree.
    We mirror the step $a_k \tored{}{}{\PP} a_{k+1}$
    with $\nonprobDP(\PP)$ and then use the same construction again until we reach term $t_2^\#$, etc.
    This construction creates a sequence $t_0^\#, t_1^\#, \ldots$ of terms such that 
    \[t_0^\# \; \to_{\nonprobDP(\PP)} r_0^\#\sigma_0 \;  \to_{\nonprob(\PP)}^* \; t_1^\# \;
    \to_{\nonprobDP(\PP)} r_1^\#\sigma_1\;  \to_{\nonprob(\PP)}^*\; \ldots \]
    Therefore, $(\nonprobDP(\PP),\nonprob(\PP))$ is not terminating.
\end{myproof}

\begin{example}
    A counterexample for the ``only if'' part of \Cref{theorem:prob-NPP}
    for $\mathtt{bAST}$ is the basic ADP problem $(\emptyset, \PP)$
    with $\PP$ containing the ADPs $\tf(\ta,\tb,x) \to
    \{1:\tF(x,x,x)\}^{\ttrue}$,
    $\th(x,y) \to \{1:x\}^{\ttrue}$, and $\th(x,y) \to \{1:y\}^{\ttrue}$,
    which is based on the well-known TRS from \cite{DBLP:journals/ipl/Toyama87}.
    It is $\mathtt{bAST}$, but the DP problem $(\mathcal{D}, \R)$ with 
    $\mathcal{D} = \{\tF(\ta,\tb,x) \to \tF(x,x,x)\}$ and $\R = \{\th(x,y) \to x, \th(x,y) \to y\}$ 
    considers termination w.r.t.\ 
    arbitrary start terms again, hence it is non-terminating.
\end{example}
}

\end{document}